\documentclass[Afour,sagev,times]{sagej}

\usepackage{moreverb,url}

\usepackage[colorlinks,bookmarksopen,bookmarksnumbered,citecolor=red,urlcolor=red]{hyperref}
\usepackage{multirow}
\usepackage{caption}
\usepackage{subcaption}
\usepackage{booktabs}
\usepackage{amsfonts}
\usepackage{fontenc}
\usepackage{inputenc}
\usepackage{calc}
\usepackage{fancyhdr}
\usepackage{graphicx}
\usepackage{epstopdf}
\usepackage{amsmath}
\usepackage{amssymb}
\usepackage{titlesec}
\usepackage{xcolor}
\usepackage{colortbl}
\usepackage{soul}
\usepackage{tabularx}
\usepackage{cleveref}
\usepackage{balance}
\usepackage{leftidx}
\usepackage[portuguese,english]{babel}

\newcommand\BibTeX{{\rmfamily B\kern-.05em \textsc{i\kern-.025em b}\kern-.08em
T\kern-.1667em\lower.7ex\hbox{E}\kern-.125emX}}

\def\volumeyear{2025}

\begin{document}

\runninghead{Giberna \textit{et~al.}}

\title{\LARGE \bf On Digital Twins in Defence: Overview and Applications}

\author{Marco Giberna\affilnum{1}, Holger Voos\affilnum{1}\affilnum{2},  Paulo Tavares\affilnum{3}, João Nunes\affilnum{3}, Tobias Sorg\affilnum{4}, Andrea Masini\affilnum{5}, and Jose Luis Sanchez-Lopez\affilnum{1}}

\affiliation{\affilnum{1}Interdisciplinary Center for Security Reliability and Trust (SnT), University of Luxembourg, 1855 Luxembourg, Luxembourg\\
\affilnum{2}Faculty of Science, Technology, and Medicine (FSTM), University of Luxembourg, 1359 Luxembourg, Luxembourg\\
\affilnum{3}INEGI, 4200-465 Porto, Portugal\\
\affilnum{4}Hensoldt, 88090 Immenstaad/Bodensee, Germany\\
\affilnum{5}FlySight, 57121 Livorno, Italy}

\corrauth{Marco Giberna, Interdisciplinary Center for Security Reliability and Trust (SnT), University of Luxembourg, 
1855 Luxembourg, Luxembourg.}

\email{marco.giberna@uni.lu}

\begin{abstract}
Digital twins have emerged as a transformative technology for modeling and simulation in various industries, including defense. 
This paper provides a comprehensive review of digital twin applications in defense modeling and simulation, focusing on how digital twins can enhance simulation fidelity, interoperability, and decision support within defense systems.
We consolidate existing research into a unified framework that links digital twin concepts, simulation-driven application, and real-world deployment in defense scenarios. 
We discuss the role of digital twin in applications like planning, training, execution and monitoring, and debriefing.
We introduce a standardized digital twin characterization framework suitable for defense application that aligns with industrial modeling and simulation standards, and present a taxonomy of defense specific use cases, highlighting recurring requirements. 
Additionally, practical evidence is provided from a targeted questionnaire distributed to defense stakeholders and Ministries of Defense, revealing current challenges in digital twin integration and deployment.
Finally, we conclude by identifying key gaps in digital twins application for defense modeling and simulation, including interoperability, security, and system integration, and we outline future research directions and development opportunities.
This review aims to inform defense modeling and simulation practitioners and researchers, guiding future work on digital twin design, implementation and deployment across defense applications.
\end{abstract}



\maketitle

\section{Introduction}\label{sec:introduction}
Digital Twin (DT) technology represents a groundbreaking advancement that merges real-time data integration, artificial intelligence (AI), and simulation technologies to create high-fidelity and constantly up-to-date virtual replicas, or "twins," of physical systems, processes, and environments. 
These digital models are dynamic, interactive, and capable of real-time monitoring, analysis, control, and optimization of their physical counterparts, allowing organizations to conduct in-depth "what-if" scenario analyses\cite{collins_past_2021, langreck_modeling_2019}. 
By leveraging the power of Digital Twin technology, organizations can gain valuable insights, improve operational efficiency, and make informed decisions without costly physical interventions\cite{singh_digital_2021, mendi_digital_2022}.

Over recent years, Digital Twins have gained significant attention across multiple sectors, including manufacturing, healthcare, aerospace, and energy. The technology has become a central component of various advanced technological trends, with Gartner, an American technological research and consulting firm, forecasting that by 2027, over $40\%$ \cite{panetta_top_2017, groombridge_top_2023} of large organizations worldwide will integrate Digital Twin technology. 
Despite this growing attention, the adoption of Digital Twin technology in sectors like defense remains complex, driven by technological limitations, regulatory barriers, and the need for cross-domain integration and collaboration\cite{singh_digital_2021}.

The military sector stands to benefit immensely from Digital Twin technologies\cite{mendi_digital_2022}, and their implementation has already been initiated in some countries \cite{west_digital_2017}. 
By creating digital replicas of military assets, operations, and infrastructure, defense organizations can perform detailed simulations, optimize resources, and conduct predictive evaluations\cite{zweber_digital_2017}. 
This technology enables real-time monitoring of military equipment and facilities, supports proactive maintenance, and optimizes logistics, offering substantial improvements in asset availability and efficiency. 
The concurrent integration of technologies such as IoT and AI further enhances Digital Twins by enabling the prediction of system failures, anomaly detection, and providing actionable insights based on real-time data. 
However, as the technology evolves from individual system models to interconnected, multi-system simulations, new challenges in standardization, interoperability, and cross-domain applicability must be addressed.

After an initial overview of digital twinning technology, this paper first explores how digital twin technologies can be characterized and standardized across various defense domains, providing a unified framework for their broader adoption. 
It then presents a comprehensive investigation into their use in the defense sector, addressing both the technological aspects and operational aspects that influence their deployment. 
Furthermore, we include the results of a targeted questionnaire, which was distributed to industrial stakeholders and European Ministries of Defense (MoDs), providing insights on the current practice of defense digital twins applications and the key challenges faced by private companies and public institutions.
Based on this evidence and reviewed literature, we identify gaps and limitations that hinder the successful development and deployment of digital twins in defense scenarios.

Summarizing, this paper provides a defense modeling-and-simulation-oriented review of digital twin systems: how they are modeled, simulated and composed, and what this implies for efficacy, interoperability and operational use in defense.
The main contributions are as follows. 
\begin{itemize}
    \item \textbf{Review synthesis}: A defense modeling-and-simulation-centered synthesis of digital twin concepts, architectures, and application patterns in current practice and emerging directions, clarifying what their implementation implies for simulation use (e.g., analysis, training, planning, and monitoring).
    \item \textbf{Reference model for modeling and simulation practice}: A characterization framework intended to unify and standardize defense digital twins across domains and use cases, supporting consistent discussion of structure, interfaces, and data flows.
    \item \textbf{Defense use-case mapping}:A cross-domain mapping of defense use cases to digital twin functions that are directly relevant to modeling and simulation, highlighting recurring requirements and design trade-offs.
    \item \textbf{Evidence and gaps}: An analysis of a questionnaire with industrial stakeholders and MoDs, translated into modeling-and-simulation-relevant gaps, practical challenges, and requirements that inform future work.
\end{itemize}

\section{Methodology}\label{sec:methodology}
This work combines (i) a structured literature review focused on defense-relevant digital twin modeling and simulation and (ii) a targeted stakeholder questionnaire. 

The review synthesized literature covering digital twin architectures, simulation integration, interoperability and standardization, and application in defense context.
Sources were identified using keywords searches (\textit{digital twin}, \textit{defense}, \textit{modeling}, \textit{simulation}, \textit{training}, \textit{mission planning}, \textit{decision support}, \textit{application}) and backward/forward citation tracking.

The questionnaire used in this study was designed to gather insights from key stakeholders involved in the development and implementation of Digital Twin technologies within the defense sector. 
The focus was to explore their views on the current applications and operational implications, technological challenges, and gaps related to Digital Twins in military operations.
The survey was distributed to two primary groups: industrial stakeholders engaged in the defense technology sector and representatives from European MoDs. 
These groups were selected due to their direct involvement in the adoption and deployment of cutting-edge technologies within the military domain.
A total of 32 responses were collected, providing valuable data from both industry experts and governmental bodies. 
The respondents were asked to provide input on a range of topics, including current applications, technological limitations, integration with emerging technologies, and barriers to the widespread adoption of Digital Twin solutions in defense.
Once the data was collected, it was categorized based on thematic areas, with each section focusing on a specific aspect of Digital Twin technology in the defense sector.

\section{Digital Twin Overview}\label{sec:digital_twin_overview}
The concept of a virtual equivalent of a physical asset was first introduced in 2003 by Michael Grieves at the University of Michigan Executive Course on Product Lifecycle Management (PLM). 
Grieves later introduced the term “Digital Twin” in \cite{grieves_virtually_2011}, attributed it to John Vickers of NASA and explained its broader usage throughout the entire product lifecycle.
Finally, Grieves formalized his digital twin model in a white paper \cite{grieves_digital_2014}, as well as in a later work \cite{kahlen_digital_2017} that focused on the predictability of the behavior of complex systems. 

\begin{figure}[!htbp]
\centering
\includegraphics[width=0.99\columnwidth]{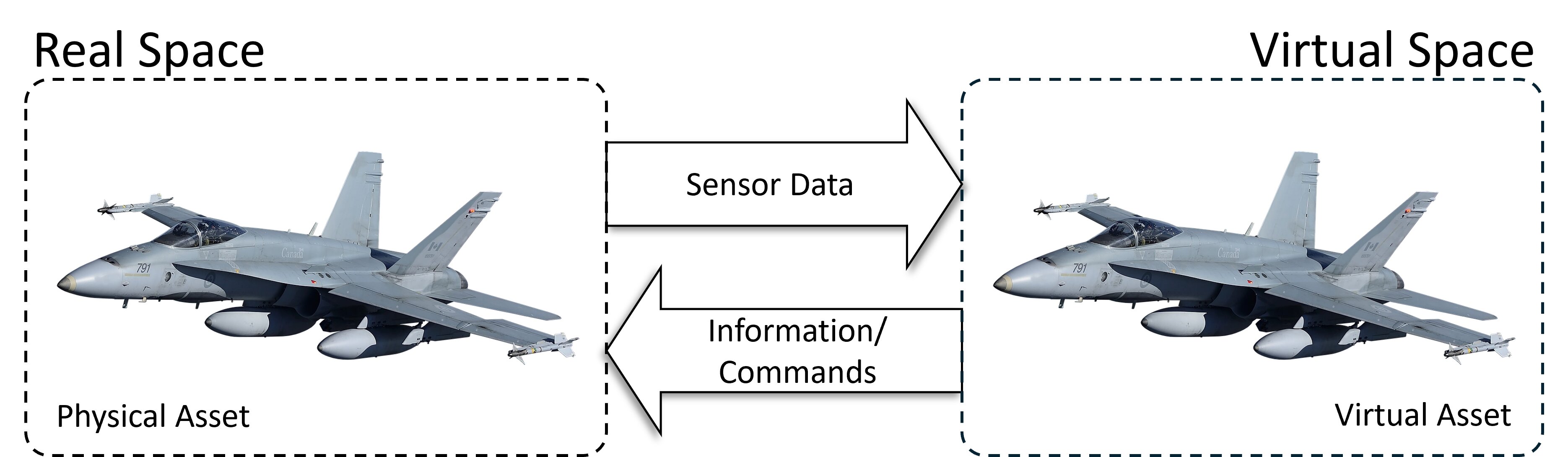}
\caption{The conceptual model of Digital Twin. The virtual asset in the virtual space replicates with high fidelity the physical asset in the real space, automatically exchanging information in both directions. }\label{fig:digital_twin_concept_model}
\end{figure}

The Digital Twin concept model, depicted in Figure \ref{fig:digital_twin_concept_model}, contains three main components: 

\begin{itemize}
    \item physical assets in the real space; 
    \item virtual assets in the virtual space;
    \item bidirectional automatic channels for data and information exchange between these spaces. 
\end{itemize}

In this model, the virtual asset maintains a high-fidelity resemblance to the real asset, staying synchronized over time. 
The bidirectional communication ensures data flows from physical to virtual assets while enabling control signals and insights to influence physical operations. 
This distinguishes a digital twin from a “digital shadow”, in which data flows automatically only from the real to the virtual space, or “digital model”, in which data and information are manually exchanged in both directions \cite{kritzinger_digital_2018, errandonea_digital_2020}.  

The conceptual framework of a  Digital Twin  has been standardized in the ISO 23247 series “Automation System and Integration – Digital Twin Framework for Manufacturing” \cite{noauthor_iso_2021-2} in 2021.
Although this standard primarily targets manufacturing applications, it thoroughly explains the digital twin model, possibly allowing for further adaptation to different use cases. 
ISO 23247 provides general principle and requirements for creating digital twins (ISO 23247-1 \cite{noauthor_iso_2021-2}), presents the digital twin reference model’s architecture both in terms of domains and entities with functional views (ISO 23247-2 \cite{noauthor_iso_2021-1}), enlists basic information attributes for the observable manufacturing elements (ISO 23247-3 \cite{noauthor_iso_2021}) and explains the network architecture within the reference model (ISO 23247-4 \cite{noauthor_iso_2021-3}). 
Two more parts of the standard are currently under development, specifically focusing on how digital threads enable the creation, connectivity, management and maintenance of a manufacturing digital twin across the product lifecycle (ISO 23247-5 \cite{noauthor_isocd_nodate-1}) and on the digital twin composition by defining principles and showing methodologies (ISO 23247-6 \cite{noauthor_isocd_nodate}). 

\begin{figure}[!ht]
    \centering
    \includegraphics[width=0.99\columnwidth]{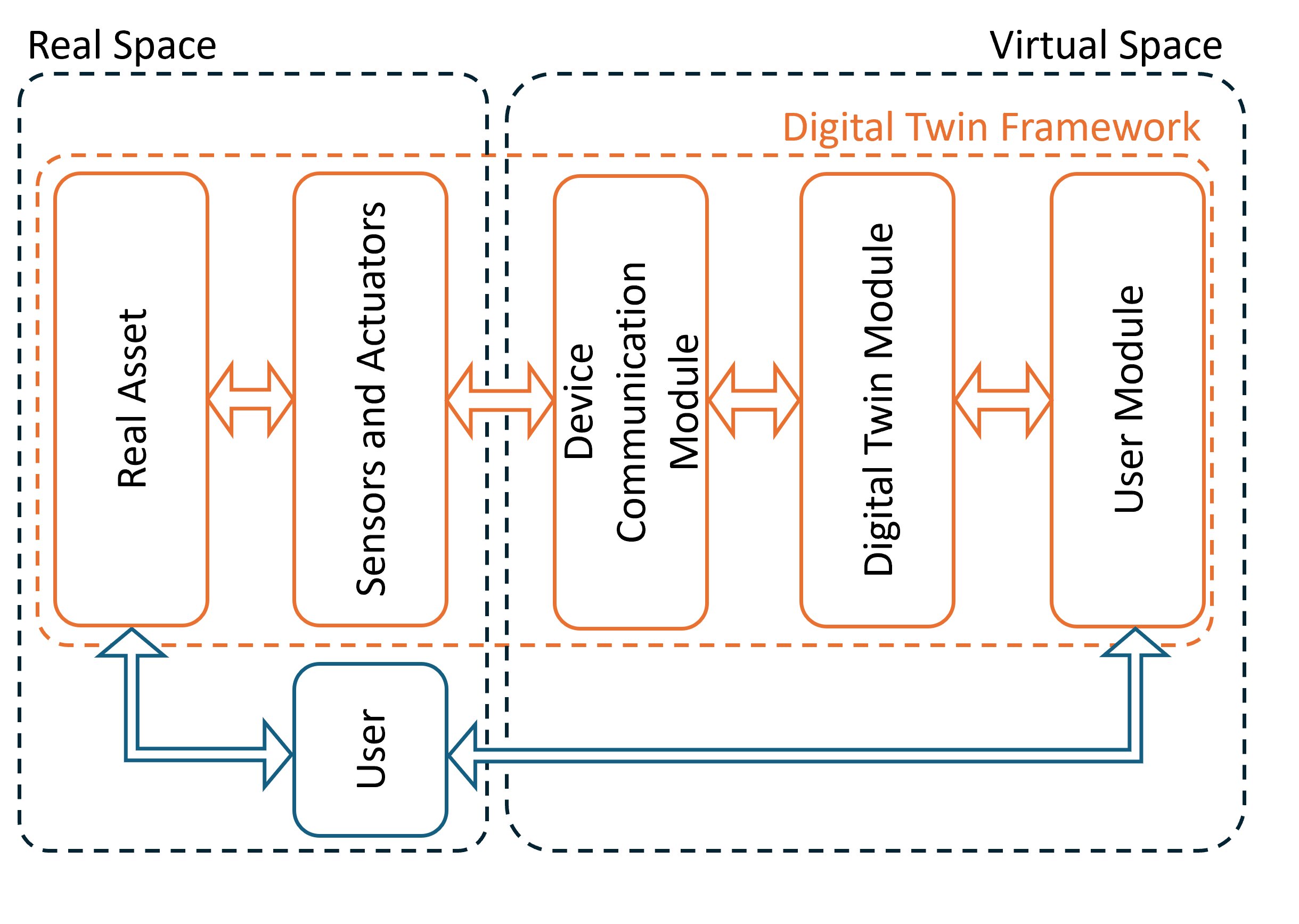}
\caption{The general modular framework of Digital Twins.}\label{fig:digital_twin_reference_model}
\end{figure}

Drawing inspiration from the ISO series, we can outline a general framework for digital twinning systems that formalizes the common core structure, as shown in Figure ~\ref{fig:digital_twin_reference_model}.
This general framework partitions a digital twin into functional modules, each with specific interfaces and tasks.
The framework also distinguishes between real and virtual spaces:
\begin{itemize}
    \item \textbf{Real Space}
    \begin{itemize}
        \item \textbf{Physical Asset}: The asset we desire to digitalize, deployed in the real space.
        \item \textbf{Sensors and Actuators}: collect data from, and control, the physical asset.
    \end{itemize}
    \item \textbf{Virtual Space}
    \begin{itemize}
        \item \textbf{Device Communication Module}: Collects and preprocesses sensor data; manages actuator commands to the real asset.
        \item \textbf{Digital Twin Module}: Maintains synchronization between real and virtual assets during a session and hosts services like simulation and analytics. It also handles data gaps, noisy inputs or other inaccuracies.
        \item \textbf{User Module}: Receives the digital twin's output states for further analysis (e.g., visualization, reporting) through a human-machine interface. It can also reinitialize a session with a stored prior state of the digital twin.
    \end{itemize}
\end{itemize}

\begin{figure*}[!ht]
\centering
\includegraphics[width=0.9\textwidth]{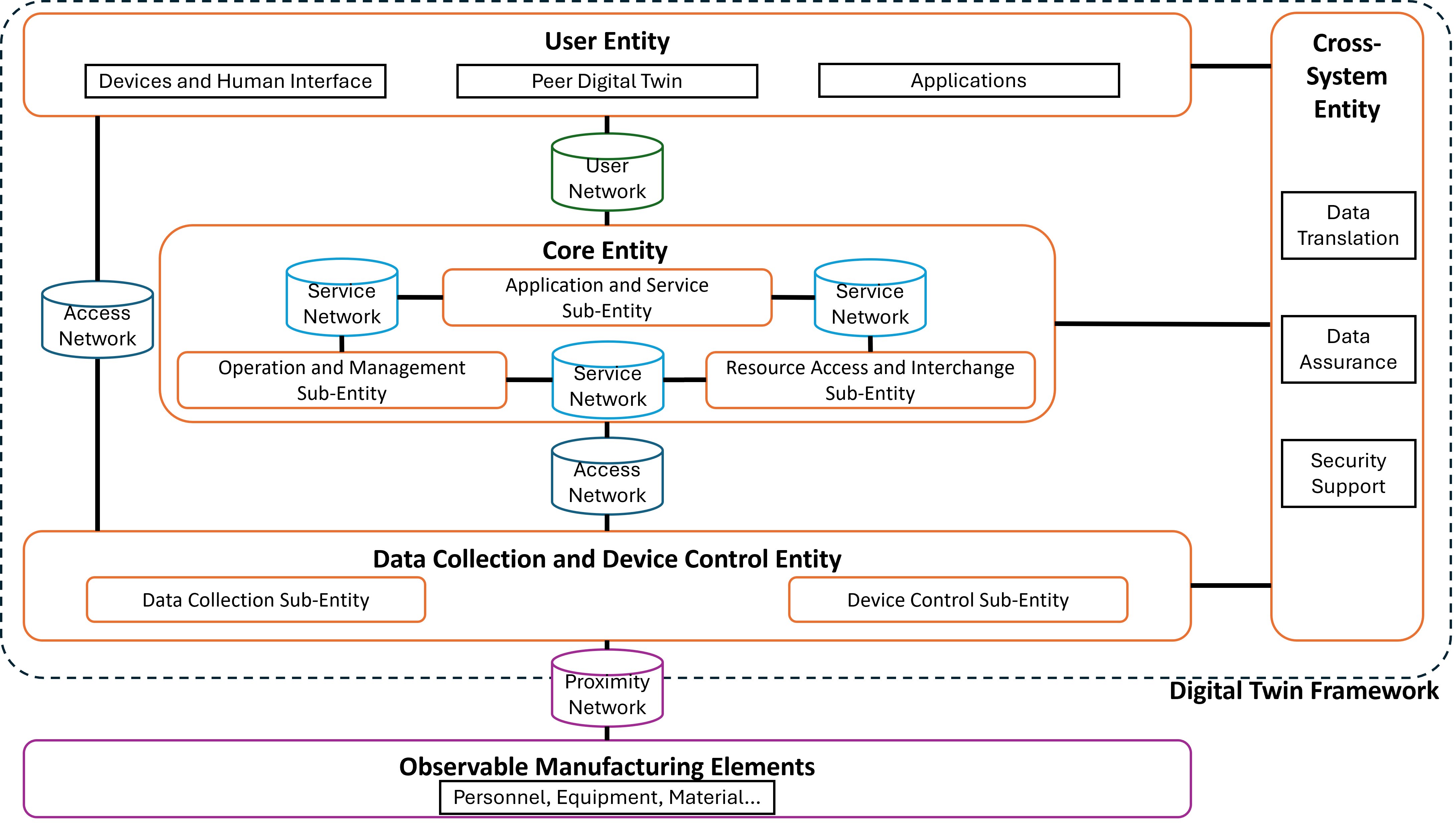}
\caption{Entities-Based Digital Twin Reference Model. Entities and Sub-Entities are reported in rounded rectangles, some attributes and functionalities in rectangles and networks in cylinders.}\label{fig:digital_twin_entity_model}
\end{figure*}

With reference to the ISO standard, the entity-based reference model (Figure~\ref{fig:digital_twin_entity_model}) partitions the system into Entities and Sub-Entities:
\begin{itemize}
    \item \textbf{Observable Manufacturing Elements}: the physical components to be modeled.
    \item \textbf{Data Collection and Device Control Entity}: Handles data acquisition and device actuation tasks, bridging real-virtual communication.
    \begin{itemize}
        \item \textbf{Data Collection Sub-Entity}: collects data from the observable manufacturing elements.
        \item \textbf{Device Control Sub-Entity}: sends control commands to the physical asset.
    \end{itemize}
    \item \textbf{Core Entity}: Maintains digital twin instances, synchronizes data, and includes sub-entities for modeling, simulation, analytics and interoperability. 
    \begin{itemize}
        \item \textbf{Operation and Management Sub-Entity}:  handles functionalities such as digital modeling, synchronization, maintenance, and representation.
        \item \textbf{Application and Service Sub-Entity}: Hosts simulation, reporting, and analytic functionalities.
        \item \textbf{Resource Access and Interchange Sub-Entity}: Manages interoperability support, plug-and-play integration, and access control.
    \end{itemize}
    \item \textbf{User Entity}: Incorporates devices, human–machine interfaces, and business applications to leverage digital twins for real-world process optimization.
    \item \textbf{Cross-System Entity}: Provides data translation, security, and assurance capabilities across all layers.
\end{itemize}
Data communication within this model is also well-defined, using four networks (User Network, Service Network, Access Network, and Proximity Network) to connect entities and sub-entities. Commonly, these employ wired or wireless IP-based solutions.

Overall, this standardized approach (and its ongoing development) provides a foundational framework that can be extended beyond manufacturing into various domains—defense included—by tailoring the modules, entities, sub-entities, and communication networks to meet specific operational requirements.

\subsection{Dual-Use and Military Applications}\label{sec:dual_use_and_military_applications}
Dual-use technologies are those with both civilian and military applications. 
Within the defense sector, such technologies offer cost-effective solutions, interoperability, and accelerated innovation by leveraging advances from commercial industries\cite{lu_evaluating_2016, zhang_civil_2024}. 
In practice, dual-use technologies span various systems, sensors, and applications specifically tailored to meet unique military operational requirements.

Beyond their direct impact on operational effectiveness, dual-use technologies can profoundly shape military capabilities across multiple domains\cite{carillo_commercial_2017, mazal_dual_2024}. 
Communication and network capabilities, often propelled by civilian demands, influence how missions are conducted. These developments introduce valuable new capabilities but also potential threats and vulnerabilities that must be addressed, particularly concerning information sovereignty. 
Secure and reliable communication systems ensure seamless coordination among military personnel, while sensor technologies (e.g., radar, surveillance) enable real-time data acquisition, situational awareness, and threat detection. 
Such information forms a critical input to decision-making processes and underpins mission success.

Leveraging dual-use technologies also fosters compatibility with existing civilian standards. 
This, in turn, smoothens integration across sectors, granting defense organizations access to state-of-the-art technologies, economies of scale, and faster deployment\cite{akimkina_technology_2021}. 
The exchange of ideas, expertise, and resources between military and commercial sectors drives innovation, keeping defense capabilities current with rapidly evolving technological frontiers. 
Moreover, interoperability benefits extend beyond departmental and national borders; shared standards and common platforms enable collaborative operations, data exchange, and joint missions among different military branches and international allies\cite{mazal_dual_2024, pereira_dual_2018}. 
This robust interoperability reinforces command and control, optimizes resource allocation, and enhances collective defense readiness.

\subsection{Technological Enablers}\label{sec:technological_enablers}
A Digital Twin can be considered a foundational tool on which different additional technological instruments are implemented to enhance capabilities, for example by automating data flow, optimizing data analytics, and extending functionality.
Common identified technological enablers to this scope are:
\begin{itemize}
    \item \textbf{Internet of Things (IoT) and Sensors}, which provide consistent and continuous data acquisition from the physical asset to the virtual counterpart;
    \item \textbf{Artificial Intelligence (AI) and Machine Learning (ML)}, which enable advanced data processing and predictive analytics;
    \item \textbf{Simulation Technologies and Advanced Computing}, necessary for high-fidelity modeling and reliable virtual evaluation;
    \item \textbf{Cloud Computing} solutions that address data accessibility and storage challenges;
    \item \textbf{Extended Reality (XR)}, which enables more intuitive and efficient user-digital twin interaction.
\end{itemize}
Although not exhaustive, the enlisted topics are among the most representative within the evolving set of data assimilation and interoperability enablers for digital twins.

\subsubsection{IoT and Sensors}\label{sec:iot_and_sensors}
The current landscape of military Internet of Things (IoT) and sensor technologies is characterized by rapid advancement and broad integration into defense applications\cite{fraga-lamas_review_2016}. 
These technologies are central to enhancing operational capabilities, improving situational awareness, and ensuring robust defense mechanisms within the military sector\cite{suri_analyzing_2016}. 
Military IoT encompasses a network of interconnected devices that collect, exchange, and process data to facilitate real-time decision-making and strategic planning. 
In practice, it spans everything from wearable devices that monitor soldiers' health and performance to sensors in vehicles and aircraft that provide critical operational and environmental data.

Sensor technologies within the military domain also evolved significantly in the recent years. 
Modern sensors are capable of detecting a wide range of physical, chemical, and biological stimuli with high precision. 
These sensors are embedded in all existing military platforms, including aerial, naval, and ground vehicles. 
Advanced imaging sensors, for example, enable night vision and thermal imaging, crucial for surveillance and reconnaissance missions conducted in low-visibility conditions. 
Table ~\ref{tab:sensors} enlists commonly used sensors in defense digital twinning, specifying their main domains of application and providing explicative examples.  

Integrating sensors with AI and ML algorithms further amplifies their utility\cite{blasch_machine_2021}. 
Large datasets generated by these sensors can be processed in real time to identify patterns, predict equipment failures, and detect anomalies. 
This capability is instrumental for predictive maintenance, enabling higher operational readiness and extending the lifespan of critical assets.

Nevertheless, increased connectivity also broadens the potential attack surface for cyber threats, making security of paramount importance \cite{toth_internet_2021, pasdar_cybersecurity_2024}. 
Military organizations invest heavily in securing IoT networks and devices to protect sensitive data and ensure operational integrity, other than readiness from cyber-attacks\cite{bagrodia_using_2023}. 
Additionally, communication failures pose a significant risk when numerous sensors are in use or the system relies heavily on them; this frequently necessitates redundancy or fallback approaches. 
In low-connectivity environments or stealth scenarios, connectivity may be deliberately limited, requiring systems to be both autonomous and resilient to potential disruptions, leveraging employed simulation solutions to fill data gaps\cite{grumazescu_wsn_2016}.
Interoperability remains another pressing challenge, as differing protocols and standards can prevent seamless integration of diverse systems. 

\begin{table*}[ht]
\caption{List of sensors commonly used in defense digital twinning technologies.}\label{tab:sensors}
\centering
\renewcommand{\arraystretch}{1.2}
\scriptsize 
\resizebox{\textwidth}{!}{
\begin{tabular}{l|l|llll|lll|l}
\toprule
\multirow{2}{*}{\textbf{Sensor}} & \multirow{2}{*}{\textbf{Measurement}}                         & \multicolumn{4}{c|}{\textbf{Main Domain(s)}}                                                                        & \multicolumn{3}{c|}{\textbf{Main Type(s)}}                                       & \multirow{2}{*}{\textbf{Examples}}     \\ 
                                 &                                                               & \multicolumn{1}{l|}{\rotatebox{90}{\textbf{Land}}} & \multicolumn{1}{l|}{\rotatebox{90}{\textbf{Maritime}}} & \multicolumn{1}{l|}{\rotatebox{90}{\textbf{Air}}} & \rotatebox{90}{\textbf{Space}} & \multicolumn{1}{l|}{\rotatebox{90}{\textbf{Environment}}} & \multicolumn{1}{l|}{\rotatebox{90}{\textbf{Asset}}} & \rotatebox{90}{\textbf{Human}}             
                                 \\ \midrule
\textbf{IMU}                     & Velocities, Accelerations, Yaw Angle                          & \multicolumn{1}{l|}{$\checkmark$}  & \multicolumn{1}{l|}{$\checkmark$}      & \multicolumn{1}{l|}{$\checkmark$} &                & \multicolumn{1}{l|}{$\checkmark$}         & \multicolumn{1}{l|}{$\checkmark$}   & $\checkmark$   & MPU-6050              \\
\textbf{GPS}                     & Absolute Position                                             & \multicolumn{1}{l|}{$\checkmark$}  & \multicolumn{1}{l|}{$\checkmark$}      & \multicolumn{1}{l|}{$\checkmark$} &                & \multicolumn{1}{l|}{$\checkmark$}         & \multicolumn{1}{l|}{$\checkmark$}   & $\checkmark$   & u-blox NEO-M8N        \\
\textbf{Barometer}               & Atmospheric Pressure                                          & \multicolumn{1}{l|}{}              & \multicolumn{1}{l|}{}                  & \multicolumn{1}{l|}{$\checkmark$} &                & \multicolumn{1}{l|}{$\checkmark$}         & \multicolumn{1}{l|}{$\checkmark$}   &                & Bosch BMP280          \\
\textbf{Altimeter}               & Altitude                                                      & \multicolumn{1}{l|}{}              & \multicolumn{1}{l|}{}                  & \multicolumn{1}{l|}{$\checkmark$} & $\checkmark$   & \multicolumn{1}{l|}{$\checkmark$}         & \multicolumn{1}{l|}{$\checkmark$}   &                & BMP390                \\
\textbf{RFID}                    & Absolute Position (using electromagnetic field) & \multicolumn{1}{l|}{$\checkmark$}  & \multicolumn{1}{l|}{$\checkmark$}      & \multicolumn{1}{l|}{$\checkmark$} &                & \multicolumn{1}{l|}{}                     & \multicolumn{1}{l|}{$\checkmark$}   &                & DecaWav0e DVM1000     \\
\textbf{RGB Camera}              & Visible Light                                                 & \multicolumn{1}{l|}{$\checkmark$}  & \multicolumn{1}{l|}{$\checkmark$}      & \multicolumn{1}{l|}{$\checkmark$} &                & \multicolumn{1}{l|}{$\checkmark$}         & \multicolumn{1}{l|}{$\checkmark$}   &                & IDS uEye LE           \\
\textbf{RGB-D Camera}            & Visible Light, Depth                                          & \multicolumn{1}{l|}{$\checkmark$}  & \multicolumn{1}{l|}{}                  & \multicolumn{1}{l|}{$\checkmark$} &                & \multicolumn{1}{l|}{$\checkmark$}         & \multicolumn{1}{l|}{$\checkmark$}   &                & Intel RealSense D435i \\
\textbf{IR Camera}               & Infrared Radiation                                            & \multicolumn{1}{l|}{$\checkmark$}  & \multicolumn{1}{l|}{$\checkmark$}      & \multicolumn{1}{l|}{$\checkmark$} &                & \multicolumn{1}{l|}{$\checkmark$}         & \multicolumn{1}{l|}{$\checkmark$}   & $\checkmark$   & FLIR Lepton           \\
\textbf{Hyperspectral Camera}    & Electromagnetic Spectrum                                      & \multicolumn{1}{l|}{}              & \multicolumn{1}{l|}{}                  & \multicolumn{1}{l|}{}             & $\checkmark$   & \multicolumn{1}{l|}{$\checkmark$}         & \multicolumn{1}{l|}{}               &                & FX-10e VNIR           \\
\textbf{Event Camera}            & Brightness log-intensity changes                              & \multicolumn{1}{l|}{$\checkmark$}  & \multicolumn{1}{l|}{}                  & \multicolumn{1}{l|}{}             &                & \multicolumn{1}{l|}{$\checkmark$}         & \multicolumn{1}{l|}{}               &                & Sony IMX636ES         \\
\textbf{LIDAR}                   & Metric distances and angles of scene points                   & \multicolumn{1}{l|}{$\checkmark$}  & \multicolumn{1}{l|}{$\checkmark$}      & \multicolumn{1}{l|}{$\checkmark$} &                & \multicolumn{1}{l|}{$\checkmark$}         & \multicolumn{1}{l|}{}               &                & Velodyne VLP-16       \\
\textbf{Radar}                   & Distance, Direction, Velocity (using radio waves)             & \multicolumn{1}{l|}{$\checkmark$}  & \multicolumn{1}{l|}{$\checkmark$}      & \multicolumn{1}{l|}{$\checkmark$} & $\checkmark$   & \multicolumn{1}{l|}{$\checkmark$}         & \multicolumn{1}{l|}{}               &                & AWRL6432AOP           \\
\textbf{Magnetometer}            & Magnetic Field                                                & \multicolumn{1}{l|}{$\checkmark$}  & \multicolumn{1}{l|}{$\checkmark$}      & \multicolumn{1}{l|}{$\checkmark$} & $\checkmark$   & \multicolumn{1}{l|}{}                     & \multicolumn{1}{l|}{$\checkmark$}   &                & HMC5883L              \\
\textbf{Proximity Sensor}        & Electromagnetic Field Changes 
& \multicolumn{1}{l|}{$\checkmark$}  & \multicolumn{1}{l|}{}                  & \multicolumn{1}{l|}{}             &                & \multicolumn{1}{l|}{$\checkmark$}         & \multicolumn{1}{l|}{$\checkmark$}   &                & OPT8241               \\
\textbf{Thermometer}             & Temperature                                                   & \multicolumn{1}{l|}{$\checkmark$}  & \multicolumn{1}{l|}{$\checkmark$}      & \multicolumn{1}{l|}{$\checkmark$} & $\checkmark$   & \multicolumn{1}{l|}{$\checkmark$}         & \multicolumn{1}{l|}{$\checkmark$}   & $\checkmark$   & HDC3022-Q1            \\
\textbf{Humidity Sensor}         & Humidity                                                      & \multicolumn{1}{l|}{$\checkmark$}  & \multicolumn{1}{l|}{$\checkmark$}      & \multicolumn{1}{l|}{$\checkmark$} &                & \multicolumn{1}{l|}{$\checkmark$}         & \multicolumn{1}{l|}{$\checkmark$}   & $\checkmark$   & HDC3022-Q1            \\
\textbf{Chemical Sensor}         & Chemical composition or properties                            & \multicolumn{1}{l|}{$\checkmark$}  & \multicolumn{1}{l|}{$\checkmark$}      & \multicolumn{1}{l|}{$\checkmark$} & $\checkmark$   & \multicolumn{1}{l|}{$\checkmark$}         & \multicolumn{1}{l|}{$\checkmark$}   & $\checkmark$   & H10-14                \\
\textbf{pH Sensor}               & Acidity or alkalinity                                         & \multicolumn{1}{l|}{$\checkmark$}  & \multicolumn{1}{l|}{$\checkmark$}      & \multicolumn{1}{l|}{$\checkmark$} &                & \multicolumn{1}{l|}{}                     & \multicolumn{1}{l|}{$\checkmark$}   & $\checkmark$   & GF Signet 3-2751-1    \\
\textbf{Pressure Sensor}         & Pressure in gases or liquids                                  & \multicolumn{1}{l|}{$\checkmark$}  & \multicolumn{1}{l|}{$\checkmark$}      & \multicolumn{1}{l|}{$\checkmark$} & $\checkmark$   & \multicolumn{1}{l|}{}                     & \multicolumn{1}{l|}{$\checkmark$}   & $\checkmark$   & ESI-GS4000            \\
\textbf{Sound Sensor}            & Audio intensity or specific frequency                         & \multicolumn{1}{l|}{$\checkmark$}  & \multicolumn{1}{l|}{}                  & \multicolumn{1}{l|}{}             &                & \multicolumn{1}{l|}{$\checkmark$}         & \multicolumn{1}{l|}{$\checkmark$}   & $\checkmark$   & RAYMING PCB           \\
\textbf{Ultrasonic Sensor}       & Distance (using ultrasound waves)                             & \multicolumn{1}{l|}{}              & \multicolumn{1}{l|}{}                  & \multicolumn{1}{l|}{$\checkmark$} &                & \multicolumn{1}{l|}{$\checkmark$}         & \multicolumn{1}{l|}{$\checkmark$}   &                & HC-SR05               \\
\textbf{Force and Torque Sensor} & Force and Torque applied to object                            & \multicolumn{1}{l|}{$\checkmark$}  & \multicolumn{1}{l|}{$\checkmark$}      & \multicolumn{1}{l|}{$\checkmark$} & $\checkmark$   & \multicolumn{1}{l|}{}                     & \multicolumn{1}{l|}{$\checkmark$}   & $\checkmark$   & HPS-FT060S            \\
\textbf{Vibration Sensor}        & Vibrations                                                    & \multicolumn{1}{l|}{$\checkmark$}  & \multicolumn{1}{l|}{$\checkmark$}      & \multicolumn{1}{l|}{$\checkmark$} &                & \multicolumn{1}{l|}{}                     & \multicolumn{1}{l|}{$\checkmark$}   &                & WTVB01-485            \\
\textbf{Rotary Encoder}          & Rotational Position                                           & \multicolumn{1}{l|}{$\checkmark$}  & \multicolumn{1}{l|}{$\checkmark$}      & \multicolumn{1}{l|}{$\checkmark$} & $\checkmark$   & \multicolumn{1}{l|}{}                     & \multicolumn{1}{l|}{$\checkmark$}   & $\checkmark$   & Omron E6B2-CWZ6C      \\
\textbf{Linear Encoder}          & Linear Position                                               & \multicolumn{1}{l|}{$\checkmark$}  & \multicolumn{1}{l|}{$\checkmark$}      & \multicolumn{1}{l|}{$\checkmark$} & $\checkmark$   & \multicolumn{1}{l|}{}                     & \multicolumn{1}{l|}{$\checkmark$}   & $\checkmark$   & Heidenhain LC 483     \\
\textbf{ECG Sensors}             & Cardiac Electrical Potential Waveforms                        & \multicolumn{1}{l|}{$\checkmark$}  & \multicolumn{1}{l|}{$\checkmark$}      & \multicolumn{1}{l|}{$\checkmark$} & $\checkmark$   & \multicolumn{1}{l|}{}                     & \multicolumn{1}{l|}{}               & $\checkmark$   & AD8232 ECG Sensor \\
\bottomrule
\end{tabular}}
\end{table*}

\subsubsection{Artificial Intelligence and Machine Learning Technologies}\label{sec:artificial_intelligence_and_machine_learning}
have witnessed substantial progress in recent years, transforming a variety of sectors, including defense. 
These technologies excel at addressing complex problems, extracting actionable insights from large datasets, and enhancing decision-making processes\cite{prasad_machine_2022, davis_artificial_2022}.

In the defence sector, AI technologies can analyze extensive data streams, facilitate pattern recognition, and improve situational awareness\cite{szabadfoldi_artificial_2021}. Techniques such as deep learning and neural networks have proven effective for image and speech recognition, image processing, natural language processing, and autonomous operations. Key algorithmic approaches include supervised and unsupervised learning, reinforcement and incremental learning,  and anomaly detection, all of which enable systems to learn from historical data and adapt to evolving conditions.

The application of AI technologies in defence encompasses a wide range of use cases ranging from mission planning, training and execution (accounting for terrain data, weather, and threat assessments) \cite{van_lent_applications_2022, stevens_machine_2021} to enhanced intelligence, surveillance, and reconnaissance (ISR) capabilities through automated sensor data analysis and real-time decision support\cite{rashid_artificial_2023}. AI also plays a pivotal role in cybersecurity, assisting with threat identification, anomaly detection, and proactive defense solutions\cite{clark_detection_2021}.
Successfully deploying AI within defense contexts requires careful attention to ethical considerations, robust data management practices, and interpretability. Addressing bias, ensuring fairness, and safeguarding data integrity are paramount. Furthermore, explainability in AI models fosters trust, allows for human oversight, and aligns with legal and ethical standards. Finally, validating and verifying AI systems for operational contexts remains a critical challenge.

\subsubsection{Simulation Technologies and Advanced Computing}\label{sec:simulation_technologies_and_advanced_computing}
have witnessed remarkable advancements, revolutionizing the modeling, analysis, and visualization of complex systems. 
High-Performance Computing (HPC) and Graphics Processing Units (GPUs) have played a pivotal role in enabling sophisticated simulations, real-time rendering, and data-processing capabilities\cite{petrea_hpc_2024}.

Simulation tools allow the replication of real-world scenarios in controlled virtual environments, enabling the evaluation of strategies, factors, and outcomes without risking resources or personnel\cite{collins_past_2021}. 
In the military context, this means replicating complex operations, testing hypotheses, and refining decision-making processes\cite{davis_artificial_2022}. 
High-fidelity simulations offer critical insights into the performance of personnel\cite{alim_measuring_2024}, equipment\cite{felix_real-time_2021}, and systems under a variety of conditions\cite{van_der_zwet_promises_2022, langreck_modeling_2019}.

HPC systems, with their parallel-processing capabilities, efficiently manage massive amounts of data and complex calculations, resulting in faster simulations and improved accuracy\cite{petrea_hpc_2024}. 
Meanwhile, GPUs excel at real-time rendering and high-fidelity visualizations, enabling immersive and realistic training environments. 
Defense organizations capitalize on these technologies for mission planning, wargaming, and assessing environmental factors, ultimately uncovering vulnerabilities, refining tactics, and optimizing resource allocation.

Furthermore, integrating advanced simulation with digital twinning unlocks even greater potential\cite{bagrodia_using_2023}. 
While digital twins model physical assets in real time, simulation helps address incomplete or noisy data to maintain an accurate replica in the virtual space. 
Military planners can experiment with variable conditions, stress-test systems, and optimize operational decisions dynamically, offering profound benefits in both effectiveness and efficiency.

\subsubsection{Cloud Computing}\label{sec:cloud_computing}
extends many of the advantages offered by advanced simulation and computing. 
By offloading computation and storage requirements to cloud infrastructures, defense organizations overcome the hardware limitations of physical assets, benefiting from on-demand scalability and greater flexibility\cite{zaerens_enabling_2011}. 
Cloud solutions also promote broader interoperability among systems, since diverse platforms can interact seamlessly via standardized interfaces in the cloud environment\cite{tiganus_cloud_2023}.

In practical terms, this means that the massive data generated by sensors, simulations, or AI-driven analytics can be processed centrally, reducing the need for localized high-performance hardware. 
This centralized approach to data management also streamlines collaboration between different branches or allied entities, enabling secure data sharing and common situational awareness.

When considered into a digital twin framework, cloud computing can store and manage digital twins of multiple systems concurrently\cite{stergiou_digital_2022}. 
This arrangement not only eases data accessibility but also permits real-time updates, advanced analytics, and continuous improvements to models, even when end users operate in low-connectivity or resource-constrained environments. 
Overall, the integration of cloud computing with digital twins offers scalable, resilient, and cost-effective solutions that support defense capabilities in an ever-evolving technological landscape.

\subsubsection{Extended Reality}\label{sec:extended_reality}
comprises a spectrum of technologies, from Augmented Reality (AR), which overlays digital content onto the real world, to Virtual Reality (VR), which immerses users in entirely synthetic environments.
By replacing or enhancing traditional displays with headsets or laser-based projections, XR solutions integrate digital information more seamlessly into human perception\cite{boyce_enhancing_2022}.
 
XR-based training has been particularly impactful in defense, enabling immersive skill development for tasks such as maintenance and marksmanship, as well as larger-scale tactical exercises across multiple domains\cite{stanney_performance_2021}. Systems like the Small Unit Immersive Training (SUIT), used by the Netherlands Ministry of Defense \cite{brouwer_met_2017}, demonstrate the efficacy of XR in education and training. Operationally, XR can facilitate mission planning, remote assistance, and augmented operator field-of-view with labels or annotations \cite{stacchio_empowering_2022}, enhancing both situational awareness and team coordination.

Given XR’s ability to deliver intuitive, real-time interactions, it is exceptionally well suited for integration with digital twins. 
This synergy elevates how defense personnel can monitor, analyze, and interact with complex systems, ultimately driving innovation and operational improvements across multiple defense domains.

\section{Digital Twins Ontologies}\label{sec:digital_twin_characterizations}
\begin{figure*}[!ht]
\centering
\captionsetup[subfigure]{justification=centering}
\begin{subfigure}{0.3\textwidth}
\centering
  \includegraphics[height=0.9\textwidth]{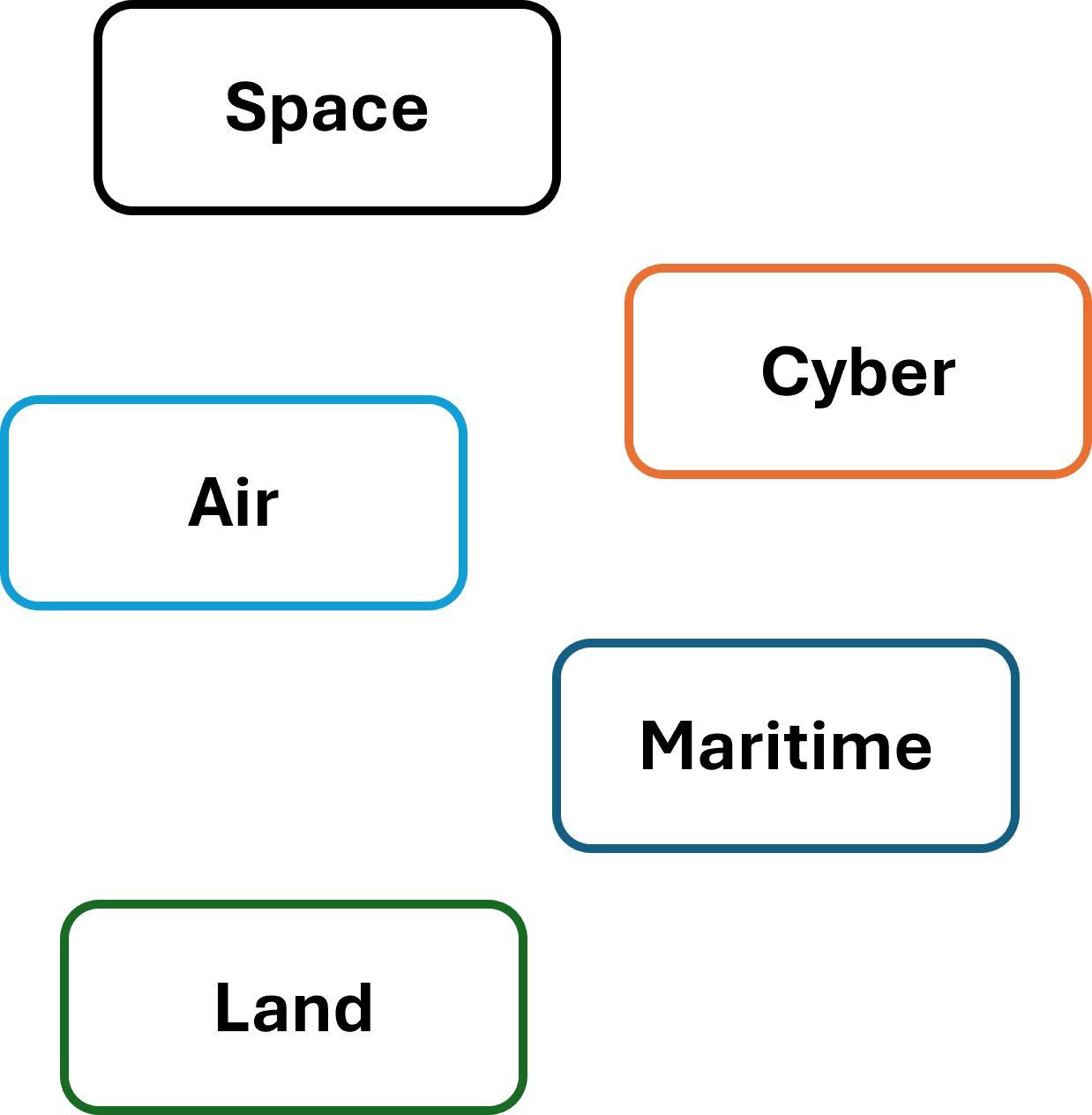}
  \caption{Thematic}\label{fig:thematical_characterization}
\end{subfigure}
\begin{subfigure}{0.3\textwidth}
\centering
  \includegraphics[height=0.9\textwidth]{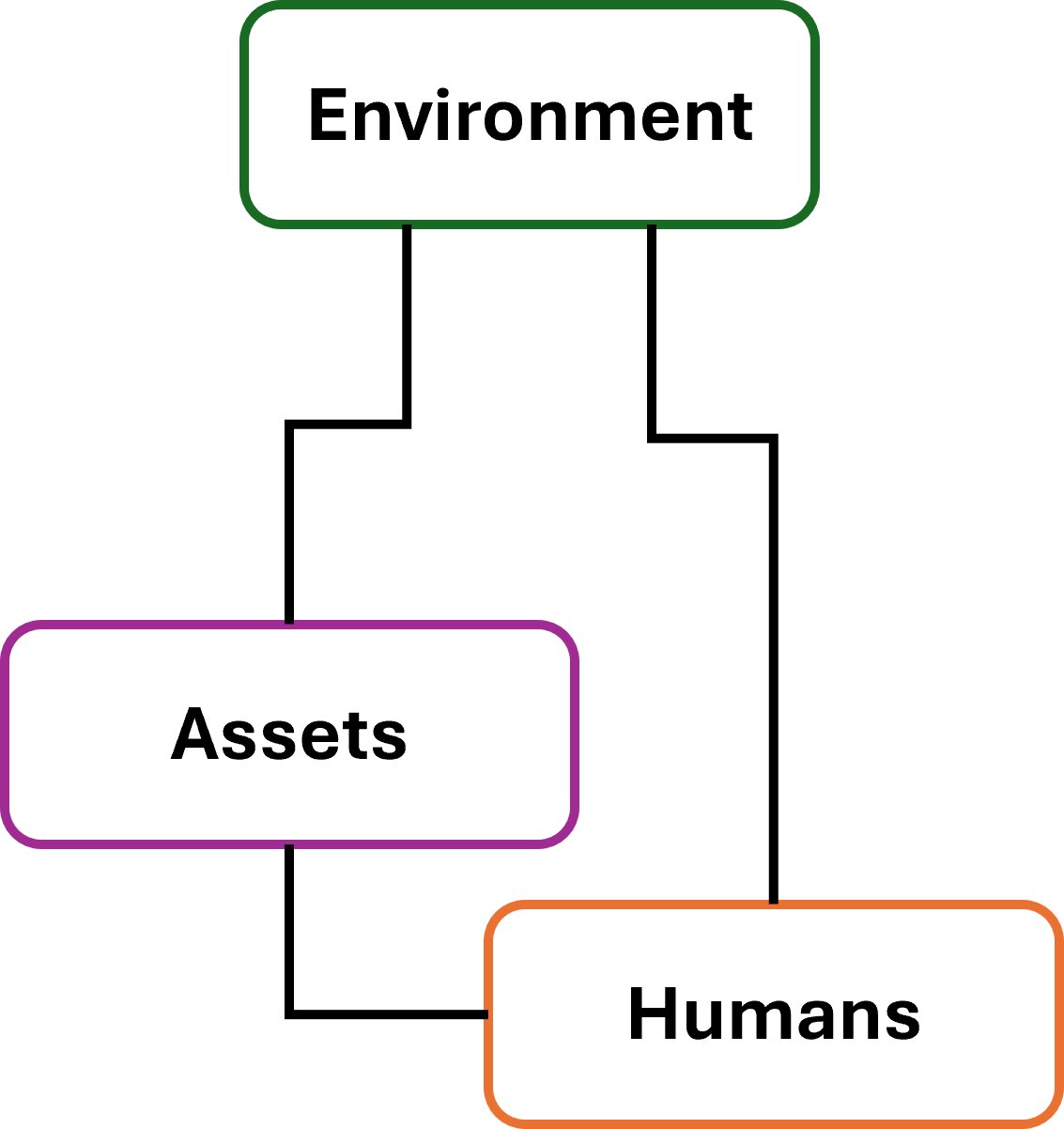}
  \caption{Technical}\label{fig:technical_characterization}
\end{subfigure}
\begin{subfigure}{0.3\textwidth}
\centering
\includegraphics[height=0.9\textwidth]{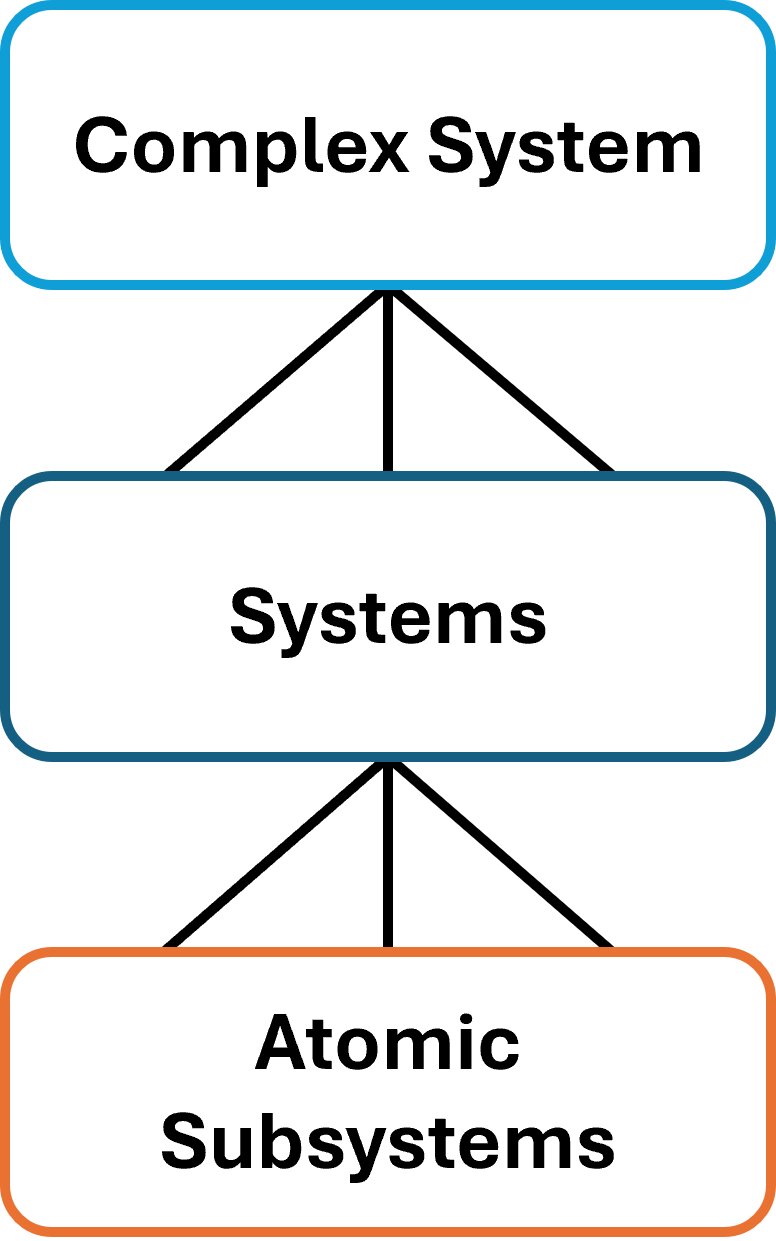}
  \caption{System of Systems}\label{fig:system_of_systems_characterization}
\end{subfigure}
\caption{Identified characterizations of digital twins in the military field.}\label{fig:characterizations}
\end{figure*}

The ability to group digital twins by common characteristics is crucial for both understanding and developing them effectively. 
By identifying shared attributes, be they functional, technical, or contextual, organizations can more readily implement and interoperate various digital twin solutions. 
This holds true across diverse operational environments, including military-specific scenarios, purely civilian applications, or dual-use contexts that span both domains.

Multiple frameworks can be employed to characterize digital twins, each highlighting a different aspect of their design or usage: some focus on thematic domains, others on the types of elements involved, and still others on hierarchical system-of-systems structures. 
In the subsections that follow, we present three complementary approaches that, taken together, provide a robust foundation for understanding how digital twins can be deployed, interconnected, and scaled across a variety of operational needs.

\subsection{Thematic Characterization}\label{sec:thematic_characterization}
A common way to categorize digital twins within military contexts is to group them by operational domain, aligned with the branches of defense organizations and ministries. 
Accordingly, digital twins can be classified into the \textbf{land}, \textbf{maritime}, \textbf{air}, \textbf{space}, and \textbf{cyber} domains, each reflecting a distinct operational environment and set of mission objectives. 

The \textbf{land domain} covers operations and activities that occur on the Earth’s surface. It includes activities such as ground warfare, troop movements, military installation, and infrastructure protection, while operations can involve infantry, armored vehicles, artillery and in general other ground-based assets. Also, land environments themselves are considered as part of this domain.

The \textbf{maritime domain} involves operations and activities which take place on or nearby bodies of water, which can be either oceans, seas, rivers, or lakes. It includes naval warfare, maritime surveillance, maritime patrol, amphibious operations, and maritime environments, while operations in the maritime domain can involve naval vessels, underwater systems, and marines. 

The \textbf{air domain} encompasses operations and activities conducted in the Earth’s atmosphere, including air warfare, aerial reconnaissance, air transport and support for ground and maritime operations, and the aerial environment itself. Operations can be carried out by military aircraft, such as fighters and drones.

The \textbf{space domain} covers operations and activities conducted in outer space. It includes satellite operations, space exploration, space surveillance and space-based communication, other than the space environment itself. Within this domain, operations employ satellites, spacecrafts, space stations, astronauts, and equipment.

The \textbf{cybersecurity domain} involves protecting and defending operations systems, networks and digital infrastructure from cyber threats and attacks. It includes activities such as cybersecurity monitoring, threat detection and response, and vulnerability assessment. Operations within this domain aim to safeguard sensitive information and preserve critical systems and services. 

\subsection{Technical Characterization}\label{sec:technical_characterization}
Thematic characterization can rapidly become outdated and limitative when examining the real-world applications of complex systems. 
Digital twin solutions, in particular, often overlap multiple domains and apply seamlessly to different subsystems and components. 
Therefore, a technical characterization offers a more robust approach by recognizing the cross-domain nature of digital twin architectures.

Digital twins can be categorized into the following three types: \textbf{humans}, \textbf{assets}, and \textbf{environments}. 

\textbf{Human digital twins} focus on physical, technical, and cognitive aspects of human beings. 
In healthcare, for instance, human digital twins monitor, predict, and optimize treatment outcomes. 
In human performance optimization, often applied in the sports domain, they help measure and enhance both hard and soft skills. 
The military domain is beginning to adopt human digital twins as well, expanding beyond traditional physical and technical competencies to include cognitive and tactical dimensions.
Modeling human response under all predictable situations for a specific mission may provide the necessary setting for the implementation of a human digital twin when the user is actually confronted in real world operations.

\textbf{Asset digital twins} represent physical objects or systems—such as machinery, vehicles, or infrastructure. 
In manufacturing, for example, a digital twin of a machine can monitor performance, predict maintenance needs, and optimize operations. 
This helps in extending the asset’s lifespan, enhancing efficiency, and reducing operational costs.

\textbf{Environmental digital twins} involve larger-scale systems such as cities, farms, or even entire ecosystems. 
An environmental digital twin could simulate weather patterns, crop growth, or city traffic patterns to help in planning, managing resources, and improving sustainability efforts.

A key advantage of this technical classification is its inherent inter- and intra-connectivity. 
Digital twins within the same category (e.g., two asset digital twins) can interface more readily because of similar adopted standards and designs, facilitating integration and shared analytics. 
Moreover, human or asset digital twins can be deployed within environmental digital twins to capture a broader operational picture, bridging multiple types and scales of systems. 
By focusing on humans, assets, and environments, this technical approach reflects the flexible, cross-domain architecture at the heart of modern digital twin solutions.

\subsection{System-of-Systems Characterization}\label{sec:system_of_systems_characterization}

\begin{figure*}[!ht]
\centering
  \includegraphics[width=0.9\textwidth]{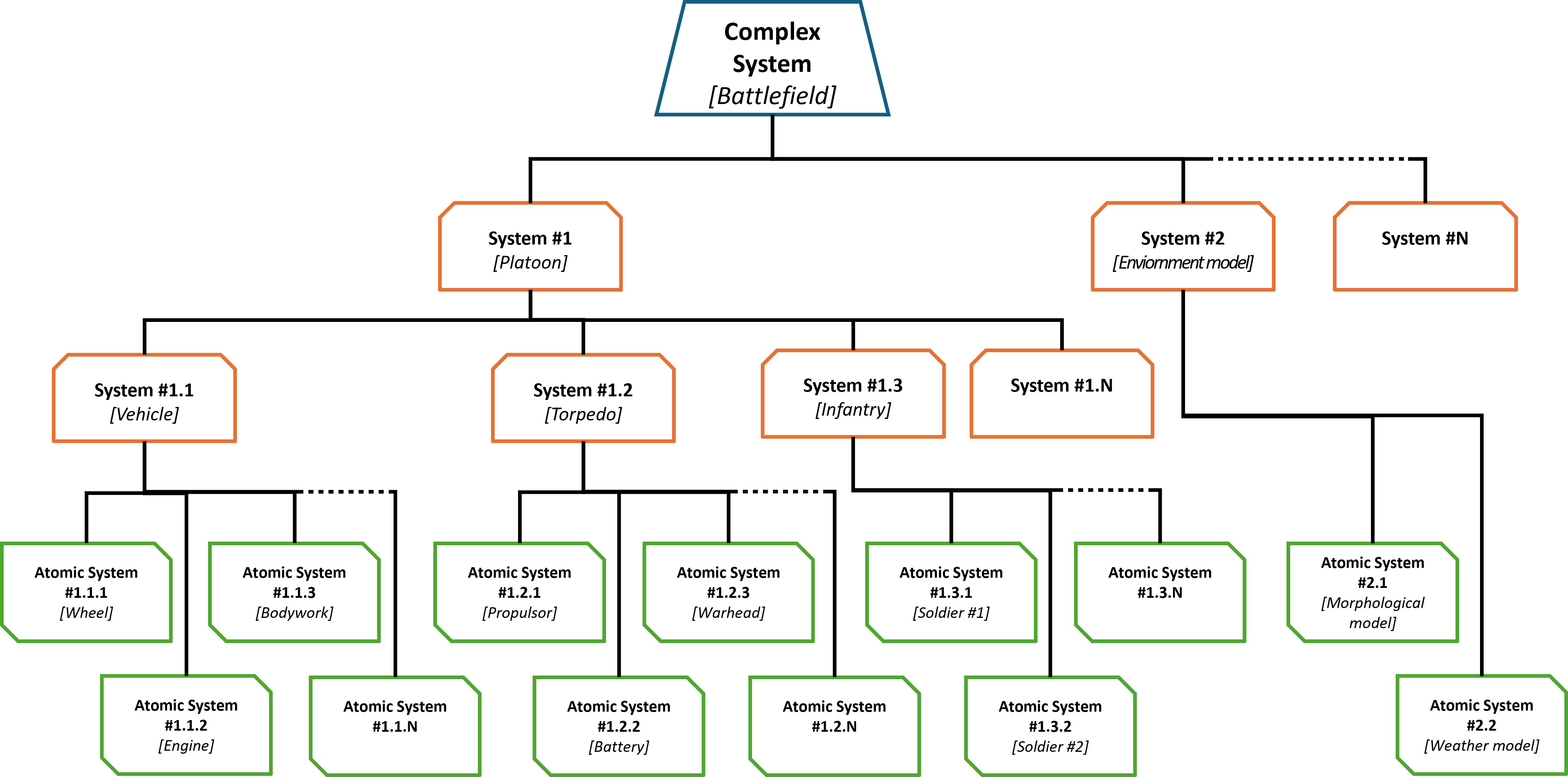}
  \caption{Example of system-of-systems digital twin architecture.}\label{fig:digital_twin_system_of_systems}
\end{figure*}

A final way to characterize digital twins is from a purely technical, more abstract, system-of-systems perspective, following a hierarchical approach in which digital twins can be linked together to form progressively larger and more complex structures. Figure \ref{fig:digital_twin_system_of_systems} illustrates this concept, which is common within Architecture Framework designs\cite{benkamoun_architecture_2014, klein_systematic_2013}.
Proceeding from the bottom up, we identify three primary layers: \textbf{unit}, \textbf{system}, \textbf{complex system}. 

At the lowest level, \textit{units} or \textit{atomic subsystems} represent atomic entities that either cannot—or need not—be subdivided into smaller, stand-alone instances. 
Each digital twin here is fully capable of operating independently. For example, a battery, an engine, or a vehicle’s bodywork could each be modeled as a separate digital twin at this level.

The next layer comprises \textit{systems} or \textit{subsystems} which are larger entities potentially composed of multiple lower-level digital twins. For instance, a weapon system might be built from a battery digital twin and additional component twins, while a vehicle system might integrate engine and bodywork twins (along with other subsystems). 
Depending on the complexity, multiple sub-layers may exist within a system, forming systems of subsystems. 
Communication at this level can be both decentralized (between unit twins) and centralized (to and from the system-level digital twin).

At the highest layer is the \textit{complex system} or \textit{environment} acting as a parent node for the underlying system-level digital twins. It coordinates data exchange and overall control, while still allowing decentralized communication among lower-level twins where beneficial.

This hierarchical structure greatly enhances interoperability. Digital Twins may receive data from real-world assets or from other Digital Twin instances, enabling them to form a network, shaped as a \textit{system-of-systems}, that can itself be considered a higher-level Digital Twin. 
Several works have examined this idea of interconnecting Digital Twin subsystems, focusing on lifecycle interoperability, segmentation of larger systems into smaller ones, and composability across various domains \cite{silvera_navantias_2020, peter_felstead_knds_2024, skinner_taking_nodate, altamiranda_system_2019,noauthor_defence_2022, dietz_digital_2020}.
The created network easily highlights active interfaces and communication channels, allowing for enabling or disabling them at occurrence. 

The Digital Twin Consortium’s whitepaper presents detailed requirements and characteristics for a “Digital Twin System Interoperability Framework” \cite{budiardjo_anto_digital_2021}. 
According to them, each system, either digital, physical, or cyber-physical, must be inherently composable and connectable, supporting:
Dynamic multi-domain connections across hierarchical levels.
Scalability from simple to complex use cases.
Implementation-agnostic protocols and data formats.

A guiding principle is to begin with a clear objective or purpose for each Digital Twin, such as physics-based modeling, simulation, information sharing, or a model of models. 
Systems and subsystems can then interconnect, forming larger federated compositions that amplify the distributed, heterogeneous, and accessible nature of the overall architecture \cite{silvera_navantias_2020}.

This architecture simplifies system evolution and maintenance. An individual Digital Twin can be replaced or operated on without disrupting the entire system; the remainder need not even be aware of the change if interoperability guidelines are followed. 
Similarly, this architecture is beneficial in terms of preventing cyber attacks by promptly isolating and prohibiting communication to infected subsystems.
At the highest level, the complex system node can provide centralized control where desired, while the decentralized communications within or across layers ensure local autonomy, rapid data exchange, and redundancy.

Finally, the user interface can be tailored to specific operational roles: a high-level operator could log into the complex system layer for an entire overview, whereas another lower-level user might connect exclusively to a system-level or unit-level twin for focused insight on that subsystem alone. 
In this way, the architecture provides flexible access control, aligns with defense security and data protection requirements, and helps scale digital twin deployments in step with organizational and operational demands.

\section{Digital Twins in the Military Field}\label{sec:digital_twins_in_the_military_field}
This section explores the most relevant types of digital twin within the military domain, focusing on their technical characteristics. 
Rather than grouping digital twins by application area or branch of service, we choose to classify them based on the nature of what they represent, therefore following the aforementioned technical characterization. 
This choice facilitates interoperability across different defense contexts.
Accordingly, we organize digital twins into three main groups: environment, asset, and human. 
Each group is technically analyzed to highlight how digital twin technologies can enhance performance, decision-making, and strategic outcomes in their respective military applications. 

\subsection{Environment}\label{sec:environment}
Environment digital twins involve creating a virtual representation of environmental systems and their conditions. 
They enable real-time monitoring, analysis and simulation of the natural or artificial environment.
By continuously collecting and processing data from the real world, these digital twins can evaluate “what-if” scenarios and support decision-making through a bidirectional information exchange. 
This distinguishes them from static mapping systems or pure virtual replicas, as the virtual model gets continuously updated and allows for sending commands or feedback to the physical realm.

The primary advantage of environmental digital twins is the enhanced situational awareness they provide. 
They can function as stand-alone models, such as those simulating natural phenomena in specific areas, or serve as foundational layers for other digital twins, for instance, to support pilot training or mission simulations. 
In military scenarios, environmental digital twins might represent battlefields, training fields, military facilities, operational areas, or even entire cities. 

Multiple sensors can be employed for this scope, whereas the final choice must be tailored to the specific use case. An insight of them can be found in Table \ref{tab:sensors}.
Weather sensors are implemented to measure temperature, humidity, wind speed and precipitation. Air quality, water quality and acoustic sensors also provide useful information to be integrated in the digital twin, helping with decision making and operation planning procedures. 
Remote sensing devices, including satellites, unmanned aerial vehicles (UAVs), unmanned ground vehicles (UGVs), and unmanned underwater or surface vehicles (UUVs/USVs) equipped with cameras, LiDARs or radars, provide crucial information about the terrain morphology or building structures, or other environmental details\cite{wu_remote_2023}. 
Usually, a comprehensive mapping is performed initially to establish a baseline; subsequent updates only focus on areas where changes occur, thus reducing the frequency and cost of data acquisition.
If a map of the environment is not available from the beginning, incremental and online map updates are also possible, allowing to augment the knowledge of the environment as soon as it is explored during a mission \cite{piras_digital_2024}.  

Environment digital twins can be integrated with data coming from assets or humans operating within them, providing real-time information of their positions and states, other than exteroceptive data. 
Such data comes either directly from them, in case they are aware of their position, or can be computed by the environmental digital twin through data provided by additional remote sensing devices or other distance sensors.  

Once the collected data is integrated, also providing real-time data processing and visualization, simulation and modelling tools are normally integrated in environmental digital twins.
They use data to simulate environmental processes and scenarios, helping in predicting future states and assessing the impact of eventual possible interventions, enabling the so-called "what-if" scenarios. 
Finally, it is crucial to design a proper user interface to facilitate and at best exploit its usage. 
It can contains dashboards and visualization tools to interact with the digital twin, other than providing insights and actionable information for decision-making.  

Environmental digital twin are highly appropriate for dual-use applications, such as in smart city \cite{farsi2020digital, white2021digital, deng2021systematic, mohammadi2017smart, deren2021smart}, forest management \cite{aguiar2020localization, nie2022artificial, buonocore2022digitalforestry, tagarakis2024forestryagricolture}, healthcare \cite{hassani2022impactful, sun2023digital, liu2019novel}, construction \cite{tuhaise2023technologies, opoku2021digital}, smart farming \cite{tagarakis2024forestryagricolture, nie2022artificial, alves2019digitalfarming, verdouw2021digitalfarming, pylianidis2021introducing, nasirahmadi2022toward}, and more. 
Although these use cases may differ in focus, the underlying digital twin infrastructure often requires only minor and cheap modifications or adaptation, regarding for example specialized data processing or simulation frameworks, to align with each application's goal. 

The specific attributes modeled can vary widely. 
A generic environment may encompass atmospheric conditions, terrain, or topological information, whereas specialized domains, e.g., maritime, demand parameters such as sea state, depth, and water temperature. 
The complexity of modeling can also differ significantly depending on the phenomena of interest. Sea wave modeling, for example, can be highly intricate.

Some promising projects focused on the development of digital twins of environment are focused on applying digital twins technologies on large-scale military facilities. 
Booz Allen Hamilton Inc., jointly with ARES Security Corporation and Unity Technologies, claims to deliver a comprehensive digital twin of the Tyndall Air Force Base \cite{noauthor_building_nodate}. 
Small UASs and vehicles are used to acquire photogrammetry and LiDAR data to be integrated to the available Building Information Modeling (BIM) objects. 
Casey et al. \cite{casey_real-time_2024} developed a digital twin of a aircraft hangar facility, named the Smart Hangar project, aiming to increase security within. It employs computer vision, LiDAR, and ultra-wideband sensors within a Unity3D environment to track objects and agents in real-time. An Active Safety system then processes the data for path planning and collision avoidance, alerting the physical space through smart devices and signs. 
Wang et al. \cite{wang_digital_2021} introduced a digital twin battlefield concept with an online learning algorithm for unmanned combat. Their approach illustrates that a digital twin need not always contain full 3D or morphological details; instead, it may focus on abstract environmental cues or sensor data, depending on mission requirements.

Accurate environment digital twins often rely on detailed mapping procedures, a longstanding topic in robotics since the 1980s \cite{thrun2002robotic}. 
Early approaches employed 2D occupancy grids \cite{elfes_1989}, then shifting towards three-dimensional reconstructions, while modern techniques enable enriching them with additional semantic layers \cite{kostavelis2015semantic}. 
Sensors such as LiDARs \cite{khan2021comparative} or visual (e.g., monocular, stereo or depth) cameras \cite{tourani2022visual} can be optionally fused with GPS and IMU data for generating accurate maps according to the specific use case. 
These advancements enhance the situational awareness \cite{bavle2023slam, giberna2025dynemo} for the robot or user exploring the environment and for the user to utilize it later. 
Maps can be built incrementally or dynamically in a single or multiple sessions, capturing different types of information as required.

Fiducial markers \cite{jurado2023planar,kalaitzakis2021fiducial} are often used for localizing targets or embedding additional information. However, they can be easily spotted by adversaries. In contrast, IMarkers \cite{agha2022unclonable} exploit cholesteric spherical reflectors \cite{schwartz2018cholesteric} for transparent, difficult-to-detect markers that also provide authentication and secure data transfer \cite{schwartz2021linking}. They are particularly promising for military battlefield mapping, where stealth and security are paramount.

Environmental digital twin applications range from port infrastructures (e.g., OHB SE’s port digital twin using AI and satellite data \cite{noauthor_this_2021}) to Earth-scale modeling efforts \cite{bauer_digital_2021,defelipe_towards_2022,daya_sagar_digital_2021}.
Although this applies to all digital twins, the latter is an outstanding example as the importance of clear use-case objectives is paramount to avoid undue complexity.  
Multiple sets of data sources can be used for this scope, ranging from satellite, UAV, UGV down to UUV, depending on the use case and requirements. 
In the same way, rather than trying to recreate the whole system as it is in the real world, it is important to focus on what and why the digital twin is used for, aligning its capabilities accordingly, eventually recreating just a subsample of the environment and specific wanted phenomena.

For UAV deployment, testing and training, many works \cite{lee_digital_2021, alaez_uavradio_2024, zhu_mastering_2024, soliman_ai-based_2023} explore the design and implementation of dedicated environment digital twin. 
Closing to the robotics world, similar tools are exploited for the creation of such models. We can find Gazebo multi-body dynamic simulator \cite{koenig_design_2004}, Robotic Operating System (ROS) \cite{quigley_ros_2009}, the game engine Unity3D \cite{noauthor_unity_2023} among others. 

Overall, environmental digital twins provide a robust platform for enhanced situational awareness, mission rehearsal, and operational planning\cite{langreck_modeling_2019, collins_past_2021}. 
They can work as standalone or can seamlessly integrate with asset and human digital twins to form richer system-of-systems frameworks, ensuring a comprehensive, real-time picture of the operational environment.

\subsection{Asset}\label{sec:asset}
In this context, an \textit{asset} refers to any entity that is neither an environment nor a human. This broad category includes machinery, equipment, weapons, vehicles, and other tools or devices. Digital twins of such assets enable real-time monitoring, analysis, and optimization of performance and maintenance processes.

Data is primarily collected from sensors placed on the physical asset\cite{ochando_data_2023}. 
The type of sensor and desired metric strongly depend on the asset's nature and application. 
Most sensors for asset digital twins are proprioceptive sensors, gathering information regarding the internal state of the system, such as temperature, vibration, pressure, humidity, acceleration and orientation. 
Exteroceptive sensors may also be integrated to improve situational awareness or data processing capabilities, particularly valuable in dynamic and ever-changing environments.

Collected data must be aggregated and stored, often on cloud platforms for ease of management and scalable processing\cite{guo_technical_2025}. 
Predictive analytics tools can be applied both on historical and real-time data to forecast future states, guiding proactive maintenance and operational decision. 
Simulation models further complement these capabilities by exploring "what-if" scenarios and assessing their impact on asset performance.

Also in this case, it is essential to design proper user interfaces presenting in a clear and straightforward manner data regarding the asset performance, health, current state and condition, and operational status through dashboards and, whereas relevant, 3D visualization\cite{liu_study_2018}.
Additionally, digital twins support remote control, granting users the ability to manage or operate the asset from a distance.
Beyond manual oversight, the framework can also facilitate automated adjustments based on real-time data through the use of actuators.  
Recent studies have shown that digital twins lend themselves well to control-theory applications \cite{he2019data, stavropoulos2021robust, gehrmann2019digital}, leading to novel architectures where physical and virtual closed-loop systems run in parallel and exchange information \cite{liu_control_2024}. 

Several initiatives showcase the relevance of asset digital twins in military contexts.
Naval vessels, for example, perfectly embody system-of-systems assets involving cross-domain subsystems, and there are already ongoing projects \cite{silvera_navantias_2020, noauthor_fcx_2023} on creating comprehensive digital twins out of them.  
Navantia's approach \cite{silvera_navantias_2020} constructs a comprehensive digital twin composed of multiple interconnected subsystem twins, each of which recognizes its own functional and health status and communicates autonomously.  
The human-machine interface comprises different levels of intelligence from basic to complex, suitable to support upper-level decisions and execute them. Due to limited access to the Internet, the naval ship’s digital twin architecture requires enough on-board processing capabilities, waiting to close the loop with cloud components when closer to land. 
Lockheed Martin, on the other side, is developing structural digital twins for military vehicles and aircrafts, specifically the F35 Lightning II series \cite{noauthor_delivering_2021}. 
The structural digital twin allows visualizing all the known data of a physical asset, including material information, test data, configuration and force management results based on operational environments. 
On a smaller scale, Singh et al. \cite{singh_physical-virtual_2024} developed a robotic hand digital twin with haptic feedback, while Saab Dynamics explores meta-learning strategies \cite{modeer_towards_2023} to construct digital twins of real-world assets using adaptive neural networks in limited data-scenario, highly common in military application.

Main benefits of asset digital twins are that they can enhance performance by continuously monitoring and optimizing to ensure maximum efficiency, they can improve reliability by performing proactive management and maintenance, and they can provide operational insights which help to improve operational strategies and decision-making. 
Moreover, they can reduce operational costs through optimized performance and maintenance schedules, especially by early identifying potential issues and avoid unexpected downtime. 

These advantages strongly depend on maintaining a high fidelity representation of the physical asset, and on handling large volume of heterogeneous data, which can be achieved by relying on cloud computing. 
Moreover, it is important  to seamlessly integrate data from diverse sensors and platforms into a unified system, other than ensuring data security and protection from eventual cyber threats. 
A secure, interoperable architecture ensures that digital twins deliver lasting benefits without compromising mission-critical data.

\subsection{Human}\label{sec:human}
A Human Digital Twin (HDT) is an advanced concept that integrates multiple layers of data to enhance performance in various domains such as sports \cite{noauthor_physical_2007} \cite{desmond_why_2022}, military operations, clinical application \cite{okegbile2022human} \cite{sun2023digital}, up to daily life. 
Different fields prioritize specific aspects of humans to be modeled in a human digital twin, leading to different data sources, types of measurements, system designs and frameworks \cite{lin2024human} \cite{miller2022unified}. 
In the following we will focus on human digital twin for military and training applications, nevertheless without precluding generality and transferability.

The HDT centers around the development of four primary and interrelated skill layers: physical, technical, tactical, and mental. 
Each of these layers plays a crucial role in performance optimization and requires different sensing and assessment strategies. 

The physical layer encompasses general conditioning, strength, and power, typically developed in fitness centers. 
Metrics such as heart rate, heart rate variability (HRV), breathing rate, and speed are used to monitor physical fitness.  

The technical layer involves the specific movements required on the battlefield or in a sport, often trained through closed drills under the supervision of a coach or instructor who provides feedback.
External sensors, like cameras, measure specific actions such as shooting accuracy, response times, and motor coordination. 

The tactical layer focuses on decision-making and situational reactions, trained through open drills where individuals must make strategic decisions under varying conditions. 
Activity sensors track task performance, particularly for dismounted soldiers, covering tasks like shooting, moving, and communicating.  

The mental layer pertains to performance in mentally challenging high-pressure scenarios, including stress and cognitive load management. 
Stress levels can be estimated using cortisol or HRV, along with other psychological or behavioral indicators. 

Despite the comprehensive approach of HDTs, there are challenges, particularly the lack of established standards for the tactical and mental layers, other than legal, ethical and social implications \cite{glascoe_human_2024}. 
While there are solid training standards for the physical and technical layers, the absence of norms for the tactical and mental aspects means that assessments of warfighters' readiness often rely on assumptions rather than measurable criteria\cite{alim_measuring_2024, stanney_performance_2021}. 
Establishing solid training standards across all four layers is crucial for comprehensive performance evaluation, inviting to a multidisciplinary approach. 

Human digital twins offer several advantages for military training and operations\cite{fawkes_digital_2025}. 
They accelerate training, allowing more individuals to be ready in a shorter time frame, dealing with the current shortage of troops, by providing personalized feedback and integrating multiple performance dimensions. 
Human digital twins provide precise operational insights, from which trainers and commanders gain clear, data-driven evidence of an individual's competence and capabilities, improving decision-making for deployments and mission planning. 
Additionally, by simulating operational scenarios, human digital twins can familiarize warfighters with realistic stressful situations beforehand, reducing the mental strain during actual operations. This alleviates mental strain and supports a "deja-vu" effect during real missions. 

The successful implementation of Human Digital Twins relies on several technological enablers. 
Advanced sensing and data analytics allow collecting and processing large volumes of real-time data from wearables, biosensors, and activity trackers\cite{fawkes_digital_2025}.  
Edge computing and ad-hoc connectivity support reliable communication networks and on-device processing to handle resource-constrained, remote, or tactical environments.
Finally, extended reality provide immersive training experiences that closely replicate real-world conditions.

By integrating physical, technical, tactical, and mental data within a secure, scalable framework, Human Digital Twins represent a significant advancements in comprehensive training and performance optimization. 
However, to fully realize their potential, it is essential to develop standardized training protocols across all layers, particularly tactical and mental domains, and establish commonly agreed protocols to act in accordance with the legal, ethical, and social implications of HDTs\cite{glascoe_human_2024}. 
This holistic approach not only accelerates training but also ensures that individuals are better prepared for real-world challenges, ultimately enhancing mission success and safety. 

\section{Applications of Digital Twins in the Military Field}\label{sec:applications_of_dt_in_the_military_field}
Having established the conceptual foundations of digital twins and their core components, environments, assets, and humans, this section examines the diverse ways digital twin technology is applied across military operations. 
We focus on five primary domains: land, maritime, air, space, and cyber.
In each domain, digital twins are leveraged to enhance readiness, optimize logistics, improve situational awareness and decision-making. 
These practical use cases illustrate how digital twins can unify multiple data sources, provide realistic training simulations, enable predictive maintenance, and ultimately transform the way modern defense forces plan, execute, and learn from operations.

\subsection{Land Domain}\label{sec:land_domain}
Digital twins find strategic applications in the land domain by enhancing maintenance, situational awareness, and training of assets and humans. 
They enable real-time monitoring and simulation of battlefield conditions, support logistics and resource management, and allow detailed operational debriefing. 

A main use case for digital twins is the optimization of maintenance processes by monitoring and predicting the assets’ status exploiting the real-time collected data and deeply analyzing it in the virtual world. 
A thoroughly knowledge and accurate modeling of the monitored assets facilitates the implementation of predictive analytics and algorithms for anomaly detection, enabling an optimized and instance-specific maintenance schedule. 
The KNDS-Arquus NumCo project \cite{peter_felstead_knds_2024,knds_knds_2024} exemplifies this by developing a digital twin demonstrator for Armoured Infantry Fighting Vehicles, where health and usage data from onboard sensors feed into a virtual model. 
This allows predictive algorithms to estimate the remaining lifespan of key mechanical components based on the vehicle’s operational profile.
Conner et al. \cite{eddy_predictive_2024} similarly demonstrate how field data can be integrated into ground-vehicle digital twins for statistical failure prediction, while Song et al. \cite{song_architecture_2022} use digital twins for assessing equipment battle damage and support operational decisions.
Beyond maintenance, management and logistics functions benefit from deep digital modeling. 
Li et al. \cite{li_preliminary_2020} focus on digital twins of power systems and water supply infrastructure, applying AI to predict equipment states by combining historical and real-time data. 
Meanwhile, Booz Allen Hamilton Inc. and ARES Security Corporation \cite{noauthor_building_nodate}, along with Casey et al. \cite{casey_real-time_2024}, propose digital twins for military facilities and aircraft hangars, enhancing security, crisis response, and overall infrastructure management.
%
In general, within the predictive maintenance use case, digital twinning technologies are integrated with statistical or AI tools to forecast failures and recognize patterns in either the on-field acquired data or the synthetic data produced by the digital counterpart.  

Testing is another key use case for digital twins within the land domain. 
Virtual replicas closely mirroring real assets permit thorough design optimization without the cost or risk of physical trials.
Systems integrating AI components can be retrained on new realistic synthetic data in the simulated environment, then validated through the digital twin model.
Wang et al. \cite{wang_digital_2021} propose an online battlefield learning algorithm, demonstrating how digital twins for unmanned combat can facilitate equipment testing, performance evaluation, and strategic decision support.
This emphasis on simulation-enabled decision support aligns with established defense modeling and simulation practice, where discrete-event simulation \cite{collins_past_2021} and (increasingly AI-assisted) computer-aided wargaming \cite{davis_artificial_2022, langreck_modeling_2019} are used to explore tactics, assess trade-offs, and evaluate future capabilities.
Digital twinning can provide realistic training simulations for land-based operations, allowing military personnel to train in virtual environments that replicate actual terrains, equipment, combat scenarios, and adversary tactics\cite{stanney_performance_2021}. 
This includes tactical exercises, urban warfare training, and vehicle operation simulations\cite{collins_past_2021}. 
In 2018, Rheinmetall AG in cooperation with Rohde \& Schwarz GmbH \& Co KG founded RRS-MITCOS, which mission is to digitalize the German Army \cite{noauthor_rheinmetall_2018}, while benntec Systemtechnik GmbH \cite{noauthor_e-learning_nodate} is currently developing digital learning applications for both basic and specialized training. 
Specifically they employ digital twins of military vehicles jointly with VR technologies to technically and operationally train vehicle operators, maintenance personnel and instructors for their tasks in real operational environments. 
Nexter, a company of the KNDS group, developed Caesar Virtual Maintenance Training \cite{noauthor_knds_2024}, a fully digitalized training for maintenance tasks system providing simulated environment, vehicles and operators.  
Emerging 6G solutions in the field \cite{rohdeschwarz_magazine_looking_2024} promise even higher-fidelity simulations and gesture-based interactions, exploiting the increased bandwidth and boosted data throughput, pushing the boundaries of immersive, highly interconnection between the real and virtual realms. 
Digital twinning also permits to train data-driven models and systems, which require a large amount of data. 
But when dealing with defence application and scenario, data is often lacking. 
Whether it is for human or model training, realistic simulated environments and simulation platforms are therefore essential.  

Digital twins can significantly improve land-based operational planning by creating virtual models of deployment areas, analysing terrain features, and simulating mission scenarios. 
Commanders can optimize resource allocation, assess potential risks, and devise more effective strategies. During mission execution, the digital twin can monitor equipment status and remote operations, as exemplified by Singh et al. \cite{singh_physical-virtual_2024} in the context of hazardous duties like explosive ordnance disposal.
Wang et al. \cite{tan_construction_2024} propose a battlefield digital twin architecture where a command-and-control system operates as a central node in a system-of-systems framework, supporting mission planning, real-time monitoring, and debriefing.  
The EDF project MoSaiC \cite{noauthor_mosaic_2022} integrates drone-based monitoring and sampling for CBRN (Chemical, Biological, Radiological, or Nuclear) threat assessments, while Zhang \cite{zhang_digital_2024-1} focuses in search and rescue operations in a drone-assisted ground network scenario, illustrating the cross-domain synergies between ground and air assets.
When coming to Operational Planning, Execution and Monitoring or Debriefing, an accurate knowledge of the situation is required. 
This is achieved by reinforcing the data acquisition and communication channels, by implementing suited sensors and actuators able to achieve the system requirements, e.g. data throughput, latency, robustness, and acquisition frequency. 
It is also desired to handle situation in which acquired data go missing, thus being able to keep the operation on going by simulating the scenario. 
While storing massive datasets can be expensive, targeted data retention of essential mission elements enables thorough replay and analysis. 
User interfaces and data processing pipelines must be designed for clarity and efficiency, ensuring that relevant insights, such as equipment performance or tactical decisions, are readily accessible post-mission. 

Overall, digital twins in the land domain span predictive maintenance, realistic training simulations, and advanced planning and monitoring tools, all of which enhance operational effectiveness and resource optimization. 
As sensor technologies evolve and networking capabilities expand, these digital twin applications will continue to deepen and broaden, driving innovation and improved outcomes on the battlefield.

\subsection{Maritime Domain}\label{sec:maritime_domain}
Digital twin technology is reshaping the military maritime domain by improving operational efficiency, safety, and strategic planning. 
Key use cases include maintenance and repair of naval vessels, real-time monitoring of onboard systems and structures, training for naval personnel, operational planning and simulation, and port logistics and management. 

Digital twins enable real-time monitoring of a ship's systems and structures, facilitating predictive maintenance. 
By analyzing onboard sensor data, potential issues can be identified before they lead to system failures, or at their earliest stages.
This reduces downtime and extends the lifespan of naval assets.  
A notable example is Fincantieri S.p.A.’s FCX30 military vessel, whose digital twin solution supports the entire product life cycle, from logistics and energy optimization to predictive maintenance and online decision support \cite{noauthor_fcx_2023}.
It is designed to be fully interoperable with allied navies and to be able to be refitted with new technologies to anticipate or quickly react to new scenarios. 
The FCX30’s digital twin leverages a data-centric architecture to facilitate these processes, driving a shift in many manufacturers’ business models. 
%
In the service phase, Choi et al. \cite{choi_digital_2022} used system dynamics to integrate heterogeneous, multidisciplinary data and protocols, building digital twins of naval ships for operations and maintenance.
Similarly, Navantia, S.A. presents in \cite{silvera_navantias_2020} their ongoing projects regarding digital twin implementation on military navy. 
The F110 class frigate, in production since April 2022 \cite{noauthor_prime_2022}, already includes a digital twin able of connecting, extracting, and processing data from overboard equipment and crew members. 
It further supports maneuvering and piloting ships for training purposes in a realistic environment. 
The correlated NAVANTIS product suite further enhances training and subsystem maintenance.

Digital twins also serve as dynamic and interactive training platforms, allowing personnel to practice real-world procedures under simulated conditions. 
By replicating various scenarios, including emergencies, trainees develop safer and more effective responses at sea.
%
For instance, Major et al. \cite{major_real-time_2021} demonstrated an online ship digital twin for remote crane operations, integrating data from GPS, wind sensors, Motion Reference Units (MRUs) and Wave Sensors, and crane PLCs. 
This setup transmits real-time information to an onshore mirror via a 4G connection, allowing onshore operators to pilot missions safely through a simulator.
Digital twins find effective application in model training scenarios, even with limited data availability. 
Blachnik et al. \cite{blachnik_development_2023} create a digital twin of an underwater environment containing unexploded ordnance (UXO) and non-UXO objects for generating datasets to be used for training machine learning models within the task of scanning underwater areas using magnetometers. 
Digital twins find themselves extremely useful when it comes to training, whether it is for personnel or for models. 
They drastically cut costs, providing training environment which are ready to use and endlessly reusable. 
This requires appropriate software and simulation platform, as well as low latency equipment and sensors to provide a realistic experience for human training, and HPC platforms to speed up and achieve successful model trainings. 

Operational and tactical planning benefit significantly from maritime digital twins. 
Commanders can simulate multi-ship maneuvers, test strategies, and anticipate possible outcomes. 
This capacity to explore different operational strategies in a safe, controlled environment improves decision-making and helps develop more robust contingency plans. 
As with training, these simulations demand accurate modeling for reliable, mission-critical results.
Maritime surveillance and monitoring can similarly be enhanced via digital twin solutions. 
General Dynamics Mission Systems–Italy’s BlueSHIELD platform \cite{noauthor_superior_nodate}, for example, collects and fuses data from numerous sensors to create a unified maritime picture that improves situational awareness and informs decision-making through built-in AI analytics.

Digital twins in the maritime sector also facilitate ship design and construction. 
By virtually modeling a vessel during the design phase, engineers can predict its performance under diverse conditions and make adjustments before physical manufacturing begins. 
This leads to more efficient ship designs, optimized for performance and resilience.
Saab Dynamics applies this concept when designing torpedoes and underwater weapon systems, including Li-ion battery digital twins \cite{modeer_towards_2023}. 
In military contexts, data scarcity is a persistent challenge, often due to GNSS-denied environments and silent operations. Consequently, developing and validating these models under such constraints can be challenging but crucial for optimizing the final system performance.

Beyond onboard applications, digital twins can optimize port operations by simulating vessel traffic, cargo handling, and infrastructure usage. 
OHB SE, for example, developed a digital twin of the port of Bremen combining satellite data, smart sensors and AI, integrating the digital twin with indirect tracking \cite{noauthor_this_2021}. 
Although their system acts more like a “digital shadow” than a full bidirectional digital twin, it highlights how diverse data sources, satellites, UAVs, and on-machine sensors, can be integrated for situational awareness and streamlined workflows. 
Further refinements, such as adding more data layers or incorporating predictive functions, would align more closely with a comprehensive digital twin architecture.

Overall, digital twins in the maritime domain provide a versatile foundation for predictive maintenance, personnel training, strategic planning, and logistics optimization. 
As sensor integration, data-processing capabilities, and connectivity solutions advance, the depth and efficacy of these applications will continue to grow, underscoring digital twins’ pivotal role in modern naval operations.

\subsection{Air Domain}\label{sec:air_domain}
In the air domain, digital twinning revolves around modeling aircraft systems and flight operations to optimize maintenance schedules, predict failures, and boost pilot training through highly realistic simulations. 
By providing safe, cost-effective environments for testing new tactics and technologies\cite{lunsford_evaluation_2021}, digital twins allow operators to experiment without the risks and overhead of real-world trials.

Digital twins can be used to monitor and analyse the performance of aircraft systems and state of their components, from the manufacturing stages to the deployment and end-of-service. 
By creating virtual replicas of aircraft subsystems, operators can track system health, predict maintenance needs, and optimize maintenance schedules, resulting in increased operational readiness. 
Many works address these use cases. 
Lockheed Martin Corporation declares to be developing structural digital twins of military vehicles, specifically in their F35 Lighting II series, helping optimizing their manufacturing process \cite{noauthor_delivering_2021}. 
The use of digital partnership technologies in the manufacturing process helps validate the design, manage and optimize production lines, improve the performance of the product, and test the flight software, in addition to allowing for online anomaly detection and optimized fleet maintenance planning.  
The SAMAS 2 project \cite{panagiotopoulou_samas_2024}, coordinated by the European Defense Agency, targets military helicopters, leveraging digital twins to monitor corrosion and ballistic damage during missions and thus boost aircraft availability.
Kraft \cite{kraft_air_2016} integrates fluid dynamics into digital twins, merging physics-based modeling with experimental data to form a robust representation of air vehicle life cycles. 
Agrawal et al. \cite{agrawal_deep_2024} apply deep learning to data from multiple combat aircraft systems for health assessment. 
These examples underscore the high accuracy essential for predictive maintenance, as even minor model inaccuracies can significantly impact flight safety and performance.

Digital twins enable immersive flight simulation and training by recreating virtual cockpits, aircraft models, and flight environments. Pilots can practice complex maneuvers, emergency procedures, and mission scenarios in near-realistic conditions. Enhanced situational awareness and predictive capabilities further support tasks such as remote or autonomous control of unmanned aircraft, route optimization, and ground-decision processes.
Italian defense company Leonardo S.p.A. uses digital twin technologies for drones and helicopters, accelerating production cycles and improving operational performance and predictive maintenance \cite{noauthor_leonardo_2022}. 
Airbus SE and BAE Systems plc, in collaboration with Leonardo, plan to integrate XR technology into the Eurofighter Typhoon for more efficient maintenance and crew training \cite{noauthor_technology_2022}. Meanwhile, Wang et al. \cite{wang_research_2021} emphasize cloud computing as a foundation for large-scale UAV digital twins, enhancing mission planning, execution, and predictive maintenance with AI-assisted command, control, and situational analysis.
Specifically, cloud computing can enhance the efficiency of virtual and real networking collaboration. 
Moreover, cloud computing allows independent and solid situational awareness through AI-assisted command, control, and assessment analysis. 
Structural integrity represents another area of focus, focusing on leveraging simulation techniques \cite{lunsford_evaluation_2021}. 
Kapteyn et al. \cite{kapteyn_predictive_2022} propose real-time structural health monitoring for UAVs, dynamically adapting missions to balance vehicle preservation, mission aggressiveness, and operational effectiveness. 
Pinello et al. \cite{pinello_preliminary_2024} develop a digital twin for a nose landing gear within Simulink, generating signal data for subsequent training of damage-detection algorithms.
Digital twinning technologies can be also used for testing UAVs and plan their missions in highly realistic replica of the environment in which to be later deployed. 
The lack of data plays a crucial role in effective training of agents and models, especially when implementing deep data-driven techniques. 
One solution involves the creation of ad-hoc testbeds \cite{darema_hardware_2020}, however, this approach is often both cost- and time-intensive. 
Therefore, environments need to be digitalized in order to use them for offline virtual training, and simultaneously they require to incorporate as many details and attributes as possible achieving high levels of reality according to the specific training requirements. 
Shen et al. \cite{shen_multi-uav_2023} use a digital twin to train multiple UAVs via multi-agent deep reinforcement learning in a dynamic environment with constrained sensing and communication. 
This approach centralizes learning while distributing execution, accelerating model evolution.
Bayesian Belief Networks have been demonstrated to be capable of preemptively identify threats coming from adversarial UAVs, even when there is a lack of available historical data \cite{middeldorp_quanitfying_2023}.
Finally, Ji et al. \cite{ji_digital_2021} propose a comprehensive digital twin for mission planning, training, predictive maintenance, and online anomaly detection. Their modeling method intertwines geometric, physical, behavioral, and rule-based aspects, ensuring consistent bidirectional feedback between the physical quadrotor UAV and its digital counterpart. 

In the long run, digital twin technology will likely support every phase of a system’s life cycle—from design to manufacturing, operational service, and eventual retirement. 
Employing a unified digital twin across these stages avoids costly model transitions and allows manufacturers to offer “digital twin as a service” for diverse user needs. 
This holistic approach brings continuity and efficiency to the entire air-domain ecosystem, elevating safety, readiness, and innovation.

\subsection{Space Domain}\label{sec:space_domain}
Digital twins in the space domain focus on replicating satellites and other orbital assets to monitor health, detect anomalies, and plan or simulate a range of missions. 
By creating virtual counterparts of satellites, ground stations, and communication networks, operators can anticipate performance issues, test potential maneuvers, and optimize maintenance schedules. 
One notable application is space debris monitoring, where digital twins help model orbital debris trajectories and predict collision risks, informing safe evasive maneuvers for active satellites.

Slingshot Aerospace, for instance, offers a digital twin of space environments that integrates real-time orbital object mappings with space weather data \cite{keskerian_slingshot_2022}. 
A built-in physics-based simulator shows how planned missions may behave in the real environment, improving collision avoidance and long-term asset safety. 
Shangguan et al. \cite{shangguan_digital_2020} similarly propose a satellite digital twin that incorporates sensor readings, i.e. pressure, temperature, altitude, and payload data, into simulations for structural health monitoring, fault diagnosis, and dynamic mission adjustments.
Efforts toward standardizing digital twins in the aerospace sector include adapting ISO 23247 (Digital Twin Framework for Manufacturing) \cite{noauthor_iso_2021} to non-manufacturing contexts like orbit collision avoidance and debris detection \cite{shtofenmakher_adaptation_2024}. 
Establishing universal frameworks facilitates interoperability, reusability, and streamlined implementation across heterogeneous systems. 
Along similar lines, Lei et al. \cite{lei_digital_2024} present a digital twin construction, evaluation, and management framework for spacecraft systems, highlighting applications such as pre-mission simulation, real-time on-orbit monitoring, and rapid operational forecasting.
In turn, a framework for digital twin construction, evaluation, management, and implementation in spacecraft systems is presented in \cite{lei_digital_2024}. 
Some applications of digital twins in spacecraft systems are also highlighted, that are pre-mission ground simulation, on-orbit real-time, comprehensive monitoring, and rapid prediction of operating status.  
A position paper by the Aerospace Industries Association (AIA) and the American Institute of Aeronautics and Astronautics (AIAA) \cite{noauthor_digital_2020} underscores the value of digital twins throughout an aerospace system’s life cycle. 
Performance monitoring, validation, and optimization, as well as design refinement, upgrades, and predictive maintenance, all benefit from accurate virtual representations. 
This end-to-end integration often yields substantial cost reductions, as real-world trials, redesigns, and maintenance schedules become more precise.  

Space-based imaging from satellites forms the backbone of many large-scale digital twin applications, including the Digital Twin Earth concept \cite{bauer_digital_2021,defelipe_towards_2022,daya_sagar_digital_2021}. 
Satellites are considered as one of the main data sources when it comes up to Earth imaging, feeding many applications such as surveillance, surface reconstruction, among others. 
Data quality, frequency and responsiveness are key requirements use-case tailored.
The created platforms are meant to simulate in space and time specific phenomena given specific parameters. 
These global reconstructions incorporate terrestrial data (e.g., terrain, weather) to simulate near-future phenomena,be it dam-break scenarios, road traffic projections, maritime security shifts, or battlefield evolutions. 
Such dual-use capabilities appeal to both defense and civilian sectors, although real-time accuracy and interoperability remain pivotal. 
In particular, when multiple organizations or agencies share infrastructure and data sources, common standards and protocols become critical to sustaining robust, integrated digital twin ecosystems.

\subsection{Cyber Domain}\label{sec:cyber_domain}
In the cybersecurity domain, digital twins model networks, systems, and processes to expose potential vulnerabilities, predict emerging threats, and validate defense mechanisms in controlled, realistic scenarios.
By offering dynamic environments for threat simulation, incident response testing, and continuous monitoring, digital twins significantly enhance the readiness and resilience of military cyber infrastructure. 
They also play a growing role in securing supply chains and other critical elements of military operations.

Threat simulation is a critical, proactive application of digital twins in military cybersecurity, wherein a virtual replica of the cyber infrastructure is subjected to various hypothetical attacks.
This controlled environment helps experts identify vulnerabilities, from phishing to multi-vector intrusions, and assess the effectiveness of existing defense measures without risking real operations. 
It also serves as a hands-on training platform, enabling cybersecurity teams to practice incident response. 
By continuously iterating simulations and refining strategies, organizations can adjust their defensive position and more effectively mitigate real-world threats.
Digital twins provide a realistic and controlled environment where security teams can practice responding to cyber incidents without the risk of compromising actual military operations.
Furthermore, digital twins offer a safe testing ground for new cybersecurity technologies, allowing developers to validate solutions against high-fidelity threat scenarios in controlled environments before deployment. 
This reduces the risks of implementing novel defenses and ensures that critical infrastructure remains robust when under actual attack.

A network digital twin targets communication infrastructure. By using real-time and historical data, it accurately models network topology, protocols, traffic flow, and physical constraints. 
Military organizations can simulate cyberattacks on this environment to test resilience, identify vulnerabilities, and validate patching strategies.
Bagrodia \cite{bagrodia_using_2023}  explores cyber ranges based on virtual machines and network digital twins, highlighting how mission-centric analyses reinforce the defensive capabilities of military systems. 
Keysight Technologies' EXata \cite{noauthor_automated_2023} exemplifies a commercial solution for developing such digital twins, offering tools for design analysis, testing, and cyber assessment. 
Enhancements may incorporate additional layers, such as blockchain-based solutions \cite{avrilionis_towards_2021}, to ensure secure on-chain representation of real-world assets and off-chain state persistence.
Maathuis et al. \cite{maathuis_decision_2021} propose a modeling and simulation solution for effects classification and proportionality assessment in military cyber operations to support targeting decision. 
This method can serve as the core of a military cyber digital twin.
Vielberth et al. \cite{barker_digital_2021} create a cyber range simulating critical systems for Security Operations Center (SOC) analysts to practice detection and prevention techniques against simulated attacks.

Digital twins function as a continuous monitoring layer, tracking both physical and virtual assets in near real time. 
By integrating diverse data sources—network traffic, system logs, user behavior—digital twins allow cybersecurity teams to maintain situational awareness, rapidly detect anomalies, and tailor response strategies\cite{bagrodia_using_2023, clark_detection_2021}.
This holistic approach drives advanced threat intelligence and predictive analytics. 
Pattern analysis on synthetic or real-world datasets can forecast likely attack vectors, enabling preemptive defenses. 
The same infrastructure supports realistic training for cybersecurity operators, reinforcing skills in intrusion detection, incident response, and strategic planning\cite{prasad_machine_2022}.

Supply chains represent a crucial yet often vulnerable element in military logistics. Digital twins can visualize end-to-end operations, spotting potential disruptions, ensuring asset integrity, and streamlining decision-making. 
Although Sani et al. \cite{shafik_utilising_2022} report limited real-world adoption of digital twins for military supply chain visibility, the potential remains considerable.
Adami et al. \cite{adami_strategic_2023} further highlight digital twins’ ability to detect anomalies, enhance redundancy strategies, and counter cyber threats by learning adversarial behaviors. 
They emphasize, however, that data availability remains a major hurdle as machine-learning-driven analytics require robust datasets, suggesting a “defense data space”, or "data pools" shared among partners for maximum efficacy.
In the United States, the Global Combat Support System–Army (GCSS–Army) \cite{winbush_gcss-army_2021} is developing digital twin capabilities for ground and aviation logistics, bringing real-time insights into inventory, transportation, and delivery statuses under one web-based system. 
This integration aims to improve tactical logistics tasks and expand commanders’ situational awareness, ultimately reinforcing combat power on the battlefield.

Overall, digital twins within the cyber domain provide simulation, monitoring, and predictive analytics that strengthen military cybersecurity and supply chain resilience. 
As data-sharing frameworks and interoperability mature, these capabilities are set to expand, safeguarding critical digital infrastructure in increasingly complex threat environments.

\subsection{Cross-Cutting Use Cases}\label{sec:applications_of_dt_in_the_military_field_cross_cutting}
From our in-depth analysis of each specific defense domain, we identified several cross-cutting use cases of digital twin technology that transcend domain boundaries and are shared across land, maritime, air, space, and cyber contexts.
The identified use cases are as follows: 
\begin{itemize}
\item \textbf{Predictive Maintenance:}
Leveraging real-time data and analytics to anticipate failures and schedule maintenance tasks proactively, maximizing asset availability and lifespan\cite{glaessgen_digital_2012}.

\item \textbf{Design and Construction:}  
Employing digital twins early in the product life cycle to validate designs, optimize manufacturing processes, and ensure consistency from concept to deployment\cite{liu_research_2024}.

\item \textbf{Planning, Testing, and Simulation:}  
Using virtual representations to explore “what-if” scenarios, test operational plans, and evaluate system performance in controlled environments before real-world deployment\cite{collins_past_2021}.

\item \textbf{Training and Optimization:}  
Providing immersive, data-driven training solutions for personnel and systems, enabling iterative skill development, performance improvements, and rapid adaptation to changing missions\cite{alim_measuring_2024}.

\item \textbf{Execution and Monitoring:}  
Enabling real-time oversight and control of ongoing missions or operations, offering situational awareness and decision support through high-fidelity and up-to-date digital replicas of physical assets or environments\cite{zhang_actor_2022}.

\item \textbf{Logistics and Management:}  
Optimizing supply chains, resource allocation, and operational coordination by integrating diverse data sources into a shared, dynamic digital system\cite{parathyras_prospects_2025}.

\item \textbf{Debriefing:}  
Reconstructing, replaying and analyzing completed missions or operations to identify successes, failures, and opportunities for refinement, ultimately guiding future strategy and training\cite{langreck_modeling_2019}.
\end{itemize}

\begin{table*}[ht]
\caption{Illustrative examples of applications of digital twins in the military domain, with respect to the most relevant use cases and thematic domains.}\label{tab:application_examples_per_usecases_and_domains}
\centering
\renewcommand{\arraystretch}{1.2}
\scriptsize 
\hyphenpenalty=10000
\resizebox{\textwidth}{!}{
\begin{tabular}{p{0.1\textwidth}|p{0.2\textwidth}p{0.2\textwidth}p{0.2\textwidth}p{0.2\textwidth}p{0.2\textwidth}}
\toprule
& \textbf{Land}                                                                                                 & \textbf{Maritime}                                                                                     & \textbf{Air}                                                                                      & \textbf{Space}                                                                                     & \textbf{Cyber}                                   \\
\midrule
\textbf{Predictive Maintenance}           & Ground-based machinery \newline predictive maintenance                                                                 & Naval vessel predictive maintenance                                                                   & Aircraft predictive maintenance                                                                   & Predictive maintenance for operating satellites                                                   & Predictive maintenance for cybersecurity systems   \\
\midrule
\textbf{Design and Construction}          & DT of ground-based vehicle shared across design and construction stages among makers and users, XR integrated & DT of naval vessel shared across design and construction stages among makers and users, XR integrated & DT of aircraft shared across design and construction stages among makers and users, XR integrated & DT of satellite shared across design and construction stages among makers and users, XR integrated & Virtual network infrastructure design            \\
\midrule
\textbf{Planning, Testing and Simulation} & Battlefield DT for ground-based operational planning                                                          & Naval vessel DT testing and mission planning using virtual sea enviornment                            & Aircraft DT testing and mission planning using virtual air environments                           & Satellite-based DT for operational planning                                                        & Cyberattack simulation in secured DT environment \\
\midrule
\textbf{Training and Optimization}        & Human DT for optimized and efficient training                                                                 & Simulation of ship and submarine systems for training                                                 & Flight training using virtual environments                                                        & Aerospace training using virtual environments                                                      & Cybersecurity operators and systems training                  \\
\midrule
\textbf{Execution and Monitoring}         & Battlefield DT for ground-based operations                                                                    & DTs for naval vessel energy optimization and operational efficiency                                   & Remote execution and monitoring of aviation and drone missions                                    & Remote execution and monitoring of satellite operations                                            & Threat Prediction and Vulnerability Assessment   \\
\midrule
\textbf{Logistics and Management}         & DT for ground-based operations management and coordination                                                    & DTs for naval vessels management and coordination                                                     & Supply chains DT                                                                                  & Supply chains DT                                                                                   & Supply chains DT                                 \\
\midrule
\textbf{Debriefing}                       & Ground operations reconstruction and replay for debriefing                                                    & Maritime operations reconstruction and replay for debriefing                                          & Mission debriefing using virtual environments and recorded DT states                              & Mission debriefing using virtual environments and recorded DT states                               & Cyberattack analysis  \\                          
\bottomrule
\end{tabular}}
\end{table*}

Table \ref{tab:application_examples_per_usecases_and_domains} provides illustrative examples of domain-specific implementations and insights for each of these generalized use cases.
The general use cases shown in the table, highlight how digital twins can offer benefits regardless of the specific domain, underscoring the universality and versatility of digital twin technology within modern defense strategies.
Embracing this transversal perspective makes it clear that cross-domain digital twins, tailored to particular needs as required, form the foundations of successful and efficient digital twin development, enabling seamless adaptation and enhanced operational effectiveness across multiple defense domains.

\begin{figure*}[!htb]
\centering
\captionsetup[subfigure]{justification=centering}
\begin{subfigure}{0.49\textwidth}
\centering
  \includegraphics[width=0.9\textwidth]{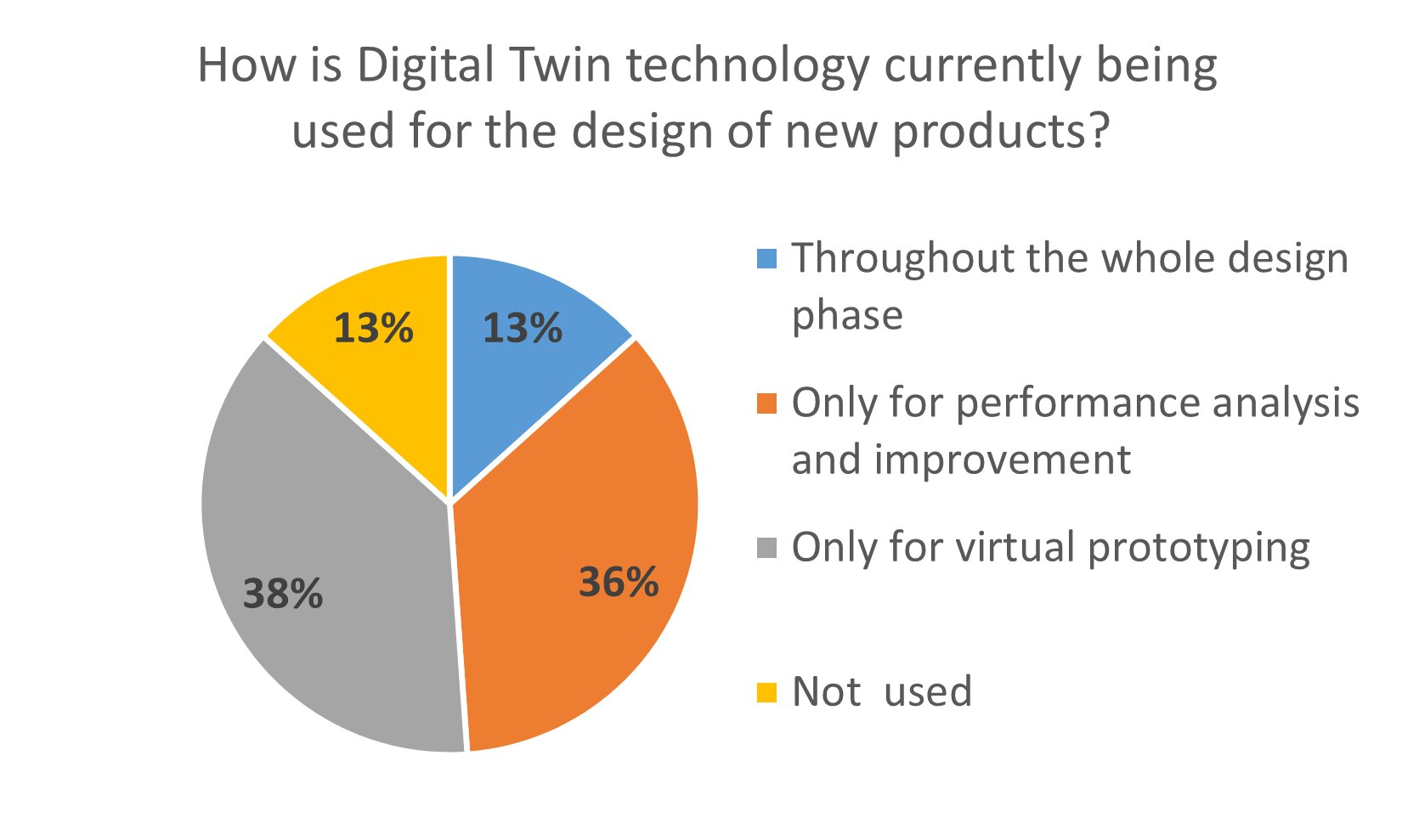}
  \caption{}\label{fig:survey_applications_design}
\end{subfigure}
\begin{subfigure}{0.49\textwidth}
\centering
  \includegraphics[width=0.9\textwidth]{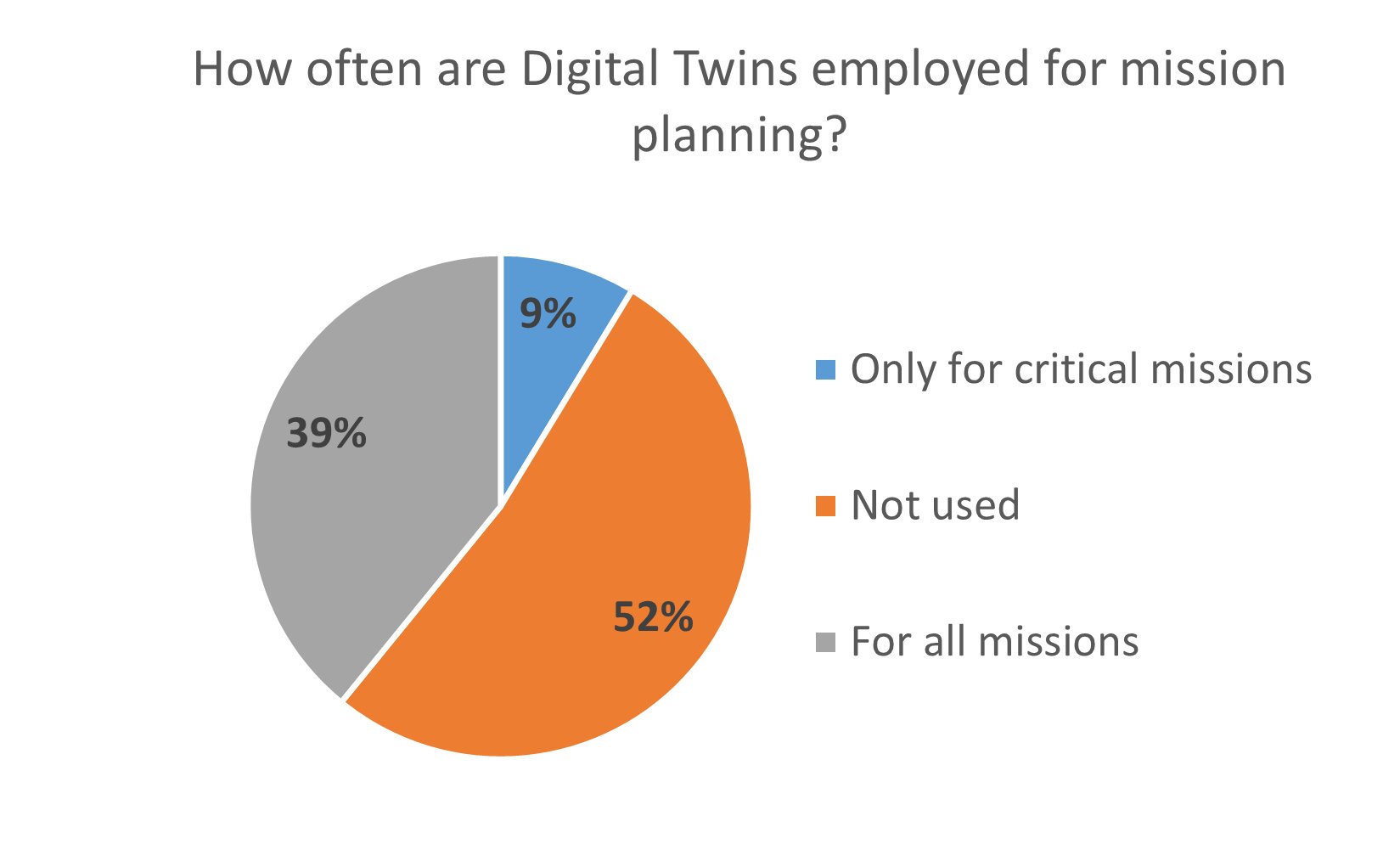}
  \caption{}\label{fig:survey_applications_planning}
\end{subfigure}
\\[2ex] 
\begin{subfigure}{0.49\textwidth}
\centering
\includegraphics[width=0.9\textwidth]{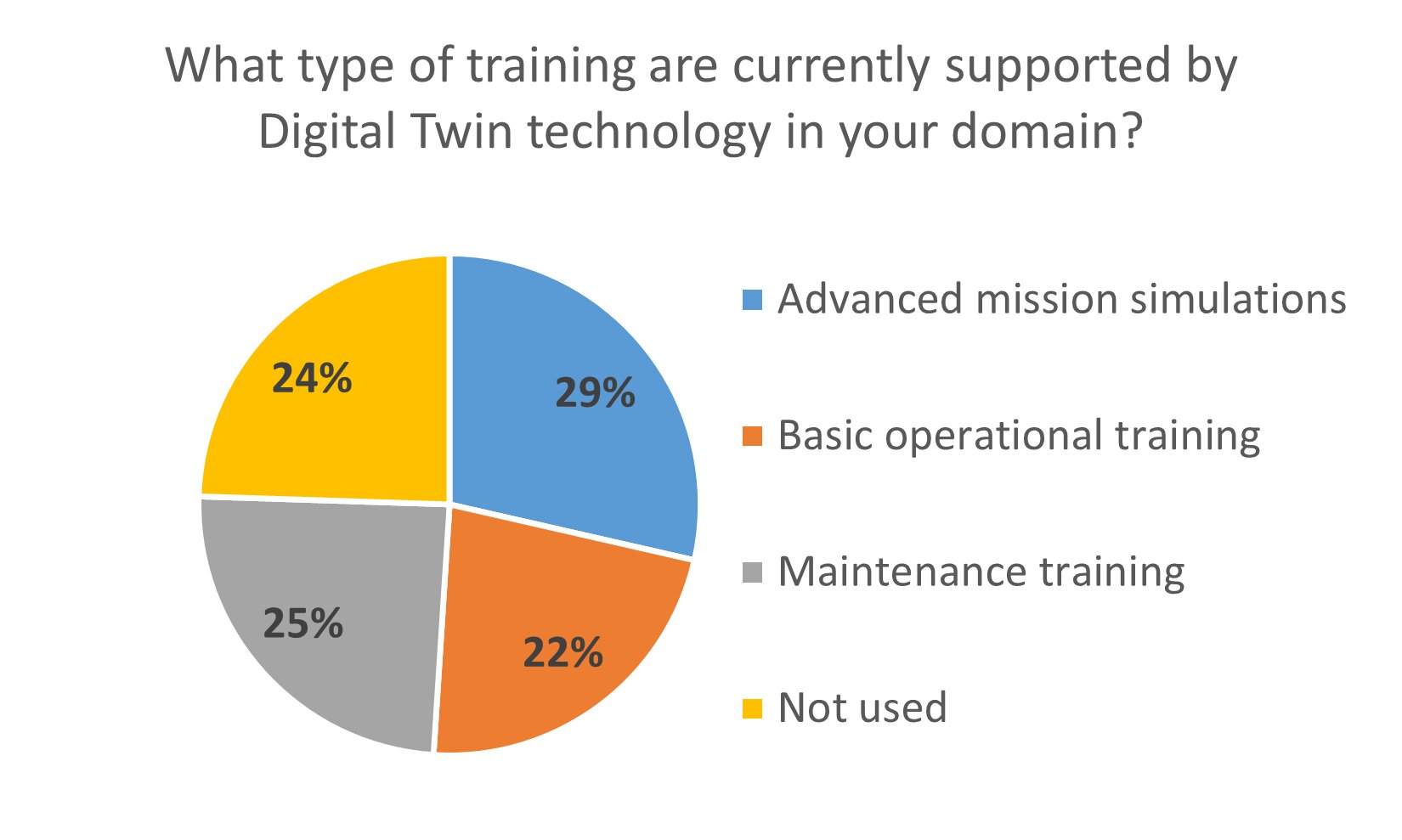}
\caption{}\label{fig:survey_applications_training}
\end{subfigure}
\begin{subfigure}{0.49\textwidth}
\centering
\includegraphics[width=0.9\textwidth]{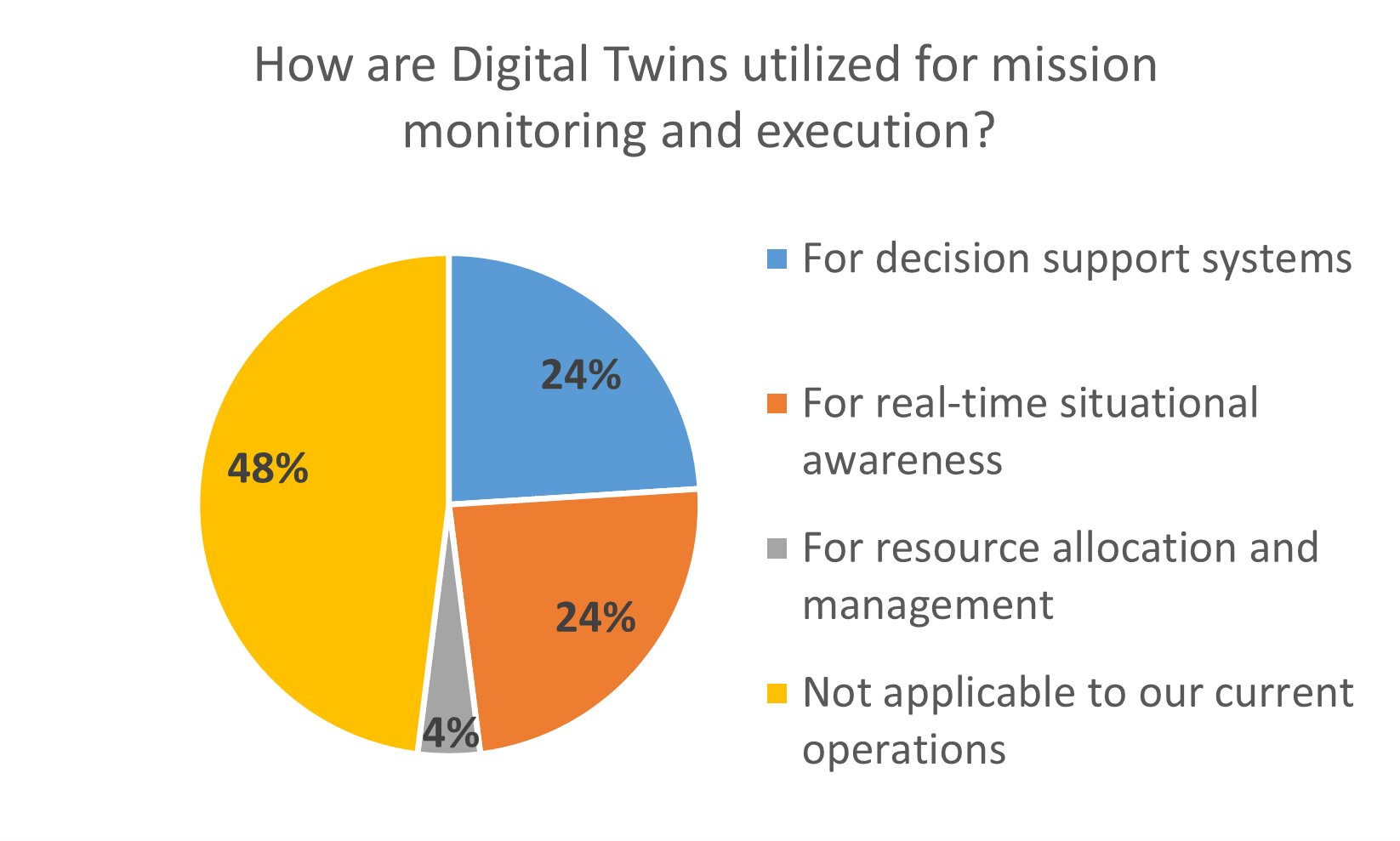}
\caption{}\label{fig:survey_applications_execution}
\end{subfigure}
\\[2ex] 
\begin{subfigure}{0.49\textwidth}
\centering
\includegraphics[width=0.9\textwidth]{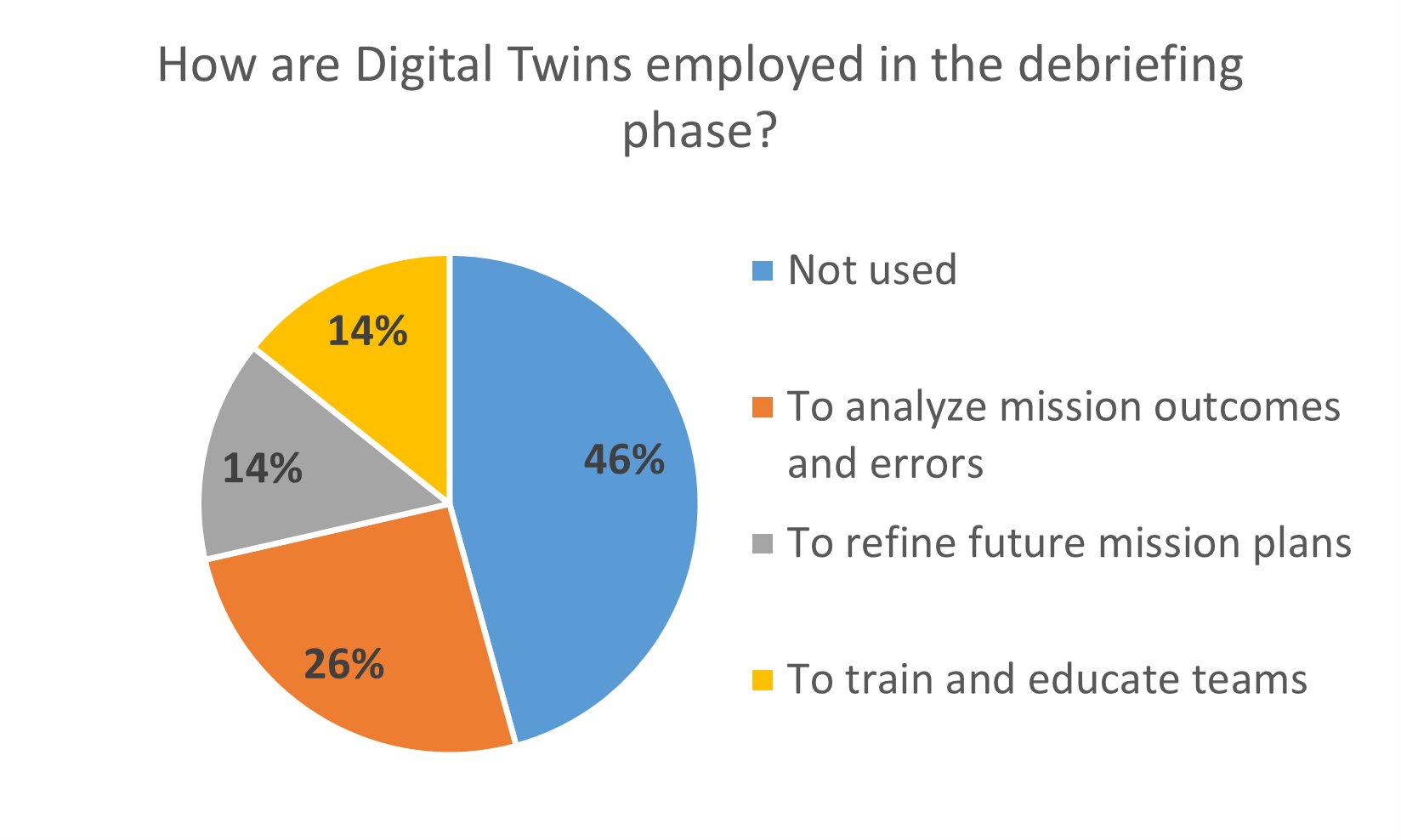}
\caption{}\label{fig:survey_applications_debrifieng}
\end{subfigure}
\begin{subfigure}{0.49\textwidth}
\centering
\includegraphics[width=0.9\textwidth]{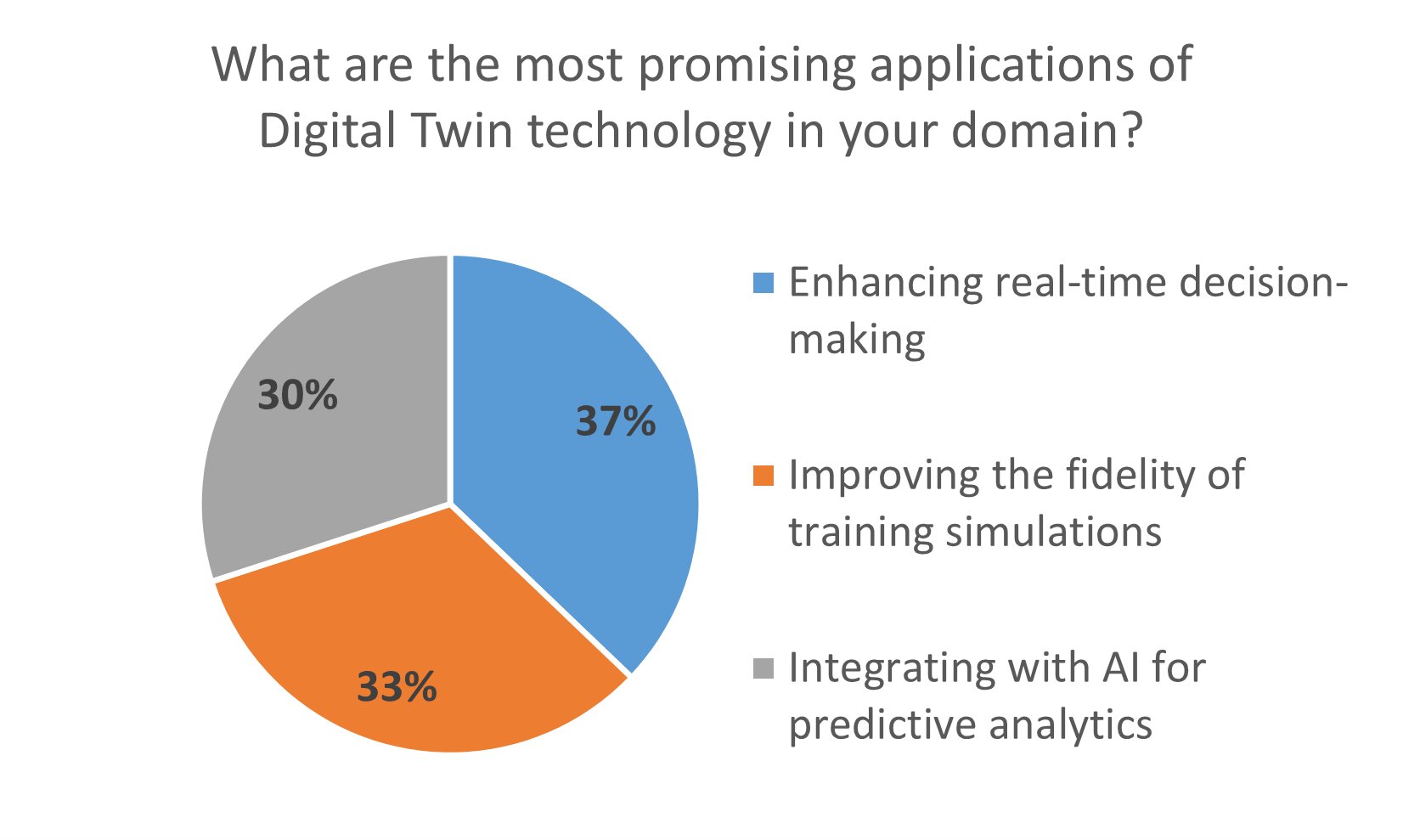}
\caption{}\label{fig:survey_applications_promising_application}
\end{subfigure}
\\[1ex] 
\caption{Results of the questionnaire related to current integration of digital twin technologies with the identified cross-domain use cases.}\label{fig:survey_applications}
\end{figure*}

\begin{figure*}[!t]
\centering
\captionsetup[subfigure]{justification=centering}
\begin{subfigure}{0.49\textwidth}
\centering
  \includegraphics[width=0.9\textwidth]{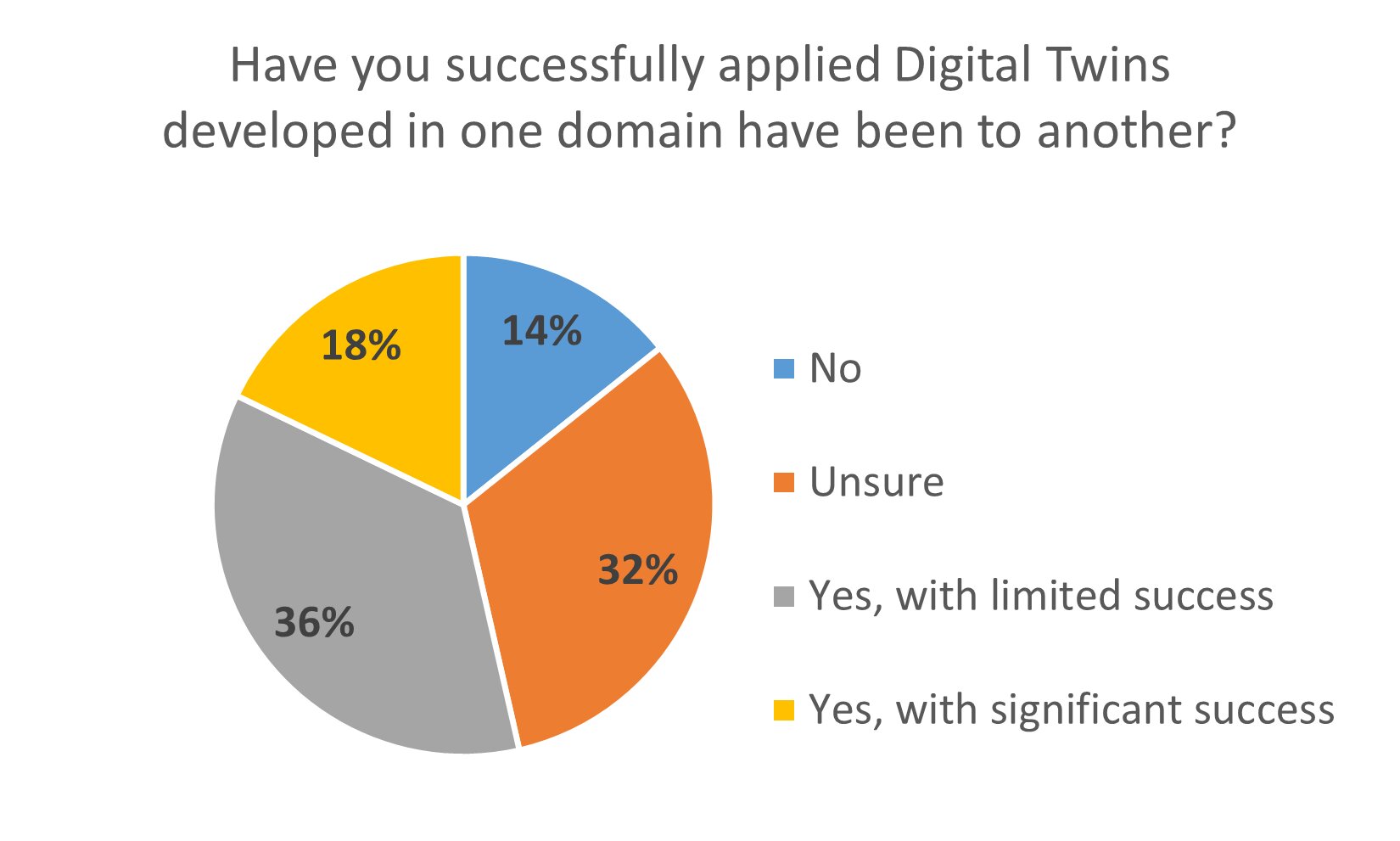}
  \caption{}\label{fig:survey_cross_domain_adaptability}
\end{subfigure}
\begin{subfigure}{0.49\textwidth}
\centering
  \includegraphics[width=0.9\textwidth]{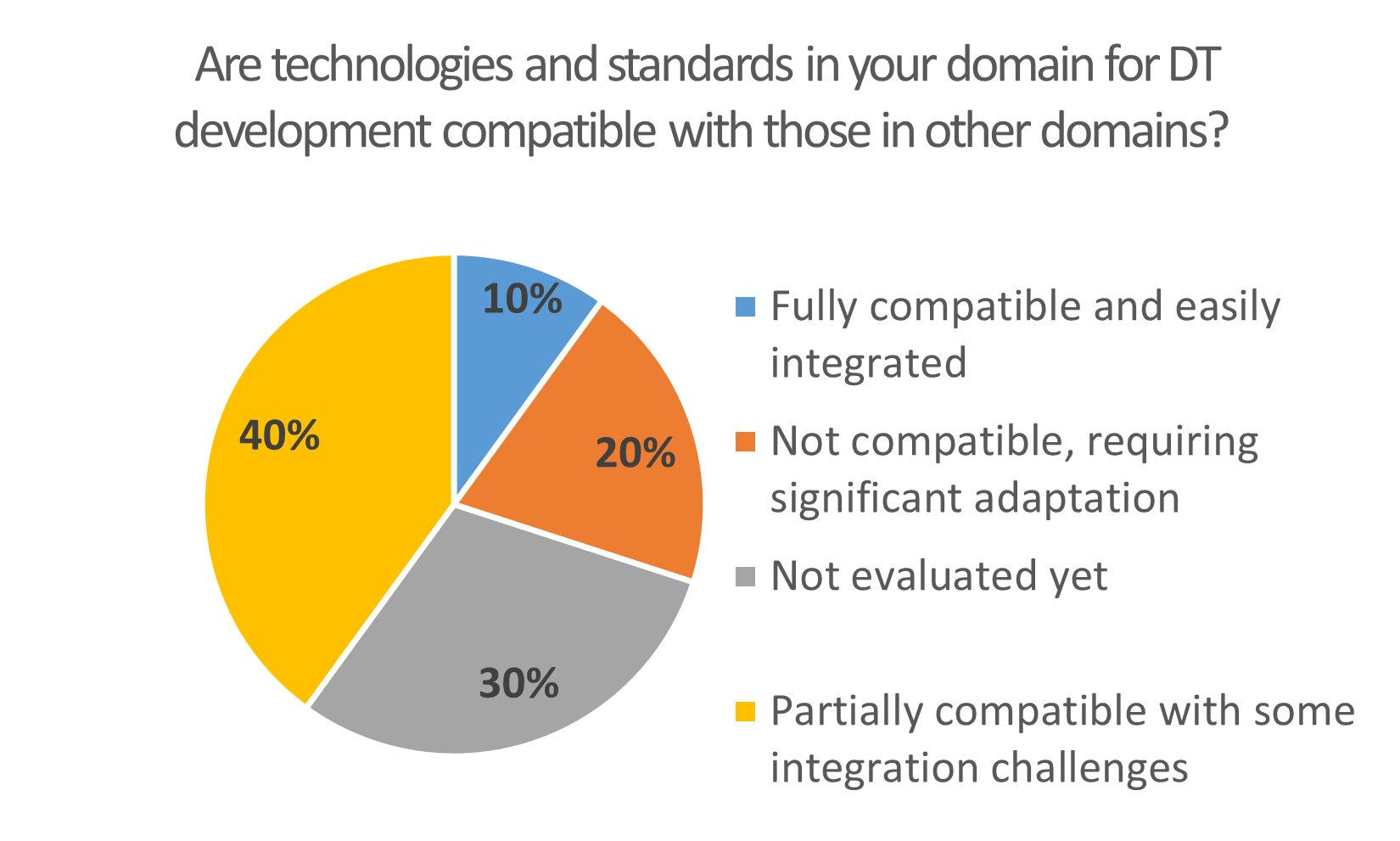}
  \caption{}\label{fig:survey_cross_domain_technologies}
\end{subfigure}
\\[2ex] 
\begin{subfigure}{0.49\textwidth}
\centering
\includegraphics[width=0.9\textwidth]{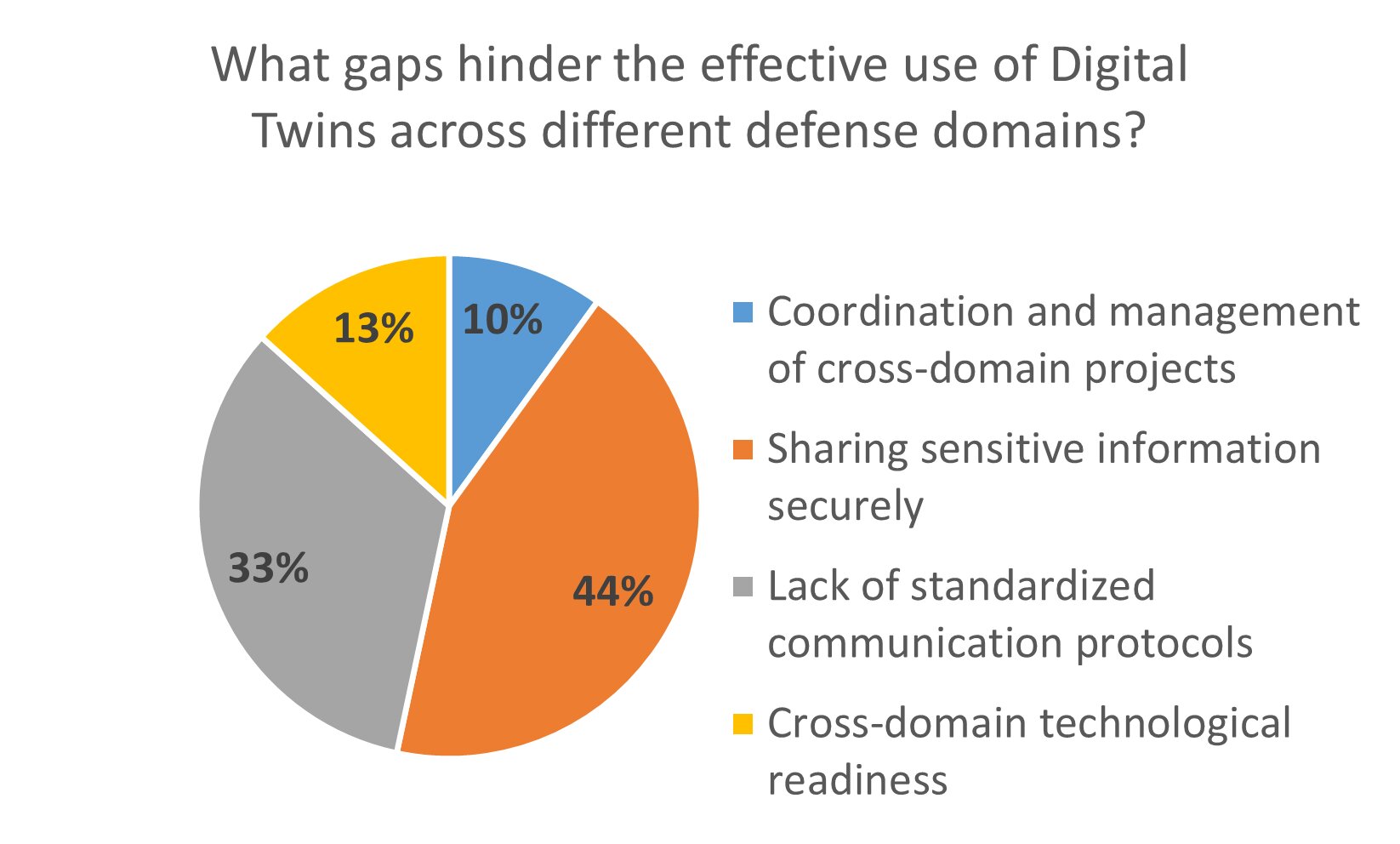}
\caption{}\label{fig:survey_cross_domain_gaps}
\end{subfigure}
\begin{subfigure}{0.49\textwidth}
\centering
\includegraphics[width=0.9\textwidth]{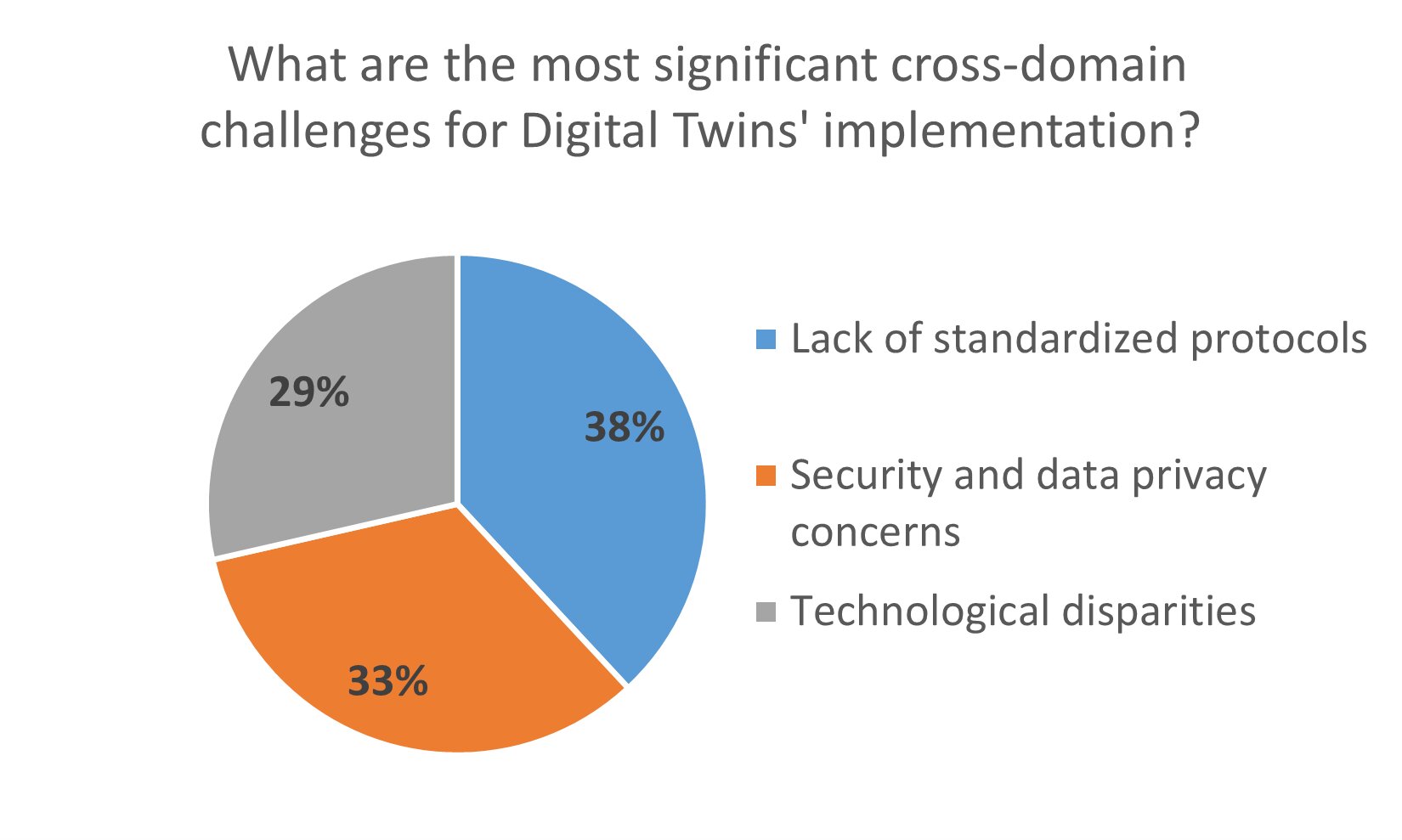}
\caption{}\label{fig:survey_cross_domain_challenges}
\end{subfigure}
\\[2ex] 
\begin{subfigure}{0.49\textwidth}
\centering
\includegraphics[width=0.9\textwidth]{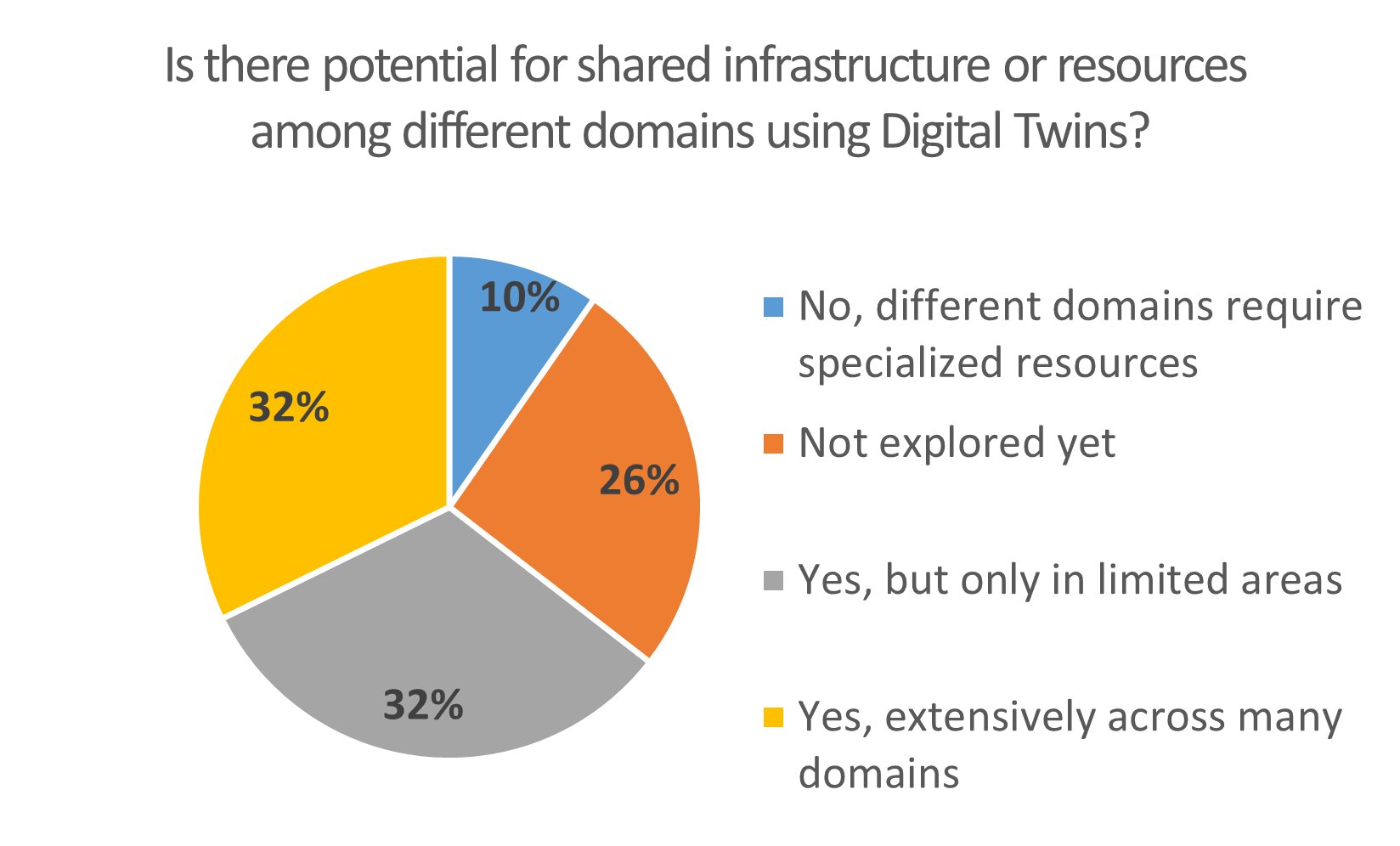}
\caption{}\label{fig:survey_cross_domain_shared_infrstructure}
\end{subfigure}
\begin{subfigure}{0.49\textwidth}
\centering
\includegraphics[width=0.9\textwidth]{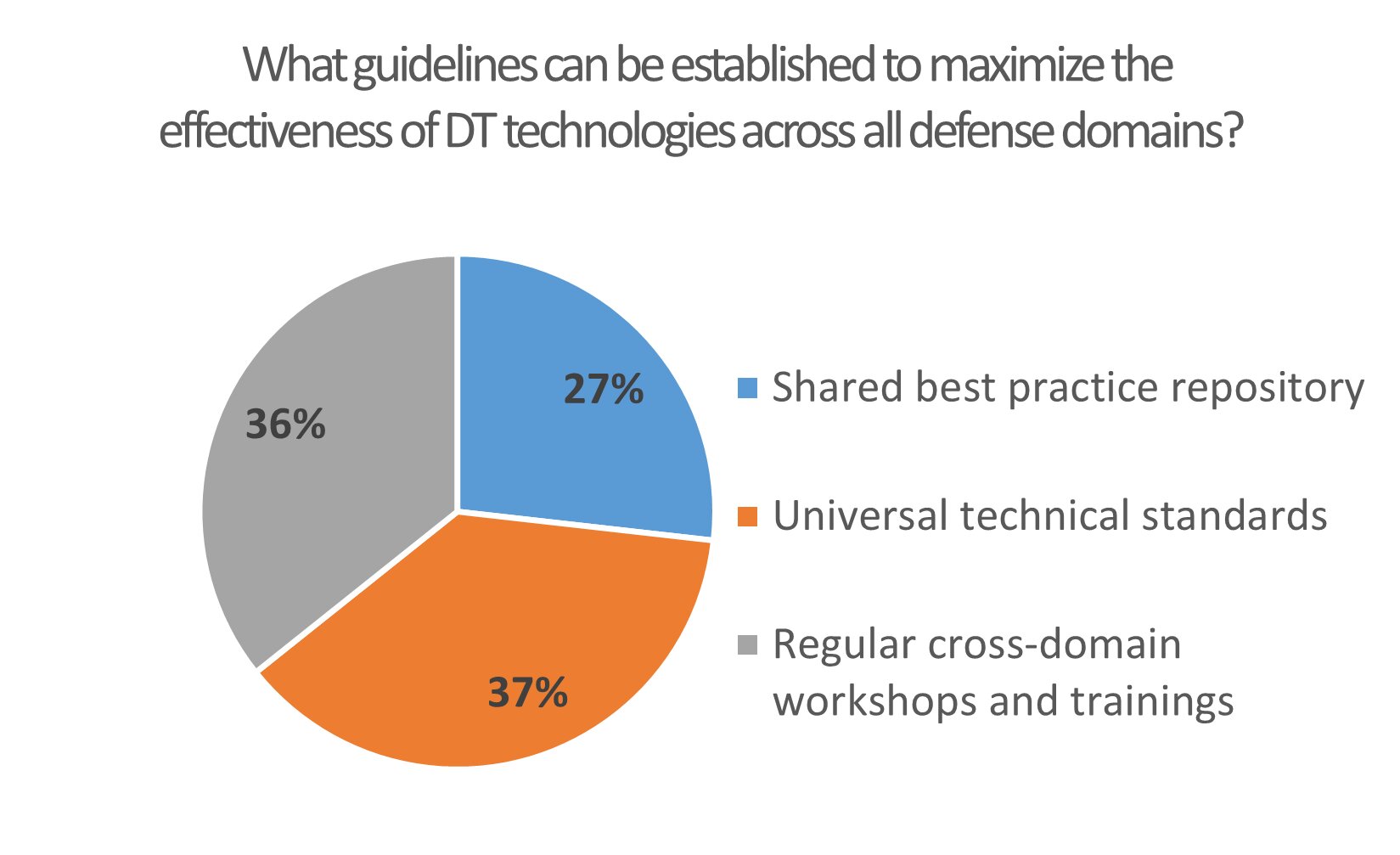}
\caption{}\label{fig:survey_cross_domain_common_guidelines}
\end{subfigure}
\\[1ex] 
\caption{Results of the questionnaire related to cross-domain usage of digital twins in the military field.}\label{fig:survey_cross_domain}
\end{figure*}

\begin{figure*}[!t]
\centering
\captionsetup[subfigure]{justification=centering}
\begin{subfigure}{0.49\textwidth}
\centering
\includegraphics[width=0.9\textwidth]{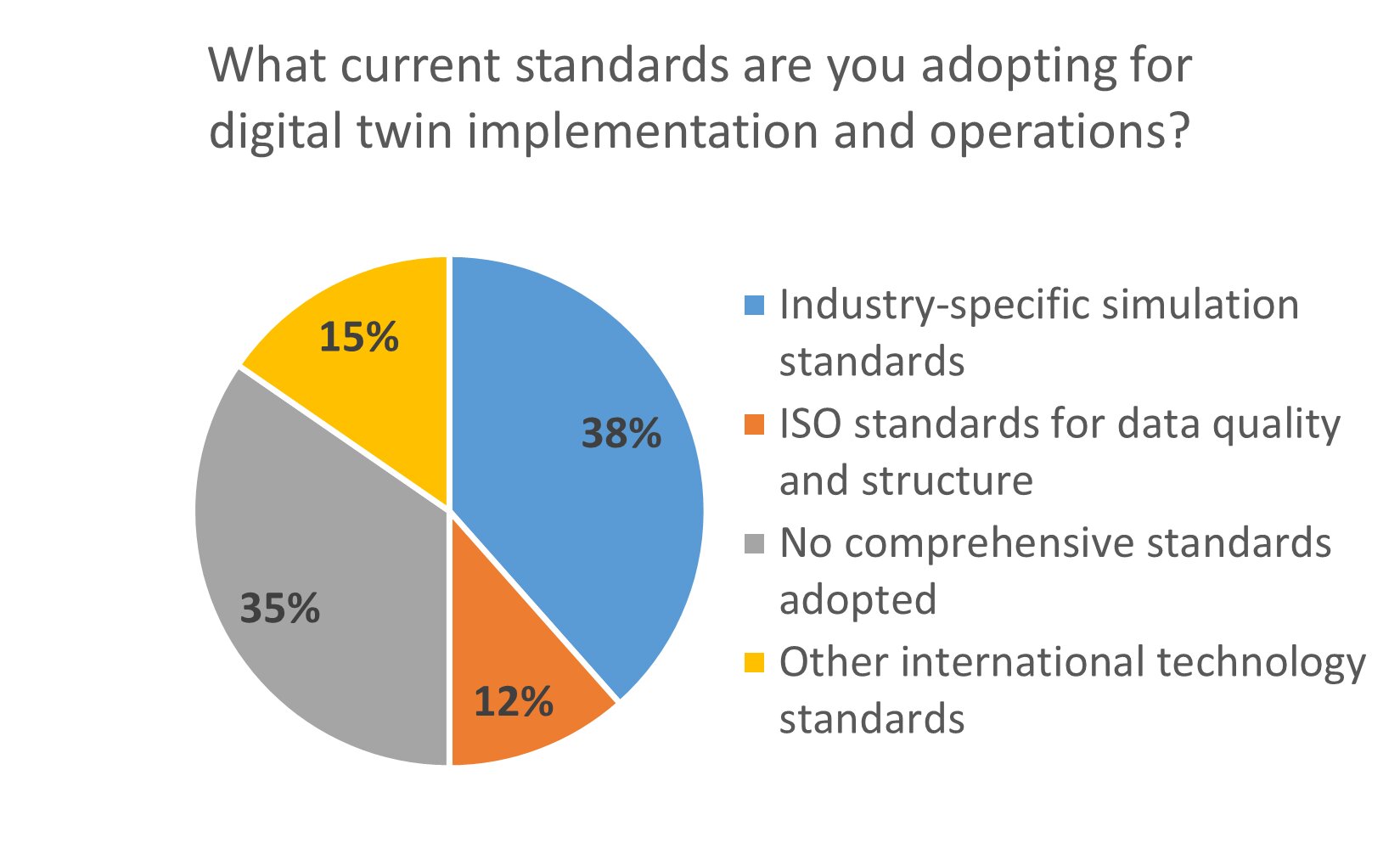}
\caption{}\label{fig:survey_standardization_currently_adopted}
\end{subfigure}
\begin{subfigure}{0.49\textwidth}
\centering
\includegraphics[width=0.9\textwidth]{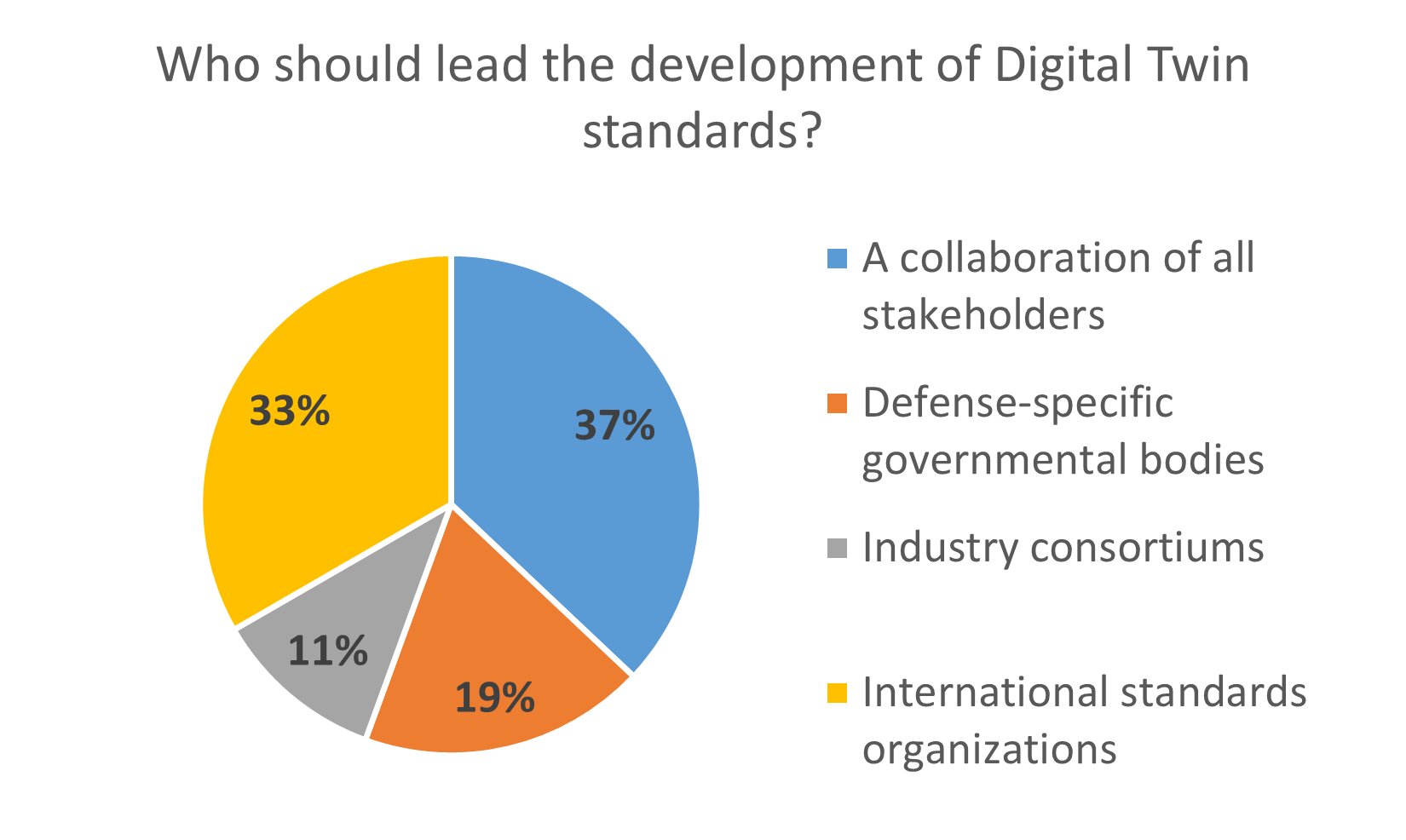}
\caption{}\label{fig:survey_standardization_leaders}
\end{subfigure}
\\[2ex]
\begin{subfigure}{0.49\textwidth}
\centering
  \includegraphics[width=0.9\textwidth]{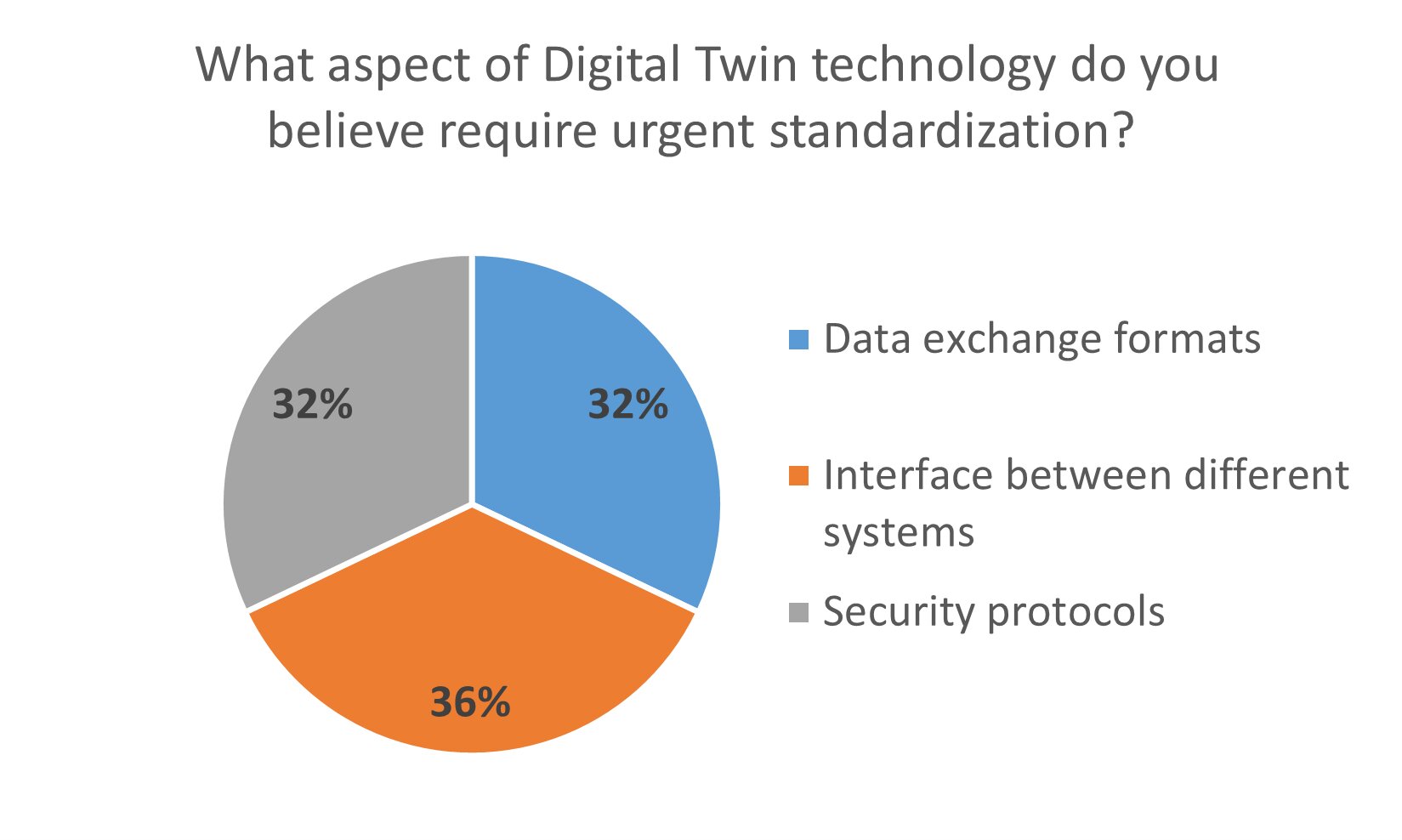}
  \caption{}\label{fig:survey_standardization_required}
\end{subfigure}
\begin{subfigure}{0.49\textwidth}
\centering
  \includegraphics[width=0.9\textwidth]{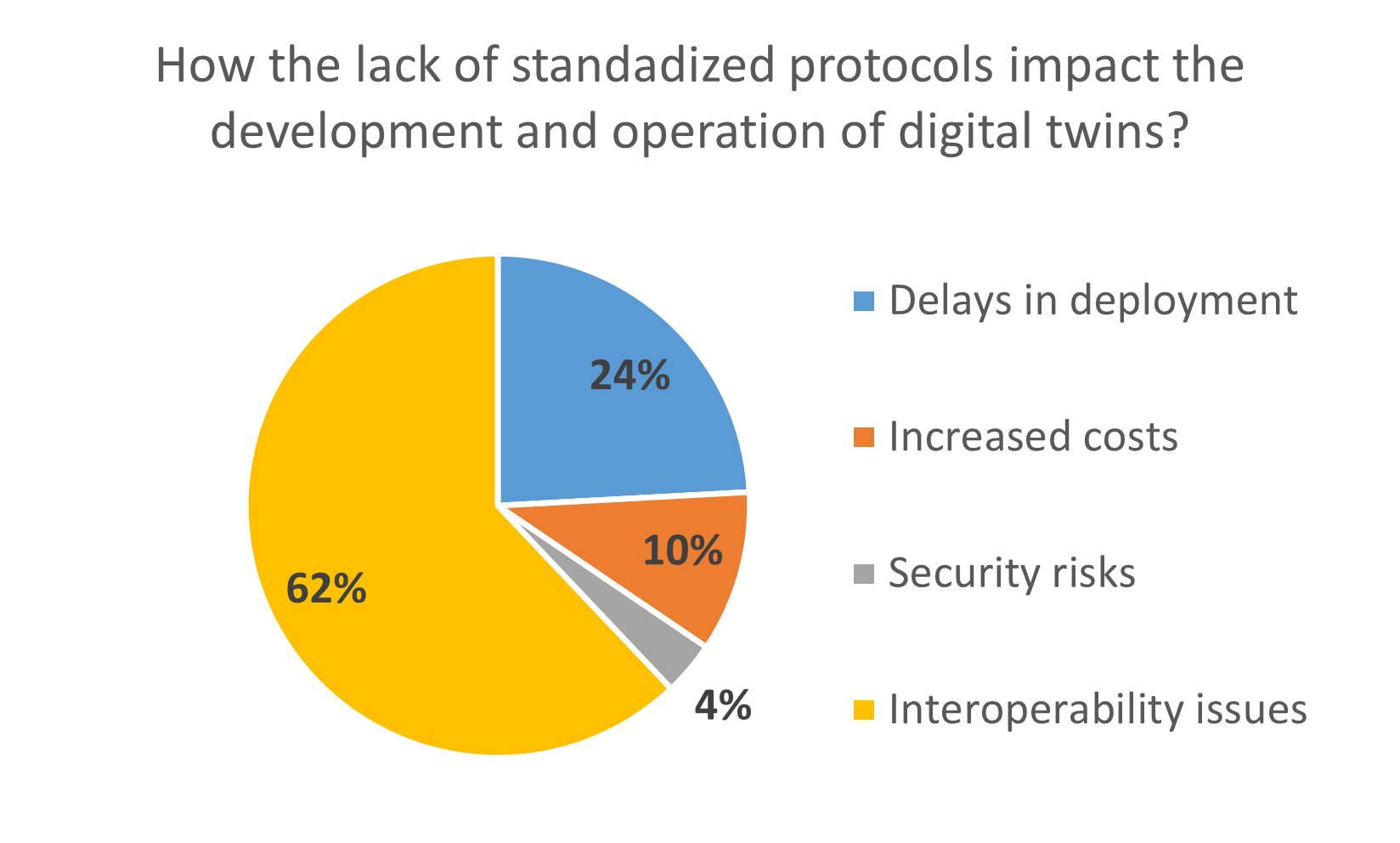}
  \caption{}\label{fig:survey_standardization_lackness_impact}
\end{subfigure}
\\[1ex] 
\caption{Results of the questionnaire related to standardization of digital twin technologies in the military field.}\label{fig:survey_standardization}
\end{figure*}
Figure ~\ref{fig:survey_applications} reports the results from the questionnaire sent to defence industries and ministries regarding the applications of digital twins across the identified general and cross-cutting use cases. 
Related to the design and construction use case (Figure ~\ref{fig:survey_applications_design}), digital twin technologies are already extensively employed by industries and governmental bodies in $87\%$ of the responses.This includes comprehensive use $(13\%)$ or partial use ($38\%$ exclusively for prototyping, $36\%$ for performance analysis).
This is consistent with the fact that digital twins were originally designed for manufacturing purposes, making them deeply integrated and effective in these contexts already.
When it comes to mission planning (Figure ~\ref{fig:survey_applications_planning}), a notable $52\%$ of respondents indicated that digital twins are not yet in use.
In contrast, $39\%$ reported using digital twins for all missions, and $9\%$ only for critical missions. This suggests that while adoption is still in progress, once implemented, there is a tendency to utilize digital twins consistently across various mission types.
In terms of training (Figure ~\ref{fig:survey_applications_training}), digital twins are already in use by $76\%$ of the respondents, with the most significant application being for advanced mission simulation ($29\%$), followed by maintenance training ($25\%$) and basic operational training ($22\%$). This demonstrates the strong potential of digital twins to enhance training efficiency and realism in military contexts.
For mission monitoring and execution (Figure ~\ref{fig:survey_applications_execution}), digital twins are less integrated, with only $52\%$ of responses indicating usage.
The primary applications are for decision support system systems and real-time situational awareness, emphasizing the need for advanced decision-making tools.
Finally, for the debriefing use case (Figure ~\ref{fig:survey_applications_debrifieng}), digital twins are not yet fully employed in $46\%$ of cases. When they are used, their main function is for analyzing mission outcomes and eventual errors, showcasing their potential in improving post-mission assessments.
Overall, when respondents were asked which digital twin applications they find most promising in their respective domains (~\ref{fig:survey_applications_promising_application}), there was a clear inclination toward enhancing real-time decision-making ($37\%$), followed by improving the fidelity of simulations ($33\%$) and predictive analytics ($30\%$). 
This highlights the growing interest in leveraging digital twins to improve decision-making, simulation accuracy, and predictive capabilities altogether.

The responses summarized in Figure~\ref{fig:survey_cross_domain} indicate that adopting a cross-domain perspective when designing and developing digital twins is generally viewed as promising. 
As shown in Figure~\ref{fig:survey_cross_domain_adaptability}, $54\%$ of participants report successfully adapting digital twins initially designed for one domain to another; of these, $67\%$ achieved significant success, while $33\%$ note limited success.
However, when shifting the focus to supporting technologies and standards related to digital twins development (Figure ~\ref{fig:survey_cross_domain_technologies}), $10\%$ consider them fully compatible across different domains, while $40\%$ claim partial compatibility that still presents certain integration challenges. 
Only $20\%$ indicate a need for significant adaptation to achieve compatibility. 
Despite these barriers, the data suggest that, with focused effort, digital twin systems can be made to function effectively across multiple domains.
When examining gaps that hinder effective cross-domain transposability (Figure ~\ref{fig:survey_cross_domain_gaps}), the most pressing issue is securely sharing sensitive information ($44\%$), followed by a lack of standardized communication protocols ($33\%$). 
This concern also appears among the major cross-domain challenges (Figure~\ref{fig:survey_cross_domain_challenges}), where $38\%$ cite an absence of standardized protocols as the main obstacle, followed by $33\%$ pointing to security and data privacy worries.
Favourably, $64\%$ of respondents (Figure~\ref{fig:survey_cross_domain_shared_infrstructure}) believe that sharing infrastructure or resources across different domains can be beneficial, either extensively across many domains or within select areas where synergy is more straightforward. 
To harness this potential, survey participants highlight the importance of universal technical standards ($37\%$) and cross-domain workshops/training ($36\%$), along with shared best-practice repositories ($27\%$)(Figure~\ref{fig:survey_cross_domain_common_guidelines}).
Overall, these results show a strong interest in achieving cross-domain digital twin adoption, with some organizations already experimenting at a basic level. 
Nonetheless, key challenges, particularly the lack of commonly adopted standards and secure information and data sharing frameworks, complicate the path toward seamless interoperability and broader success.

As depicted in Figure ~\ref{fig:survey_standardization_currently_adopted}, $35\%$ of respondents indicate that they do not follow any comprehensive digital twin standards at present. 
The remaining participants instead adapt various pre-existing standards to their digital twin projects: $15\%$ rely on industry-specific simulation standards, $12\%$ use ISO standards focused on data quality and structure, and $15\%$ employ other international technology standards.
This lack of widely adopted protocols emerges as a key impediment to digital twin development and operation (Figure~\ref{fig:survey_standardization_lackness_impact}). 
Specifically, $62\%$ of the respondents cite interoperability issues as the major consequence, while $24\%$ mention deployment delays, $10\%$ report increased costs, and $4\%$ highlight security risks.
In fact, when asked about which aspects of digital twin technology most urgently require standardization (Figure~\ref{fig:survey_standardization_required}), participants are nearly evenly split among interfaces between disparate systems, data exchange formats, and security protocols. 
Such results underscore the wideness of standardization needs across technical, procedural, and security dimensions.
Looking forward, developing commonly accepted digital twin standards is considered essential, and a collaboration of all stakeholders, including industry, governmental bodies, and international standards organizations, was indicated as the most suitable leader for this task (Figure~\ref{fig:survey_standardization_leaders}). 
This collective approach should help unify protocols, streamline interoperability, and increase security, ultimately accelerating the broader adoption of digital twin technologies.

\section{Main Gaps and Limitations}\label{sec:main_gaps_and_limitations}
The widespread adoption of digital twin technology offers immense potential for transforming the defense sector. 
Yet to ensure its effective and secure implementation, several challenges and gaps must be addressed. 
This section highlights the key issues, ranging from conceptual definitions and standardization to data sharing and integration, that stand in the way of fully realizing the benefits of digital twins in defense contexts.

\begin{figure}[!t]
    \centering
    \includegraphics[width=0.9\linewidth]{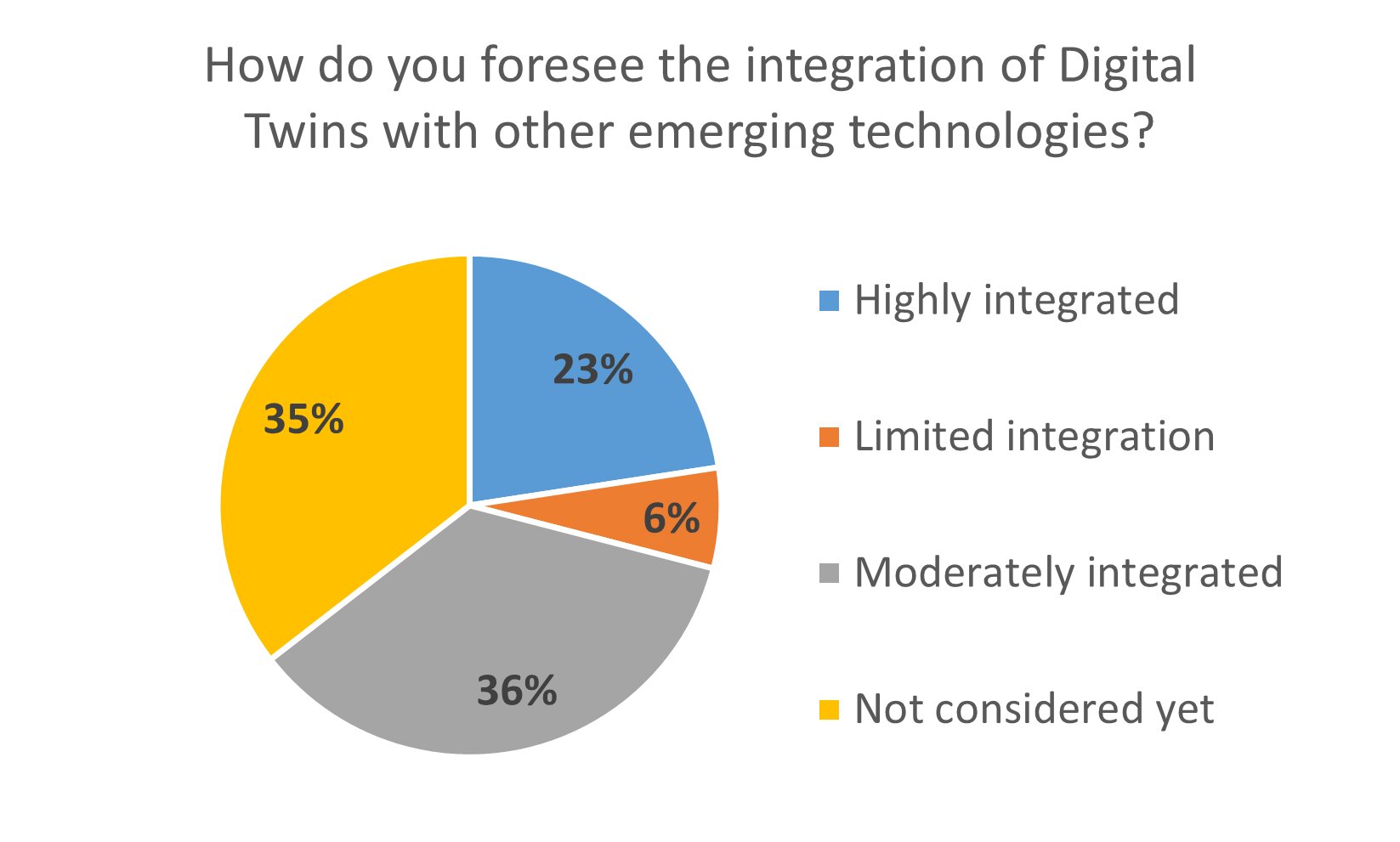}
    \caption{Result of the questionnaire related to integrating digital twins with emerging technologies.}
    \label{fig:survey_technological_enablers}
\end{figure}

The results of the questionnaire, shown in Figure ~\ref{fig:survey_technological_enablers}, reveal that there is strong interest and potential to be exploited for the integration of digital twins with other emerging technologies, as presented in \textit{Digital Twin Overview - Technological Enablers} section of this paper. 
$23\%$ of respondents foresee highly integrated solutions, while $36\%$ predict a moderate level of integration. 
Only $6\%$  expect limited integration, suggesting a broad recognition of the potential for digital twins to work synergistically with other cutting-edge technologies. 
These findings emphasize general optimism about the future capabilities of digital twin systems along with the enhanced support of innovative technological enablers.

\begin{figure*}[!htb]
\centering
\captionsetup[subfigure]{justification=centering}

\begin{subfigure}{0.49\textwidth}
\centering
  \includegraphics[width=0.9\textwidth]{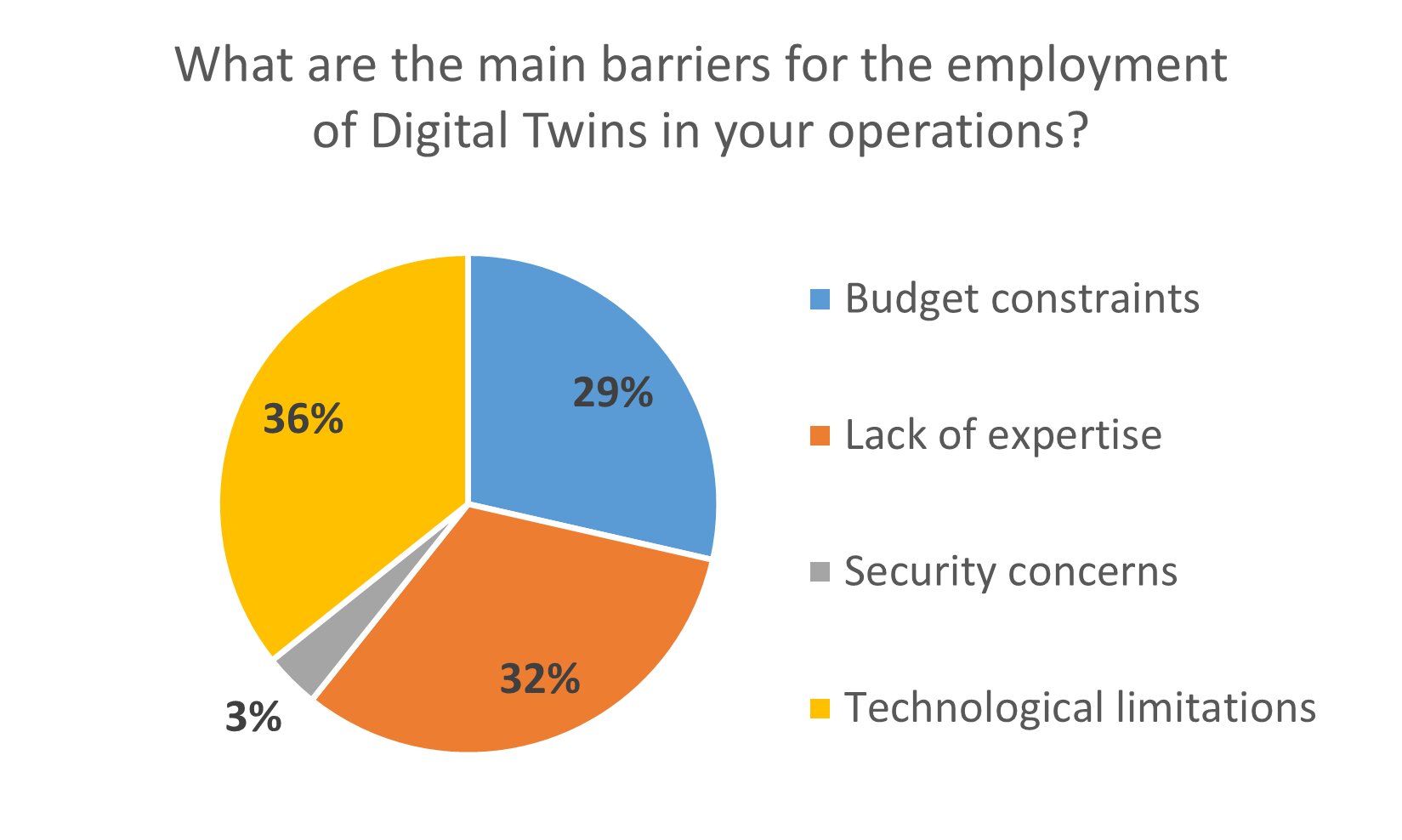}
  \caption{}\label{fig:survey_gaps_limitations_current}
\end{subfigure}
\begin{subfigure}{0.49\textwidth}
\centering
\includegraphics[width=0.9\textwidth]{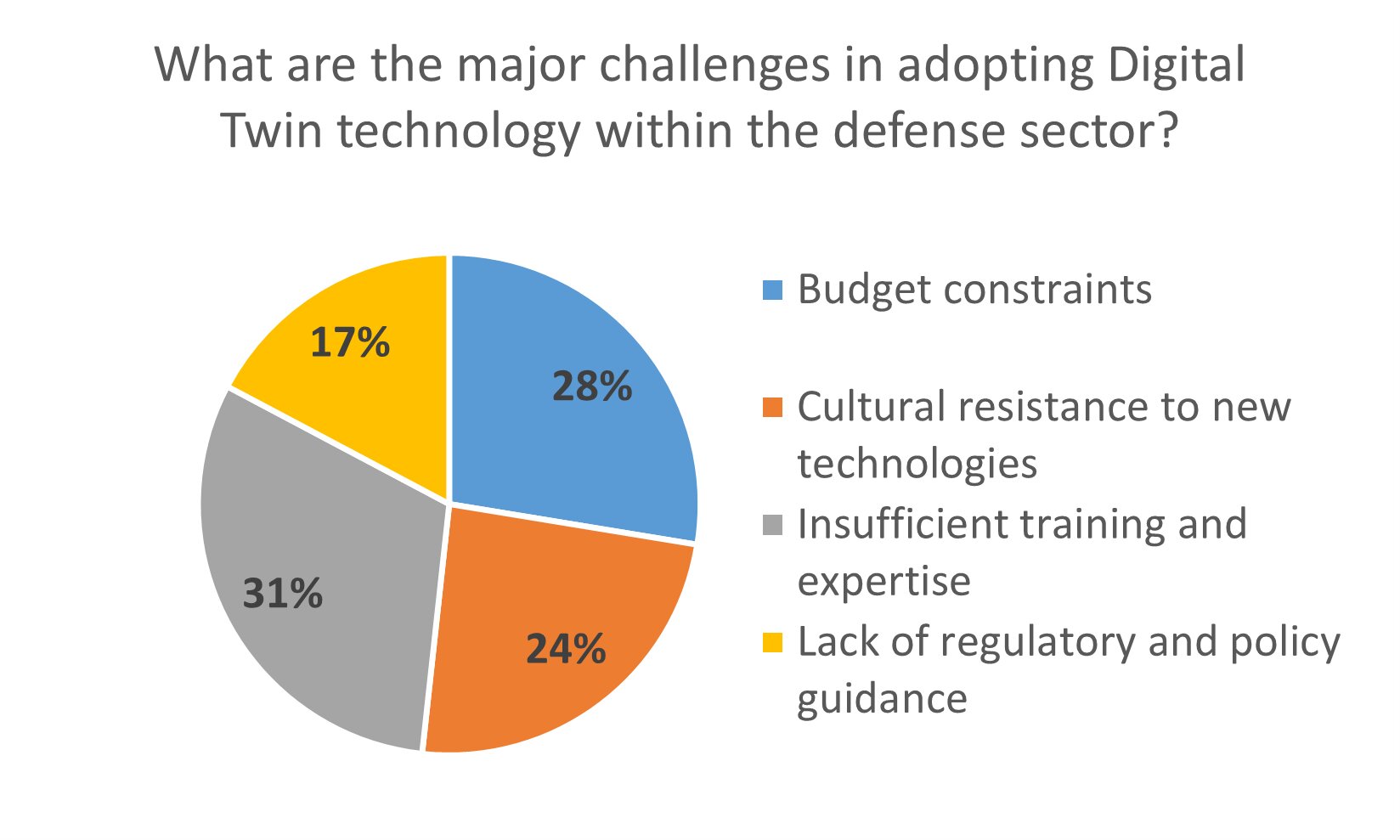}
\caption{}\label{fig:survey_gaps_limitations_major_challenges}
\end{subfigure}
\\[2ex] 
\begin{subfigure}{0.49\textwidth}
\centering
\includegraphics[width=0.9\textwidth]{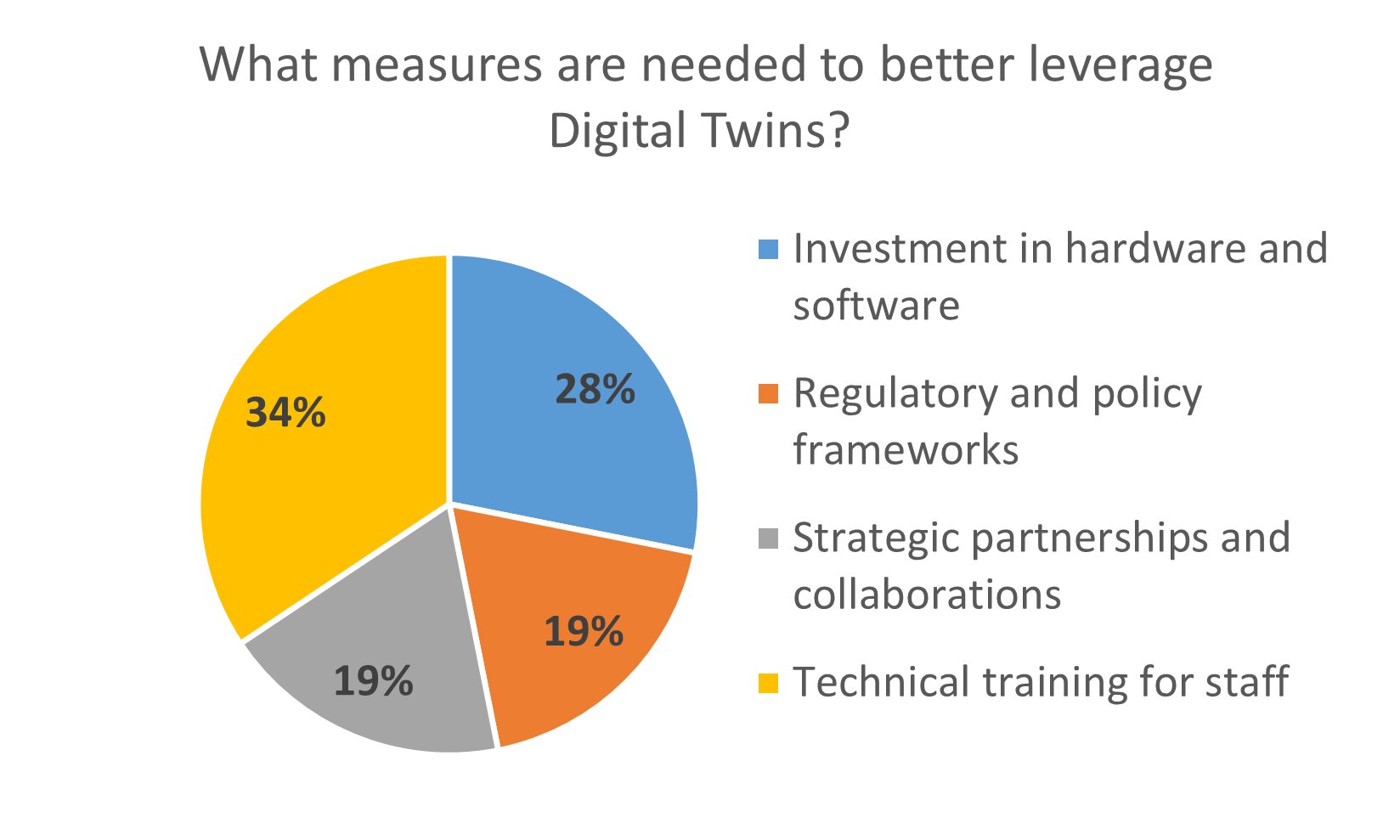}
\caption{}\label{fig:survey_gaps_limitations_future_req}
\end{subfigure}
\begin{subfigure}{0.49\textwidth}
\centering
\includegraphics[width=0.9\textwidth]{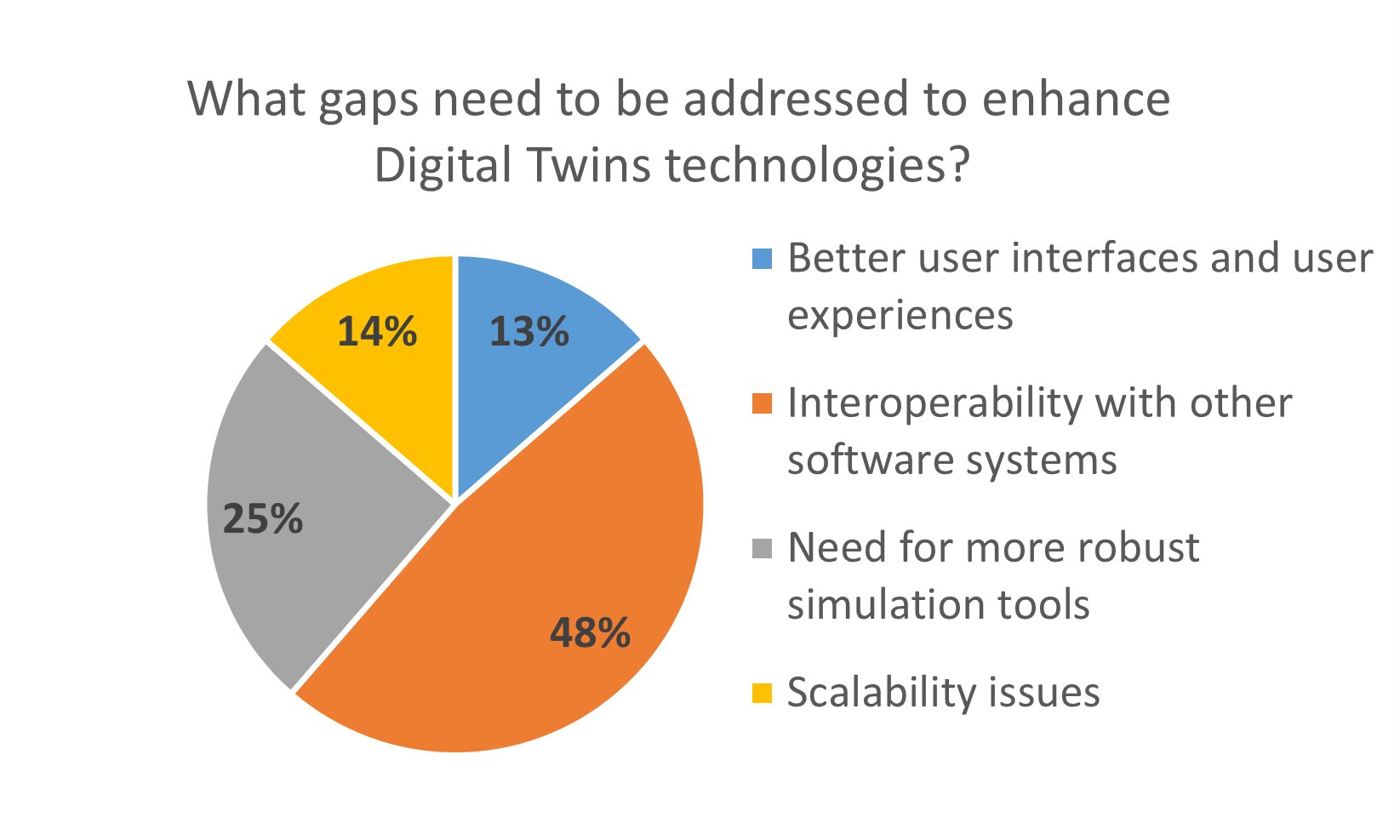}
\caption{}\label{fig:survey_gaps_limitations_gaps_tobe_addressed}
\end{subfigure}
\\[2ex] 
\begin{subfigure}{0.49\textwidth}
\centering
\includegraphics[width=0.9\textwidth]{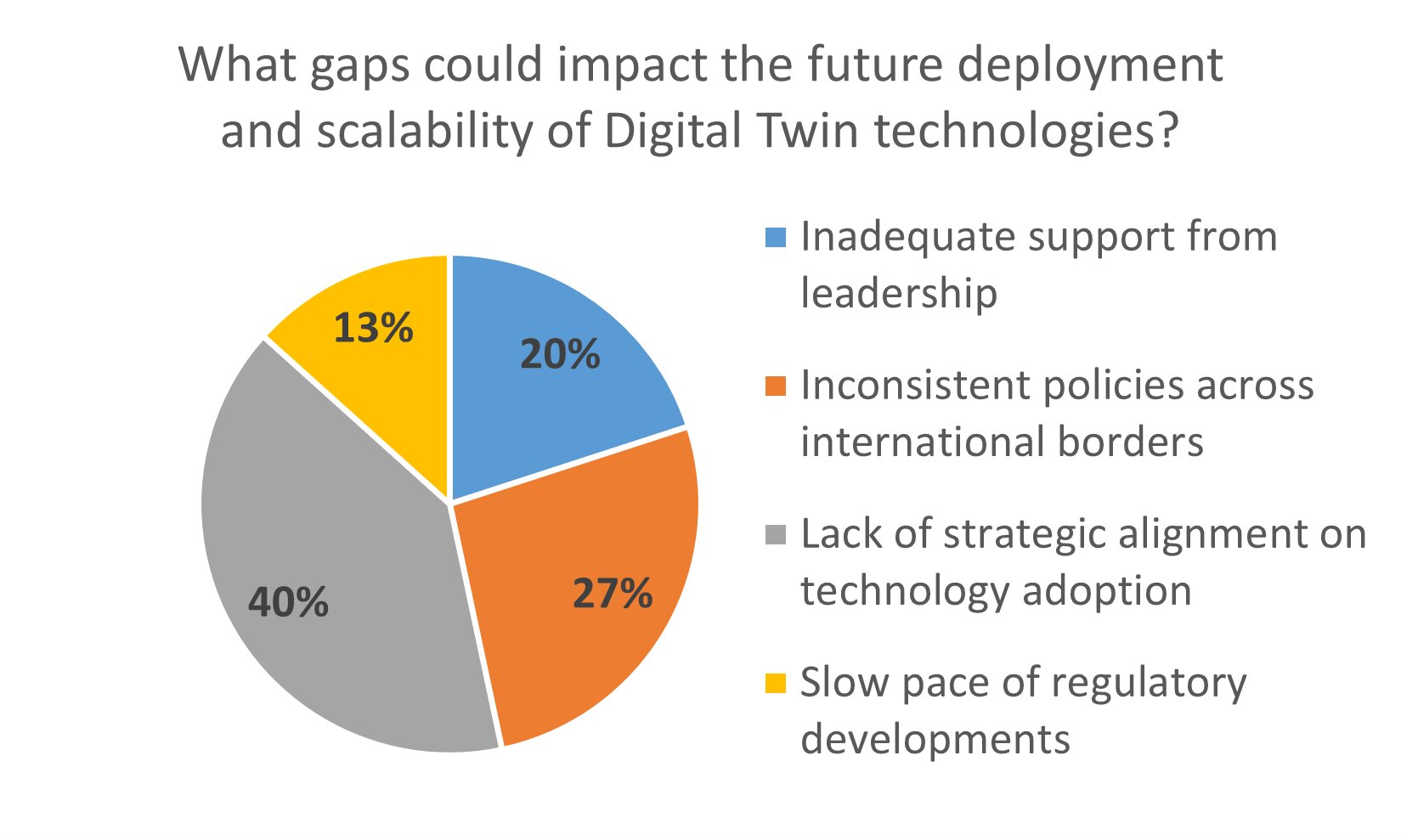}
\caption{}\label{fig:survey_gaps_limitations_impacting_scalability}
\end{subfigure}
\begin{subfigure}{0.49\textwidth}
\centering
  \includegraphics[width=0.9\textwidth]{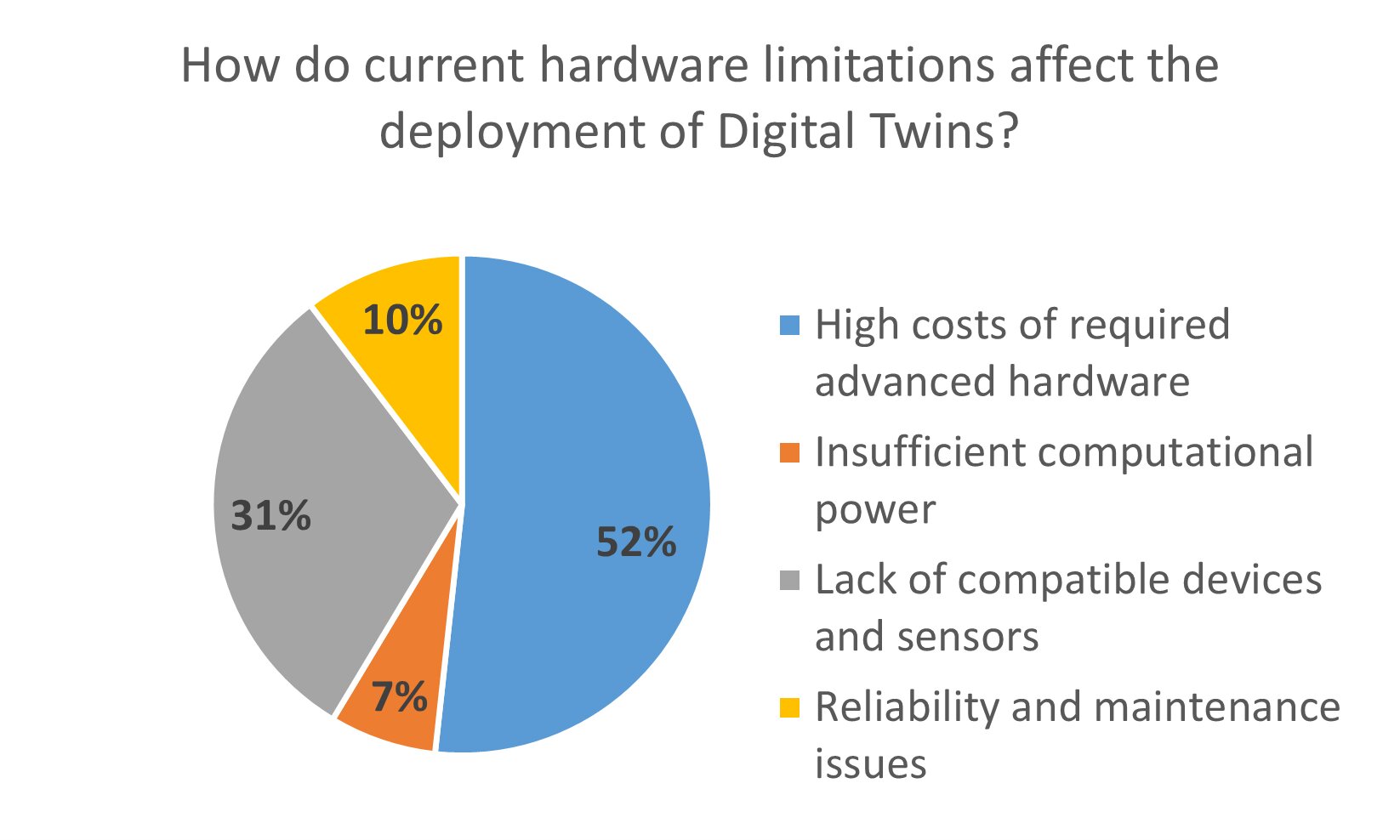}
  \caption{}\label{fig:survey_gaps_limitations_hardware}
\end{subfigure}
\\[1ex] 
\caption{Results of the questionnaire related to main gaps and current limitations for digital twin technologies development and deployment.}\label{fig:survey_gaps_limitations}
\end{figure*}

Figure ~\ref{fig:survey_gaps_limitations} illustrates the results of the questionnaire regarding the main gaps, challenges and limitations identified for the development and deployment of digital twin technologies in the defence sector. 
The main barriers to employing digital twins in current operations (Figure ~\ref{fig:survey_gaps_limitations_current}) include technological limitations, which are reported in $36\%$ of responses, followed by a general lack of expertise ($32\%$) and budget constraints ($29\%$). 
Notably, security concerns are not considered a significant barrier, suggesting as this is not considered at the moment a primary barrier for digital twin employment.
In general, the adoption of digital twin within the defense sector (Figure ~\ref{fig:survey_gaps_limitations_major_challenges}) confirms that budget constraints are a major issue ($28\%$). This is followed by insufficient training and expertise ($31\%$), and cultural resistance to new technology, identified by $24\%$ of the interviewees as a notable challenge characterizing the defence sector.   
To address these limitations and enhance the effective deployment of Digital Twins, the survey (Figure ~\ref{fig:survey_gaps_limitations_future_req}) indicates that the most crucial requirement is technical staff training, which is prioritized by $34\%$ of respondents. Other essential measures include investments in software and hardware ($28\%$), and the establishment of regulatory and policy frameworks, alongside fostering strategic partnerships and collaboration, both of which are considered significant measures by $19\%$ respondents. 
Collaboration is considered paramount also in the long run, as $40\%$  of the respondents believe that the lack of strategic alignment on technology adoption, both between industry and governmental institutions and among different defense domains, could impact the future deployment and scalability of digital twin technologies (Figure ~\ref{fig:survey_gaps_limitations_impacting_scalability}).
Similarly, inconsistent policies across allied countries pose a threat for $27\%$ of the respondents, further highlighting the growing awareness of the need for collaboration. 
This concern underscores the importance of developing a unified approach across all stakeholders, ensuring that the deployment and scalability of Digital Twin technologies are not obstructed by disjointed efforts or conflicting regulations. 

\paragraph{\textbf{Digital Twin Concept and Definition}}
Despite the extensive interest in digital twinning across various fields, there is still considerable ambiguity surrounding its precise definition. 
Simple virtual replicas of real entities are often labeled as digital twins, which can obscure the true essence of bidirectionality and continuous data flow that a genuine digital twin requires. 
Additionally, “digital threads” are sometimes mistakenly referred to as digital twins, especially in recent efforts to create virtual representations that encompass an entire product or system life cycle.
This blurring of terminology leads to miscommunication and misconception, as providers and end users may hold different expectations regarding essential requirements and specifications.
Consequently, the final commercial products often fail to meet the original vision or the operational needs they were intended to address.

\paragraph{\textbf{Missing Standards and Interoperability}} 
A significant challenge for digital twins in the defense domain lies in achieving interoperability and the integration of diverse data sources. 
In fact, $48\%$ of the interviewees identified digital twin interoperability with other software system to be the major gap to be urgently addressed to enhance digital twins in the defence domain (Figure ~\ref{fig:survey_gaps_limitations_gaps_tobe_addressed}). 
Defense systems collect vast amounts of information via multiple sensors, platforms, and networks. 
Ensuring seamless integration of these heterogeneous datasets, alongside the many platform and systems, is crucial to building a comprehensive, accurate digital twin and fully exploiting its capabilities. 
Consequently, standardized data formats, protocols, and interoperability frameworks are needed to facilitate efficient data exchange across varied defense platforms and systems.

Additionally, modern defense architectures often adhere to a system-of-systems approach, implying that multiple digital twins must be coupled. 
In a hierarchical military structure, each digital twin may require its own level of fidelity while still remaining consistent and interoperable with others. 
Although an ISO standard for digital twins (initially developed for manufacturing) exists, adapting it to other contexts, such as aerospace, like some works proposed\cite{shtofenmakher_adaptation_2024}, or defense, is crucial to facilitate interoperability among systems, applications, and organizations.
Nevertheless, establishing robust standards for inter- and intra-organizational communication remains essential, especially when looking at dual-use application. 
Following a common standard (or suitably adapting existing ones, such as ISO 23247) simplifies data sharing, speeds implementation, and ultimately leads to a more holistic understanding of the entire defense ecosystem.

\paragraph{\textbf{Lifetime Digital Twin and Use Case Specification}}
Employing standards benefits more than just interoperability among similar instances or within the same domain; it also ensures compatibility throughout the entire product life cycle, from design and production to service and beyond. 
Adopting a singular, core digital twin for use across a system’s lifetime underscores the inseparable link between physical and virtual counterparts, effectively treating them as a unified entity.

Especially in manufacturing contexts, it has been seen that adopting a lifetime digital twin, one that can be adapted to different phases of the product’s life cycle, maximizes its utility across various use cases. 
Crucially, well-defined requirements and clear application goals are needed from the outset. 
This focus on specifications ensures that the digital twin employs the right data, meets user expectations, and delivers tangible advantages, rather than merely replicating existing processes without adding value.

\paragraph{\textbf{Domain-based Approach}}
A major obstacle to successfully developing and integrating digital twin technology in the military field is the thematic domain structure traditionally entrenched within defense institutions. 
In practice, any given asset, environment, agent, or subsystem can be involved in operations that span multiple thematic domains. 
This reality contrasts with the rigid domain labels (land, maritime, air, space, and cyber) often found in military organizational structures. 
Although many defense contractors already operate across several of these domains simultaneously, such domain-based distinctions fade away when switching to technical implementations and applications.
As observed in section \textit{Applications of Digital Twins in the Military Field - Cross-Cutting Use Cases}, borders of thematic domains become much thinner and tend to overlap when addressing specific instances or subsystems, as digital twinning is leading to, according to the system of systems view. 

A technical characterization, focusing on the intrinsic properties of digital twins rather than thematic domains, thus offers a more versatile alternative. 
Grouping digital twins by their “essence” (e.g., humans, assets, and environments) simplifies integration across different organizations and use cases without requiring changes to the digital twin’s core design. 
Freed from domain constraints, these cross-domain digital twins are more adaptable to dual-use applications, military or civilian.

However, relying solely on off-the-shelf tools and frameworks can complicate interoperability across different fields, as they each have their own ecosystems of familiar tools. 
An innovative yet widely adopted approach involves a hierarchical, domain-agnostic structure that prioritizes the fundamental sets (humans, assets, and environments). 
Once a core digital twin is established for each set, it can be readily customized to specific use cases, be they military or civilian, without risking cross-compatibility. 
This approach naturally supports a complete system-of-systems perspective, where everything can be represented in the virtual realm, interlinked, and synchronized with real-world counterparts. 
Major industrial players have already shifted to such comprehensive solutions, reflecting their readiness to serve multiple thematic domains through an overarching, integrated framework.

\paragraph{\textbf{Data Sharing, Integration and Intellectual Property}}
Digital twins and their associated technologies handle sensitive defense data, making robust security and privacy measures paramount. 
Often, defense systems comprise multiple subsystems from different providers, resulting in a heterogeneous environment that can impede seamless data integration and effective exploitation of digital twin capabilities. 
Such incompatibilities may arise between subsystems, between entire systems, or even among different organizations, particularly when civilian and military institutions must interface.

Dual-use scenarios further complicate matters. 
Manufacturers and defense institutions often share a mutual desire to build high-fidelity digital twins, yet intellectual property (IP) and data ownership disputes can emerge, especially when defense stakeholders demand rights to use privately owned data or systems. 
Additionally, privacy regulations and classification protocols must be established to protect sensitive or personal information within digital twin ecosystems. 
Moreover, limited publicly available data, due to confidentiality, security, or secrecy, creates challenges for data-driven approaches and AI model training, as well as for the high-fidelity requirement, restricting the accuracy and effectiveness of digital twins.

Nevertheless, digital twin technologies also present potential solutions. 
By creating synthetic data that realistically mimics real-world scenarios, digital twins can generate large volumes of training data at minimal cost—expanding options for analysis and reducing dependence on scarce proprietary datasets. 
However, these approaches also underscore the importance of data reliability. 
In a typical digital twin environment, inputs may include sensor data of varying precision or AI-generated forecasts with uncertain accuracy. 
For instance, thermal measurements from ground sensors may be more trustworthy than outputs from a predictive model. 
Adopting reliability metrics at each data-processing stage is thus critical, and these metrics often involve multiple parameters (e.g., time, sensor quality, or model performance). 
Future research into high-fidelity digital twin frameworks should aim to operate under low-data conditions while generating realistic, large-scale datasets to facilitate broader defense applications.

\paragraph{\textbf{IoT and Cybersecurity}}
The Internet of Things (IoT), which potential is often blocked by the need for silent military operations, significantly broadens the attack surface for potential cyberattacks, making security a primary concern\cite{bagrodia_using_2023, prasad_machine_2022}. 
In the realm of digital twinning, which depends heavily on IoT and sensor data, these risks become even more pronounced. 
Because digital twins extend into the virtual space, they may introduce new vulnerabilities and be targeted by both centralized and decentralized cyber threats\cite{bagrodia_using_2023}. 
Alcaraz and Lopez \cite{alcaraz_digital_2022} provide a comprehensive perspective by investigating vulnerabilities and threat models across each layer of a digital twin.

Military systems are especially appealing targets for cyberattacks, given the high-stakes nature of defense operations. 
Any security breach can have severe and wide-reaching implications.
Moreover, the additional attack surface created by digital twins demands rigorous protective measures, including robust encryption, authentication, and access control to maintain data integrity, confidentiality, and availability.

On the other hand, digital twins can also serve as powerful tools for enhancing cybersecurity. 
Through realistic simulations, they can identify potential vulnerabilities, predict attack patterns, and enable proactive testing of security controls. 
In effect, digital twins offer both a greater security risk and a strategic advantage in safeguarding complex defense infrastructures.

\paragraph{\textbf{Real-Time Capabilities}}
Digital twins depend on real-time data acquisition and processing to deliver continuously updated representations of physical assets and operational environments. 
However, acquiring and analyzing the large data volumes typical of dynamic, complex defense scenarios poses both technical and logistical issues. 
Efficient data-capture systems, advanced signal-processing techniques, and edge computing solutions are essential to handle real-time data streams and preserve fidelity in digital twin models.

In practice, this may involve applying well-established mathematical methods alongside more experimental AI approaches that handle highly-dynamic data. 
Regardless of the methodology, seamless data integration and instant feedback loops are fundamental to unlocking the full potential of real-time digital twins, particularly when operational decisions need to be made with minimal delay.

\paragraph{\textbf{Human-Machine Interaction}}
Digital twins often integrate simulation, advanced analytics, and sometimes AI-driven predictive capabilities, generating complex outputs that are not always readily accessible or interpretable by human operators. 
Although the theoretical digital twin model does not specify requirements for human interfaces or human-machine interactions methods, effective user interfaces are crucial for tackling these insights.

Technologies such as extended reality (XR) have proven valuable in many applications, enabling immersive visualization of digital twin data. 
Likewise, intuitive dashboards and GUI can present complex information in more understandable formats. 
These tools reduce the cognitive burden on operators, allowing decision-makers to quickly grasp system statuses, predict outcomes, and respond to evolving situations with greater confidence.

\paragraph{\textbf{User Training and Adoption}}
Successfully integrating digital twin technology in the defense sector requires personnel who are skilled in developing, maintaining, and operating these complex systems. 
Consequently, robust training programs should be established to equip staff with the necessary technical competencies and to ensure that defense personnel can effectively exploit digital twin capabilities.
Although some companies, such as those mentioned in \cite{silvera_navantias_2020} and \cite{noauthor_fcx_2023}, already provide specific training and guidelines with their digital twin products, widespread best practices remain underdeveloped.

Additionally, fostering a culture of continuous learning and innovation is essential for the smooth adoption and utilization of digital twin solutions within defense organizations. 
This would address one of the major identified challenges in adopting these technologies within the defence domain and current operations (Figure ~\ref{fig:survey_gaps_limitations_major_challenges} and ~\ref{fig:survey_gaps_limitations_current}), as well as the main measure to be taken to better leverage digital twins (Figure ~\ref{fig:survey_gaps_limitations_future_req}).
On the commercial side, many companies fight with the challenge of seamlessly integrating digital twin technologies into existing production processes, often due to the fundamental business model changes these solutions may entail. 
While such transformations can be costly and resource-intensive, cost–benefit analyses, for instance highlighting reduced downtime through predictive maintenance, can demonstrate tangible short-term returns on investment, thereby encouraging broader acceptance and deployment of digital twins.

\bigskip

Effectively addressing these gaps and challenges is crucial for the deep adoption of digital twin technology within the defence sector. 
Collaborative efforts among defence organizations, industry partners, and research institutions are required to develop innovative solutions, standards, and best practices that enable the successful integration and utilization of these technologies. 
By overcoming interoperability limitations, refining data management and integration, and enhancing both cybersecurity and user readiness, the defence sector will unlock the full potential of digital twins. This will pave the way for enhanced operational capabilities, informed decision-making, and improved mission outcomes—ushering in a new era of defense technology innovation.

\section{Conclusion}\label{sec:summary_and_outlook}
Digital twins are dynamic, high-fidelity virtual representations of physical objects or systems, built on bidirectional communication between real and virtual spaces. 
They are enhanced by a range of key technologies, including IoT and sensors for data acquisition, AI for data processing and analytics, high-performance computing for simulations, cloud and edge computing for scalability, and extended reality for intuitive human–machine interactions. 
In the defense sector, digital twins offer substantial benefits across all domains, supporting applications like predictive maintenance, operational planning, simulation, training, and optimization. 
Especially, they enable detailed “what-if” scenario analyses\cite{collins_past_2021}. 

Despite this potential, several challenges remain. 
Interoperability issues, cybersecurity threats, data integration, and limited data availability must be overcome to fully leverage digital twins. 
Efforts must focus on seamless integration of these technologies with existing military systems. 
This includes standardizing data formats and communication protocols for interoperability and scalability, as well as employing advanced AI to improve predictive capabilities and decision-making. 
Cybersecurity concerns are also paramount: robust security measures must protect sensitive data, maintain system integrity, and defend any additional attack surfaces introduced by digital twins. 
Meanwhile, incorporating extended reality (XR) promises more effective training programs, improved operational planning, and more responsive real-time decision-making \cite{stanney_performance_2021}.

Successful adoption further depends on collaboration among defense organizations, research institutions, and industry, supported by common standards and best practices. 
Training programs that equip personnel with the necessary skills and expertise are crucial to ensure that digital twins are effectively deployed and thoroughly exploited.

In conclusion, digital twins stand as a promising, impactful technology that can substantially enhance defense capabilities. 
Addressing current limitations while leveraging ongoing technical advancements will be crucial for widespread adoption.
As development proceeds, digital twins will play a pivotal role in shaping the future of the military by fostering greater efficiency, situational awareness, and operational effectiveness.


\begin{funding}
This work was partially funded by the European Defence Agency (EDA) under the contract REF.23.RTI.OP.121 for the provision of Explore Defence Digital Twins (EDDI) study, and the Fonds National de la Recherche of Luxembourg (FNR) under the project DEUS (Ref. C22/IS/17387634/DEUS).
\end{funding}

\begin{dci}
The authors declare no conflicts of interest.
\end{dci}


\bibliographystyle{SageV}

\begin{thebibliography}{100}
\providecommand{\url}[1]{\texttt{#1}}
\providecommand{\urlprefix}{URL }
\expandafter\ifx\csname urlstyle\endcsname\relax
  \providecommand{\doi}[1]{DOI:\discretionary{}{}{}#1}\else
  \providecommand{\doi}{DOI:\discretionary{}{}{}\begingroup \urlstyle{rm}\Url}\fi
\providecommand{\eprint}[2][]{\url{#2}}

\bibitem{collins_past_2021}
Collins AJ, Pour FSA and Jordan CA.
\newblock Past challenges and the future of discrete event simulation.
\newblock \emph{The Journal of Defense Modeling and Simulation} 2023; 20(3): 351--369.
\newblock \doi{10.1177/15485129211067175}.
\newblock \urlprefix\url{https://doi.org/10.1177/15485129211067175}.
\newblock \eprint{https://doi.org/10.1177/15485129211067175}.

\bibitem{langreck_modeling_2019}
Langreck J, Wong H, Hernandez A et~al.
\newblock Modeling and simulation of future capabilities with an automated computer-aided wargame.
\newblock \emph{The Journal of Defense Modeling and Simulation} 2021; 18(4): 407--416.
\newblock \doi{10.1177/1548512919873980}.
\newblock \urlprefix\url{https://doi.org/10.1177/1548512919873980}.
\newblock \eprint{https://doi.org/10.1177/1548512919873980}.

\bibitem{singh_digital_2021}
Singh M, Fuenmayor E, Hinchy EP et~al.
\newblock Digital twin: Origin to future.
\newblock \emph{Applied System Innovation} 2021; 4(2).
\newblock \doi{10.3390/asi4020036}.
\newblock \urlprefix\url{https://www.mdpi.com/2571-5577/4/2/36}.

\bibitem{mendi_digital_2022}
Mendi AF, Erol T and Dogan D.
\newblock Digital {Twin} in the {Military} {Field}.
\newblock \emph{IEEE Internet Computing} 2022; 26(5): 33--40.
\newblock \doi{10.1109/MIC.2021.3055153}.
\newblock \urlprefix\url{https://ieeexplore.ieee.org/document/9345490/}.

\bibitem{panetta_top_2017}
Panetta K.
\newblock Top {Trends} in the {Gartner} {Hype} {Cycle} for {Emerging} {Technologies}, 2017, 2017.
\newblock \url{https://www.gartner.com/smarterwithgartner/top-trends-in-the-gartner-hype-cycle-for-emerging-technologies-2017} [Accessed: 2025-03-21].

\bibitem{groombridge_top_2023}
Groombridge D.
\newblock Top {Strategic} {Technology} {Trends} 2023, 2023.
\newblock \url{https://emt.gartnerweb.com/ngw/globalassets/en/publications/documents/2023-gartner-top-strategic-technology-trends-ebook.pdf} [Accessed: 2024-05-06].

\bibitem{west_digital_2017}
West TD and Blackburn M.
\newblock Is digital thread/digital twin affordable? a systemic assessment of the cost of dod’s latest manhattan project.
\newblock \emph{Procedia Computer Science} 2017; 114: 47--56.
\newblock \doi{https://doi.org/10.1016/j.procs.2017.09.003}.
\newblock \urlprefix\url{https://www.sciencedirect.com/science/article/pii/S1877050917317970}.
\newblock Complex Adaptive Systems Conference with Theme: Engineering Cyber Physical Systems, CAS October 30 – November 1, 2017, Chicago, Illinois, USA.

\bibitem{zweber_digital_2017}
Zweber JV, Kolonay RM, Kobryn P et~al.
\newblock Digital {Thread} and {Twin} for {Systems} {Engineering}: {Requirements} to {Design}.
\newblock In \emph{55th {AIAA} {Aerospace} {Sciences} {Meeting}}. Grapevine, Texas: American Institute of Aeronautics and Astronautics.
\newblock ISBN 978-1-62410-447-3.
\newblock \doi{10.2514/6.2017-0875}.
\newblock \urlprefix\url{https://arc.aiaa.org/doi/10.2514/6.2017-0875}.

\bibitem{grieves_virtually_2011}
Grieves M.
\newblock \emph{Virtually perfect: driving innovative and lean products through product lifecycle management}.
\newblock Cocoa Beach, Florida: Space Coast Press, 2011.
\newblock ISBN 978-0-9821380-0-7.

\bibitem{grieves_digital_2014}
Grieves M.
\newblock Digital {Twin}: {Manufacturing} {Excellence} {Through} {Virtual} {Factory} {Replication}.

\bibitem{kahlen_digital_2017}
Grieves M and Vickers J.
\newblock Digital {Twin}: {Mitigating} {Unpredictable}, {Undesirable} {Emergent} {Behavior} in {Complex} {Systems}.
\newblock In Kahlen FJ, Flumerfelt S and Alves A (eds.) \emph{Transdisciplinary {Perspectives} on {Complex} {Systems}}. Cham: Springer International Publishing.
\newblock ISBN 978-3-319-38754-3 978-3-319-38756-7, 2017.
\newblock pp. 85--113.
\newblock \doi{10.1007/978-3-319-38756-7_4}.
\newblock \urlprefix\url{http://link.springer.com/10.1007/978-3-319-38756-7_4}.

\bibitem{kritzinger_digital_2018}
Kritzinger W, Karner M, Traar G et~al.
\newblock Digital {Twin} in manufacturing: {A} categorical literature review and classification.
\newblock \emph{IFAC-PapersOnLine} 2018; 51(11): 1016--1022.
\newblock \doi{10.1016/j.ifacol.2018.08.474}.
\newblock \urlprefix\url{https://linkinghub.elsevier.com/retrieve/pii/S2405896318316021}.

\bibitem{errandonea_digital_2020}
Errandonea I, Beltrán S and Arrizabalaga S.
\newblock Digital {Twin} for maintenance: {A} literature review.
\newblock \emph{Computers in Industry} 2020; 123: 103316.
\newblock \doi{https://doi.org/10.1016/j.compind.2020.103316}.
\newblock \urlprefix\url{https://www.sciencedirect.com/science/article/pii/S0166361520305509}.

\bibitem{noauthor_iso_2021-2}
ISO.
\newblock {ISO} 23247-2:2021 {Automation} systems and integration — {Digital} twin framework for manufacturing, 2021.
\newblock \urlprefix\url{https://www.iso.org/standard/78743.html}.

\bibitem{noauthor_iso_2021-1}
ISO.
\newblock {ISO} 23247-3:2021 {Automation} systems and integration — {Digital} twin framework for manufacturing, 2021.
\newblock \urlprefix\url{https://www.iso.org/standard/78744.html}.

\bibitem{noauthor_iso_2021}
ISO.
\newblock {ISO} 23247-4:2021 {Automation} systems and integration — {Digital} twin framework for manufacturing, 2021.
\newblock \urlprefix\url{https://www.iso.org/standard/78745.html}.

\bibitem{noauthor_iso_2021-3}
{ISO} 23247-1:2021 {Automation} systems and integration {Digital} twin framework for manufacturing, 2021.
\newblock \urlprefix\url{https://www.iso.org/standard/75066.html}.

\bibitem{noauthor_isocd_nodate-1}
ISO.
\newblock {ISO}/{CD} 23247-5 {Automation} systems and integration — {Digital} twin framework for manufacturing, 2021.
\newblock \urlprefix\url{https://www.iso.org/standard/87425.html}.

\bibitem{noauthor_isocd_nodate}
ISO.
\newblock {ISO}/{CD} 23247-6 {Automation} systems and integration — {Digital} twin framework for manufacturing, 2021.
\newblock \urlprefix\url{https://www.iso.org/standard/87426.html}.

\bibitem{lu_evaluating_2016}
Lu WM, Kweh QL, Nourani M et~al.
\newblock Evaluating the efficiency of dual-use technology development programs from the r\&d and socio-economic perspectives.
\newblock \emph{Omega} 2016; 62: 82--92.
\newblock \doi{https://doi.org/10.1016/j.omega.2015.08.011}.
\newblock \urlprefix\url{https://www.sciencedirect.com/science/article/pii/S0305048315001851}.

\bibitem{zhang_civil_2024}
Zhang A and Qi N.
\newblock Civil-to-dual-use enterprise transition in civil-military integration: a complex network game approach.
\newblock \emph{Technology Analysis \& Strategic Management} 2025; 0(0): 1--18.
\newblock \doi{10.1080/09537325.2025.2477179}.
\newblock \urlprefix\url{https://doi.org/10.1080/09537325.2025.2477179}.
\newblock \eprint{https://doi.org/10.1080/09537325.2025.2477179}.

\bibitem{carillo_commercial_2017}
Carrillo PE.
\newblock Commercial dual-use technologies in defense acquisition reform.
\newblock Technical report, R Street Institute, 2017.
\newblock \urlprefix\url{http://www.jstor.org/stable/resrep19130}.

\bibitem{mazal_dual_2024}
Mazal J.
\newblock The {Dual} {Use} of {Civilian} and {Military} {Technologies} in the {Battlefield} of the {Future}.
\newblock In Zombory K and Szilágyi JE (eds.) \emph{Studies of the {Central} {European} {Professors}’ {Network}}. Central European Academic Publishing.
\newblock ISBN 978-615-6474-64-3, 2024.
\newblock pp. 259--307.
\newblock \doi{10.54237/profnet.2024.zkjeszcodef_6}.
\newblock \urlprefix\url{https://books.ceapublishing.hu/index.php/ceaprofnet/catalog/book/7/chapter/53}.

\bibitem{akimkina_technology_2021}
Akimkina D, Khrustalev E, Baranova N et~al.
\newblock Technology transfer of the military-industrial complex as a factor in increasing the science intensity of the civilian industry.
\newblock \emph{SHS Web of Conferences} 2021; 114: 01027.
\newblock \doi{10.1051/shsconf/202111401027}.
\newblock \urlprefix\url{https://www.shs-conferences.org/10.1051/shsconf/202111401027}.

\bibitem{pereira_dual_2018}
Pereira J and Oliveira R.
\newblock Dual use cns boosts civil-military interoperability.
\newblock In \emph{2018 Integrated Communications, Navigation, Surveillance Conference (ICNS)}. pp. 5A1--1--5A1--9.
\newblock \doi{10.1109/ICNSURV.2018.8384900}.

\bibitem{fraga-lamas_review_2016}
Fraga-Lamas P, Fernández-Caramés TM, Suárez-Albela M et~al.
\newblock A review on internet of things for defense and public safety.
\newblock \emph{Sensors} 2016; 16(10).
\newblock \doi{10.3390/s16101644}.
\newblock \urlprefix\url{https://www.mdpi.com/1424-8220/16/10/1644}.

\bibitem{suri_analyzing_2016}
Suri N, Tortonesi M, Michaelis J et~al.
\newblock Analyzing the applicability of internet of things to the battlefield environment.
\newblock In \emph{2016 International Conference on Military Communications and Information Systems (ICMCIS)}. pp. 1--8.
\newblock \doi{10.1109/ICMCIS.2016.7496574}.

\bibitem{blasch_machine_2021}
Blasch E, Pham T, Chong CY et~al.
\newblock Machine learning/artificial intelligence for sensor data fusion–opportunities and challenges.
\newblock \emph{IEEE Aerospace and Electronic Systems Magazine} 2021; 36(7): 80--93.
\newblock \doi{10.1109/MAES.2020.3049030}.

\bibitem{toth_internet_2021}
T{\'o}th A.
\newblock Internet of things vulnerabilities in military environments.
\newblock \emph{Vojensk{\'e} rozhledy} 2021; 30(3): 45--58.

\bibitem{pasdar_cybersecurity_2024}
Pasdar A, Koroniotis N, Keshk M et~al.
\newblock Cybersecurity solutions and techniques for internet of things integration in combat systems.
\newblock \emph{IEEE Transactions on Sustainable Computing} 2025; 10(2): 345--365.
\newblock \doi{10.1109/TSUSC.2024.3443256}.

\bibitem{bagrodia_using_2023}
Bagrodia R.
\newblock Using network digital twins to improve cyber resilience of missions.
\newblock \emph{The Journal of Defense Modeling and Simulation: Applications, Methodology, Technology} 2023; 20(1): 97--106.
\newblock \doi{10.1177/15485129221131226}.
\newblock \urlprefix\url{http://journals.sagepub.com/doi/10.1177/15485129221131226}.

\bibitem{grumazescu_wsn_2016}
Grumazescu C, Vlăduţă VA and Subaşu G.
\newblock Wsn solutions for communication challenges in military live simulation environments.
\newblock In \emph{2016 International Conference on Communications (COMM)}. pp. 319--322.
\newblock \doi{10.1109/ICComm.2016.7528266}.

\bibitem{prasad_machine_2022}
Prasad A and Chandra S.
\newblock Machine learning to combat cyberattack: a survey of datasets and challenges.
\newblock \emph{The Journal of Defense Modeling and Simulation} 2023; 20(4): 577--588.
\newblock \doi{10.1177/15485129221094881}.
\newblock \urlprefix\url{https://doi.org/10.1177/15485129221094881}.
\newblock \eprint{https://doi.org/10.1177/15485129221094881}.

\bibitem{davis_artificial_2022}
Davis PK and Bracken P.
\newblock Artificial intelligence for wargaming and modeling.
\newblock \emph{The Journal of Defense Modeling and Simulation} 2025; 22(1): 25--40.
\newblock \doi{10.1177/15485129211073126}.
\newblock \urlprefix\url{https://doi.org/10.1177/15485129211073126}.
\newblock \eprint{https://doi.org/10.1177/15485129211073126}.

\bibitem{szabadfoldi_artificial_2021}
Szabadföldi I.
\newblock Artificial {Intelligence} in {Military} {Application} – {Opportunities} and {Challenges}.
\newblock \emph{Land Forces Academy Review} 2021; 26(2): 157--165.
\newblock \doi{10.2478/raft-2021-0022}.
\newblock \urlprefix\url{https://www.sciendo.com/article/10.2478/raft-2021-0022}.

\bibitem{van_lent_applications_2022}
van Lent M and Schmorrow D.
\newblock The applications of artificial intelligence to education and training.
\newblock \emph{The Journal of Defense Modeling and Simulation} 2022; 19(2): 127--128.
\newblock \doi{10.1177/15485129221088717}.
\newblock \urlprefix\url{https://doi.org/10.1177/15485129221088717}.
\newblock \eprint{https://doi.org/10.1177/15485129221088717}.

\bibitem{stevens_machine_2021}
Stevens RH and Galloway TL.
\newblock Can machine learning be used to forecast the future uncertainty of military teams?
\newblock \emph{The Journal of Defense Modeling and Simulation} 2022; 19(2): 145--158.
\newblock \doi{10.1177/1548512921999112}.
\newblock \urlprefix\url{https://doi.org/10.1177/1548512921999112}.
\newblock \eprint{https://doi.org/10.1177/1548512921999112}.

\bibitem{rashid_artificial_2023}
Rashid AB, Kausik AK, Al~Hassan~Sunny A et~al.
\newblock Artificial {Intelligence} in the {Military}: {An} {Overview} of the {Capabilities}, {Applications}, and {Challenges}.
\newblock \emph{International Journal of Intelligent Systems} 2023; 2023(1): 8676366.
\newblock \doi{10.1155/2023/8676366}.
\newblock \urlprefix\url{https://onlinelibrary.wiley.com/doi/10.1155/2023/8676366}.

\bibitem{clark_detection_2021}
ClarkJr GW, Andel TR, McDonald JT et~al.
\newblock Detection and defense of cyberattacks on the machine learning control of robotic systems.
\newblock \emph{The Journal of Defense Modeling and Simulation} 2024; 21(2): 181--203.
\newblock \doi{10.1177/15485129211043874}.
\newblock \urlprefix\url{https://doi.org/10.1177/15485129211043874}.
\newblock \eprint{https://doi.org/10.1177/15485129211043874}.

\bibitem{petrea_hpc_2024}
Petrea M, Neac{\c{s}}u GC and Ciubotaru BI.
\newblock Hpc computing impact in military operations.
\newblock In \emph{International Scientific Conference ``Strategies XXI''}, volume~20. ``Carol I'' National Defence University, pp. 177--181.

\bibitem{alim_measuring_2024}
Alim H, Subramaniam A, Nor NM et~al.
\newblock Measuring operational cognitive readiness of military personnel using joint theater level simulation system (jtls).
\newblock \emph{The Journal of Defense Modeling and Simulation} 2026; 23(1): 109--122.
\newblock \doi{10.1177/15485129241239669}.
\newblock \urlprefix\url{https://doi.org/10.1177/15485129241239669}.
\newblock \eprint{https://doi.org/10.1177/15485129241239669}.

\bibitem{felix_real-time_2021}
Felix D, Colwill I and Harris P.
\newblock Real-time simulation of a fragmenting explosion for cylindrical warheads.
\newblock \emph{The Journal of Defense Modeling and Simulation} 2022; 19(4): 783--797.
\newblock \doi{10.1177/1548512921995560}.
\newblock \urlprefix\url{https://doi.org/10.1177/1548512921995560}.
\newblock \eprint{https://doi.org/10.1177/1548512921995560}.

\bibitem{van_der_zwet_promises_2022}
van~der Zwet K, Barros AI, van Engers TM et~al.
\newblock Promises and pitfalls of computational modelling for insurgency conflicts.
\newblock \emph{The Journal of Defense Modeling and Simulation} 2023; 20(3): 333--350.
\newblock \doi{10.1177/15485129211073612}.
\newblock \urlprefix\url{https://doi.org/10.1177/15485129211073612}.
\newblock \eprint{https://doi.org/10.1177/15485129211073612}.

\bibitem{zaerens_enabling_2011}
Zaerens K.
\newblock Enabling the benefits of cloud computing in a military context.
\newblock In \emph{2011 IEEE Asia-Pacific Services Computing Conference}. pp. 166--173.
\newblock \doi{10.1109/APSCC.2011.42}.

\bibitem{tiganus_cloud_2023}
Țigănuș D.
\newblock Cloud {Technologies} and the {Need} for {Hybrid} {Cloud} {Implementation} in the {Military} {Environment}.
\newblock \emph{Romanian Military Thinking} 2023; 2023(2): 46--59.
\newblock \doi{10.55535/RMT.2023.2.02}.
\newblock \urlprefix\url{https://en-gmr.mapn.ro/webroot/fileslib/upload/files/arhiva%20reviste/RMT/2023/2/TIGANUS.pdf}.

\bibitem{stergiou_digital_2022}
Stergiou CL and Psannis KE.
\newblock Digital twin intelligent system for industrial internet of things-based big data management and analysis in cloud environments.
\newblock \emph{Virtual Reality \& Intelligent Hardware} 2022; 4(4): 279--291.
\newblock \doi{https://doi.org/10.1016/j.vrih.2022.05.003}.
\newblock \urlprefix\url{https://www.sciencedirect.com/science/article/pii/S2096579622000444}.
\newblock Virtual-reality and intelligent hardware in digital twins A).

\bibitem{boyce_enhancing_2022}
Boyce MW, Thomson RH, Cartwright JK et~al.
\newblock Enhancing {Military} {Training} {Using} {Extended} {Reality}: {A} {Study} of {Military} {Tactics} {Comprehension}.
\newblock \emph{Frontiers in Virtual Reality} 2022; 3: 754627.
\newblock \doi{10.3389/frvir.2022.754627}.
\newblock \urlprefix\url{https://www.frontiersin.org/articles/10.3389/frvir.2022.754627/full}.

\bibitem{stanney_performance_2021}
Stanney KM, Archer J, Skinner A et~al.
\newblock Performance gains from adaptive extended reality training fueled by artificial intelligence.
\newblock \emph{The Journal of Defense Modeling and Simulation} 2022; 19(2): 195--218.
\newblock \doi{10.1177/15485129211064809}.
\newblock \urlprefix\url{https://doi.org/10.1177/15485129211064809}.
\newblock \eprint{https://doi.org/10.1177/15485129211064809}.

\bibitem{brouwer_met_2017}
Brouwer E.
\newblock Met {SUIT} naar waar je wilt.
\newblock \emph{Materieel gezien} 2017; 5.
\newblock \url{https://magazines.defensie.nl/materieelgezien/2017/05/mg201705suit} [Accessed: 2024-05-30].

\bibitem{stacchio_empowering_2022}
Stacchio L, Angeli A and Marfia G.
\newblock Empowering digital twins with {eXtended} reality collaborations.
\newblock \emph{Virtual Reality \& Intelligent Hardware} 2022; 4(6): 487--505.
\newblock \doi{10.1016/j.vrih.2022.06.004}.
\newblock \urlprefix\url{https://linkinghub.elsevier.com/retrieve/pii/S2096579622000596}.

\bibitem{benkamoun_architecture_2014}
Benkamoun N, ElMaraghy W, Huyet AL et~al.
\newblock Architecture {Framework} for {Manufacturing} {System} {Design}.
\newblock \emph{Procedia CIRP} 2014; 17: 88--93.
\newblock \doi{10.1016/j.procir.2014.01.101}.
\newblock \urlprefix\url{https://linkinghub.elsevier.com/retrieve/pii/S221282711400359X}.

\bibitem{klein_systematic_2013}
Klein J and Van~Vliet H.
\newblock A systematic review of system-of-systems architecture research.
\newblock In \emph{Proceedings of the 9th international {ACM} {Sigsoft} conference on {Quality} of software architectures}. Vancouver British Columbia Canada: ACM.
\newblock ISBN 978-1-4503-2126-6, pp. 13--22.
\newblock \doi{10.1145/2465478.2465490}.
\newblock \urlprefix\url{https://dl.acm.org/doi/10.1145/2465478.2465490}.

\bibitem{silvera_navantias_2020}
Silvera JI, Muñoz JL, Luquero JM et~al.
\newblock Navantia’s {Digital} {Twin} {Implementation} {Perspective} in {Military} {Naval} {Platform} {Life} {Cycle}.
\newblock \emph{NATO} 2020; .

\bibitem{peter_felstead_knds_2024}
{Peter Felstead}.
\newblock {KNDS} and {Arquus} to create ‘first digital twin demonstrator for a ground combat system’, 2024.
\newblock \urlprefix\url{https://euro-sd.com/2024/02/major-news/36516/knds-arquus-vbci-digital-twin/}.

\bibitem{skinner_taking_nodate}
Skinner SG.
\newblock Taking {Simulation} {Interoperability} {Standards} to the {Next} {Level} with {Digital} {Twins}.
\newblock In \emph{{STO} {Meeting} {Proceedings} {NATO}}.
\newblock ISBN STO-MP-MSG-197.
\newblock \urlprefix\url{https://www.sto.nato.int/publications/STO%20Meeting%20Proceedings/STO-MP-MSG-197/MP-MSG-197-25.pdf}.

\bibitem{altamiranda_system_2019}
Altamiranda E and Colina E.
\newblock A {System} of {Systems} {Digital} {Twin} to {Support} {Life} {Time} {Management} and {Life} {Extension} of {Subsea} {Production} {Systems}.
\newblock In \emph{{OCEANS} 2019 - {Marseille}}. Marseille, France: IEEE.
\newblock ISBN 978-1-72811-450-7, pp. 1--9.
\newblock \doi{10.1109/OCEANSE.2019.8867187}.
\newblock \urlprefix\url{https://ieeexplore.ieee.org/document/8867187/}.

\bibitem{noauthor_defence_2022}
of~Defence UM.
\newblock Defence {Digital} {Twin} {Implementation} {Road} {Map} and {Development} {Framework}, 2022.
\newblock \url{https://www.teamdefence.info/wp-content/uploads/2022/03/20210121-Digital-Twin-Implementation-Road-Map-and-Development-Framework-White-Paper-V1.pdf} [Accessed: 2024-05-06].

\bibitem{dietz_digital_2020}
Dietz M and Pernul G.
\newblock Digital {Twin}: {Empowering} {Enterprises} {Towards} a {System}-of-{Systems} {Approach}.
\newblock \emph{Business \& Information Systems Engineering} 2020; 62(2): 179--184.
\newblock \doi{10.1007/s12599-019-00624-0}.
\newblock \urlprefix\url{http://link.springer.com/10.1007/s12599-019-00624-0}.

\bibitem{budiardjo_anto_digital_2021}
Anto B and Doug M.
\newblock Digital {Twin} {System} {Interoperability} {Framework}, 2021.
\newblock \urlprefix\url{https://www.digitaltwinconsortium.org/pdf/Digital-Twin-System-Interoperability-Framework-12072021.pdf}.

\bibitem{wu_remote_2023}
Wu X, Lu G and Wu Z.
\newblock Remote sensing technology in the construction of digital twin basins: Applications and prospects.
\newblock \emph{Water} 2023; 15(11).
\newblock \doi{10.3390/w15112040}.
\newblock \urlprefix\url{https://www.mdpi.com/2073-4441/15/11/2040}.

\bibitem{piras_digital_2024}
Piras G, Agostinelli S and Muzi F.
\newblock Digital twin framework for built environment: A review of key enablers.
\newblock \emph{Energies} 2024; 17(2).
\newblock \doi{10.3390/en17020436}.
\newblock \urlprefix\url{https://www.mdpi.com/1996-1073/17/2/436}.

\bibitem{farsi2020digital}
Farsi M, Daneshkhah A, Hosseinian-Far A et~al.
\newblock \emph{Digital twin technologies and smart cities}.
\newblock Springer, 2020.

\bibitem{white2021digital}
White G, Zink A, Codec{\'a} L et~al.
\newblock A digital twin smart city for citizen feedback.
\newblock \emph{Cities} 2021; 110: 103064.

\bibitem{deng2021systematic}
Deng T, Zhang K and Shen ZJM.
\newblock A systematic review of a digital twin city: A new pattern of urban governance toward smart cities.
\newblock \emph{Journal of Management Science and Engineering} 2021; 6(2): 125--134.

\bibitem{mohammadi2017smart}
Mohammadi N and Taylor JE.
\newblock Smart city digital twins.
\newblock In \emph{2017 IEEE Symposium Series on Computational Intelligence (SSCI)}. IEEE, pp. 1--5.

\bibitem{deren2021smart}
Deren L, Wenbo Y and Zhenfeng S.
\newblock Smart city based on digital twins.
\newblock \emph{Computational Urban Science} 2021; 1: 1--11.

\bibitem{aguiar2020localization}
Aguiar AS, Dos~Santos FN, Cunha JB et~al.
\newblock Localization and mapping for robots in agriculture and forestry: A survey.
\newblock \emph{Robotics} 2020; 9(4): 97.

\bibitem{nie2022artificial}
Nie J, Wang Y, Li Y et~al.
\newblock Artificial intelligence and digital twins in sustainable agriculture and forestry: a survey.
\newblock \emph{Turkish Journal of Agriculture and Forestry} 2022; 46(5): 642--661.

\bibitem{buonocore2022digitalforestry}
Buonocore L, Yates J and Valentini R.
\newblock A proposal for a forest digital twin framework and its perspectives.
\newblock \emph{Forests} 2022; 13(4): 498.

\bibitem{tagarakis2024forestryagricolture}
Tagarakis AC, Benos L, Kyriakarakos G et~al.
\newblock Digital twins in agriculture and forestry: A review.
\newblock \emph{Sensors} 2024; 24(10): 3117.

\bibitem{hassani2022impactful}
Hassani H, Huang X and MacFeely S.
\newblock Impactful digital twin in the healthcare revolution.
\newblock \emph{Big Data and Cognitive Computing} 2022; 6(3): 83.

\bibitem{sun2023digital}
Sun T, He X and Li Z.
\newblock Digital twin in healthcare: Recent updates and challenges.
\newblock \emph{Digital Health} 2023; 9: 20552076221149651.

\bibitem{liu2019novel}
Liu Y, Zhang L, Yang Y et~al.
\newblock A novel cloud-based framework for the elderly healthcare services using digital twin.
\newblock \emph{IEEE access} 2019; 7: 49088--49101.

\bibitem{tuhaise2023technologies}
Tuhaise VV, Tah JHM and Abanda FH.
\newblock Technologies for digital twin applications in construction.
\newblock \emph{Automation in Construction} 2023; 152: 104931.

\bibitem{opoku2021digital}
Opoku DGJ, Perera S, Osei-Kyei R et~al.
\newblock Digital twin application in the construction industry: A literature review.
\newblock \emph{Journal of Building Engineering} 2021; 40: 102726.

\bibitem{alves2019digitalfarming}
Alves RG, Souza G, Maia RF et~al.
\newblock A digital twin for smart farming.
\newblock In \emph{2019 IEEE Global Humanitarian Technology Conference (GHTC)}. IEEE, pp. 1--4.

\bibitem{verdouw2021digitalfarming}
Verdouw C, Tekinerdogan B, Beulens A et~al.
\newblock Digital twins in smart farming.
\newblock \emph{Agricultural Systems} 2021; 189: 103046.

\bibitem{pylianidis2021introducing}
Pylianidis C, Osinga S and Athanasiadis IN.
\newblock Introducing digital twins to agriculture.
\newblock \emph{Computers and Electronics in Agriculture} 2021; 184: 105942.

\bibitem{nasirahmadi2022toward}
Nasirahmadi A and Hensel O.
\newblock Toward the next generation of digitalization in agriculture based on digital twin paradigm.
\newblock \emph{Sensors} 2022; 22(2): 498.

\bibitem{noauthor_building_nodate}
Allen B.
\newblock Building {DoD}'s largest-ever {Digital} {Twin} of its kind, 2024.
\newblock \url{https://www.boozallen.com/insights/digital-twin/building-dods-largest-ever-digital-twin-of-its-kind.html} [Accessed: 2024-03-28].

\bibitem{casey_real-time_2024}
Casey L, Dooley J, Codd M et~al.
\newblock A real-time digital twin for active safety in an aircraft hangar.
\newblock \emph{Frontiers in Virtual Reality} 2024; 5: 1372923.
\newblock \doi{10.3389/frvir.2024.1372923}.
\newblock \urlprefix\url{https://www.frontiersin.org/articles/10.3389/frvir.2024.1372923/full}.

\bibitem{wang_digital_2021}
Wang P, Yang M, Zhu J et~al.
\newblock Digital {Twin}-{Enabled} {Online} {Battlefield} {Learning} with {Random} {Finite} {Sets}.
\newblock \emph{Computational Intelligence and Neuroscience} 2021; 2021: 1--15.
\newblock \doi{10.1155/2021/5582241}.
\newblock \urlprefix\url{https://www.hindawi.com/journals/cin/2021/5582241/}.

\bibitem{thrun2002robotic}
Thrun S.
\newblock Robotic mapping: A survey.
\newblock In Lakemeyer G and Nebel B (eds.) \emph{Exploring Artificial Intelligence in the New Millenium}. Morgan Kaufmann, 2002.
\newblock To appear.

\bibitem{elfes_1989}
Elfes A.
\newblock Using occupancy grids for mobile robot perception and navigation.
\newblock \emph{Computer} 1989; 22(6): 46--57.
\newblock \doi{10.1109/2.30720}.

\bibitem{kostavelis2015semantic}
Kostavelis I and Gasteratos A.
\newblock Semantic mapping for mobile robotics tasks: A survey.
\newblock \emph{Robotics and Autonomous Systems} 2015; 66: 86--103.

\bibitem{khan2021comparative}
Khan MU, Zaidi SAA, Ishtiaq A et~al.
\newblock A comparative survey of lidar-slam and lidar based sensor technologies.
\newblock In \emph{2021 Mohammad Ali Jinnah University International Conference on Computing (MAJICC)}. IEEE, pp. 1--8.

\bibitem{tourani2022visual}
Tourani A, Bavle H, Sanchez-Lopez JL et~al.
\newblock Visual slam: What are the current trends and what to expect?
\newblock \emph{Sensors} 2022; 22(23): 9297.

\bibitem{bavle2023slam}
Bavle H, Sanchez-Lopez JL, Cimarelli C et~al.
\newblock From slam to situational awareness: Challenges and survey.
\newblock \emph{Sensors} 2023; 23(10): 4849.

\bibitem{giberna2025dynemo}
Giberna M, Shaheer M, Fernandez-Cortizas M et~al.
\newblock DYNEMO-SLAM: Dynamic Entity and Motion-Aware 3D Scene Graph SLAM.
\newblock \emph{arXiv} 2025; arXiv:2503.02050.

\bibitem{jurado2023planar}
Jurado-Rodriguez D, Mu{\~n}oz-Salinas R, Garrido-Jurado S et~al.
\newblock Planar fiducial markers: a comparative study.
\newblock \emph{Virtual Reality} 2023; 27(3): 1733--1749.

\bibitem{kalaitzakis2021fiducial}
Kalaitzakis M, Cain B, Carroll S et~al.
\newblock Fiducial markers for pose estimation: Overview, applications and experimental comparison of the artag, apriltag, aruco and stag markers.
\newblock \emph{Journal of Intelligent \& Robotic Systems} 2021; 101: 1--26.

\bibitem{agha2022unclonable}
Agha H, Geng Y, Ma X et~al.
\newblock Unclonable human-invisible machine vision markers leveraging the omnidirectional chiral bragg diffraction of cholesteric spherical reflectors.
\newblock \emph{Light: Science \& Applications} 2022; 11(1): 309.

\bibitem{schwartz2018cholesteric}
Schwartz M, Lenzini G, Geng Y et~al.
\newblock Cholesteric liquid crystal shells as enabling material for information-rich design and architecture.
\newblock \emph{Advanced materials} 2018; 30(30): 1707382.

\bibitem{schwartz2021linking}
Schwartz M, Geng Y, Agha H et~al.
\newblock Linking physical objects to their digital twins via fiducial markers designed for invisibility to humans.
\newblock \emph{Multifunctional Materials} 2021; 4(2): 022002.

\bibitem{noauthor_this_2021}
OHB.
\newblock This is how the {Digital} {Twin} works at the port of {Bremen}, 2021.
\newblock \url{https://www.ohb.de/en/magazine/how-the-digital-twin-works-at-the-port-of-bremen} [Accessed: 2024-03-29].

\bibitem{bauer_digital_2021}
Bauer P, Stevens B and Hazeleger W.
\newblock A digital twin of {Earth} for the green transition.
\newblock \emph{Nature Climate Change} 2021; 11(2): 80--83.
\newblock \doi{10.1038/s41558-021-00986-y}.
\newblock \urlprefix\url{https://www.nature.com/articles/s41558-021-00986-y}.

\bibitem{defelipe_towards_2022}
DeFelipe I, Alcalde J, Baykiev E et~al.
\newblock Towards a {Digital} {Twin} of the {Earth} {System}: {Geo}-{Soft}-{CoRe}, a {Geoscientific} {Software} \& {Code} {Repository}.
\newblock \emph{Frontiers in Earth Science} 2022; 10: 828005.
\newblock \doi{10.3389/feart.2022.828005}.
\newblock \urlprefix\url{https://www.frontiersin.org/articles/10.3389/feart.2022.828005/full}.

\bibitem{daya_sagar_digital_2021}
Nativi S and Craglia M.
\newblock Digital {Twins} of the {Earth}.
\newblock In Daya~Sagar B, Cheng Q, McKinley J et~al. (eds.) \emph{Encyclopedia of {Mathematical} {Geosciences}}. Cham: Springer International Publishing.
\newblock ISBN 978-3-030-26050-7, 2021.
\newblock pp. 1--4.
\newblock \doi{10.1007/978-3-030-26050-7_457-1}.
\newblock \urlprefix\url{https://link.springer.com/10.1007/978-3-030-26050-7_457-1}.
\newblock Series Title: Encyclopedia of Earth Sciences Series.

\bibitem{lee_digital_2021}
Lee EBK, Van~Bossuyt DL and Bickford JF.
\newblock Digital {Twin}-{Enabled} {Decision} {Support} in {Mission} {Engineering} and {Route} {Planning}.
\newblock \emph{Systems} 2021; 9(4): 82.
\newblock \doi{10.3390/systems9040082}.
\newblock \urlprefix\url{https://www.mdpi.com/2079-8954/9/4/82}.

\bibitem{alaez_uavradio_2024}
Aláez D, Celaya-Echarri M, Azpilicueta L et~al.
\newblock {UAVradio}: {Radio} link path loss estimation for {UAVs}.
\newblock \emph{SoftwareX} 2024; 25: 101628.
\newblock \doi{10.1016/j.softx.2023.101628}.
\newblock \urlprefix\url{https://linkinghub.elsevier.com/retrieve/pii/S2352711023003242}.

\bibitem{zhu_mastering_2024}
Zhu J, Kuang M, Zhou W et~al.
\newblock Mastering air combat game with deep reinforcement learning.
\newblock \emph{Defence Technology} 2024; 34: 295--312.
\newblock \doi{10.1016/j.dt.2023.08.019}.
\newblock \urlprefix\url{https://linkinghub.elsevier.com/retrieve/pii/S2214914723002349}.

\bibitem{soliman_ai-based_2023}
Soliman A, Al-Ali A, Mohamed A et~al.
\newblock {AI}-based {UAV} navigation framework with digital twin technology for mobile target visitation.
\newblock \emph{Engineering Applications of Artificial Intelligence} 2023; 123: 106318.
\newblock \doi{10.1016/j.engappai.2023.106318}.
\newblock \urlprefix\url{https://linkinghub.elsevier.com/retrieve/pii/S095219762300502X}.

\bibitem{koenig_design_2004}
Koenig N and Howard A.
\newblock Design and use paradigms for gazebo, an open-source multi-robot simulator.
\newblock In \emph{2004 {IEEE}/{RSJ} {International} {Conference} on {Intelligent} {Robots} and {Systems} ({IROS}) ({IEEE} {Cat}. {No}.{04CH37566})}, volume~3. Sendai, Japan: IEEE.
\newblock ISBN 978-0-7803-8463-7, pp. 2149--2154.
\newblock \doi{10.1109/IROS.2004.1389727}.
\newblock \urlprefix\url{http://ieeexplore.ieee.org/document/1389727/}.

\bibitem{quigley_ros_2009}
Quigley M, Conley K, Gerkey B et~al.
\newblock \emph{{ROS}: an open-source {Robot} {Operating} {System}}, volume~3.
\newblock Proc. of the IEEE Intl. Conf. on Robotics and Automation (ICRA) Workshop on Open Source Robotics, 2009.
\newblock Journal Abbreviation: ICRA Workshop on Open Source Software Publication Title: ICRA Workshop on Open Source Software.

\bibitem{noauthor_unity_2023}
Technologies U.
\newblock Unity, 2023.
\newblock \url{https://unity.com/} [Accessed: 2024-06-20].

\bibitem{ochando_data_2023}
Ochando FJ, Cantero A, Guerrero JI et~al.
\newblock Data acquisition for condition monitoring in tactical vehicles: On-board computer development.
\newblock \emph{Sensors} 2023; 23(12).
\newblock \doi{10.3390/s23125645}.
\newblock \urlprefix\url{https://www.mdpi.com/1424-8220/23/12/5645}.

\bibitem{guo_technical_2025}
Guo Y and Li L.
\newblock Technical {Practice} and {Application} of {Data} {Governance} in {Military} {Exercise} and {Training} {Data} {Management}.
\newblock In \emph{Proceedings of the 2025 {International} {Symposium} on {Machine} {Learning} and {Social} {Computing}}. Hongkong China: ACM.
\newblock ISBN 9798400721274, pp. 507--513.
\newblock \doi{10.1145/3778450.3778529}.
\newblock \urlprefix\url{https://dl.acm.org/doi/10.1145/3778450.3778529}.

\bibitem{liu_study_2018}
Liu G and Su Y.
\newblock Study on big data visualization of joint operation command and control system.
\newblock In Chin FYL, Chen CLP, Khan L et~al. (eds.) \emph{Big Data -- BigData 2018}. Cham: Springer International Publishing.
\newblock ISBN 978-3-319-94301-5, pp. 372--380.

\bibitem{he2019data}
He R, Chen G, Dong C et~al.
\newblock Data-driven digital twin technology for optimized control in process systems.
\newblock \emph{ISA transactions} 2019; 95: 221--234.

\bibitem{stavropoulos2021robust}
Stavropoulos P, Papacharalampopoulos A, Michail CK et~al.
\newblock Robust additive manufacturing performance through a control oriented digital twin.
\newblock \emph{Metals} 2021; 11(5): 708.

\bibitem{gehrmann2019digital}
Gehrmann C and Gunnarsson M.
\newblock A digital twin based industrial automation and control system security architecture.
\newblock \emph{IEEE Transactions on Industrial Informatics} 2019; 16(1): 669--680.

\bibitem{liu_control_2024}
Liu GP.
\newblock Control {Strategies} for {Digital} {Twin} {Systems}.
\newblock \emph{IEEE/CAA Journal of Automatica Sinica} 2024; 11(1): 170--180.
\newblock \doi{10.1109/JAS.2023.123834}.
\newblock \urlprefix\url{https://ieeexplore.ieee.org/document/10399361/}.

\bibitem{noauthor_fcx_2023}
Fincantieri.
\newblock {FCX} series - {A} class apart, 2023.
\newblock \url{https://www.fincantieri.com/globalassets/prodotti-servizi/navi-militari/fincantieri_broch-m-22-22-fcx-series-lr.pdf} [Accessed: 2024-03-28].

\bibitem{noauthor_delivering_2021}
F35.
\newblock Delivering {Digitally} for {F}-35 {Force} {Management} {Solutions}, 2021.
\newblock \url{https://www.f35.com/f35/news-and-features/delivering-digitally-for-f35-force-management-solution.html} [Accessed: 2024-03-28].

\bibitem{singh_physical-virtual_2024}
Singh O and Ray AK.
\newblock A physical-virtual based digital twin robotic hand.
\newblock \emph{International Journal on Interactive Design and Manufacturing (IJIDeM)} 2024; \doi{10.1007/s12008-024-01773-7}.
\newblock \urlprefix\url{https://link.springer.com/10.1007/s12008-024-01773-7}.

\bibitem{modeer_towards_2023}
Modéer MR.
\newblock Towards a {Digital} {Twin} for {Underwater} {Systems} {Based} on {Meta}-{Learning}.
\newblock NATO.
\newblock ISBN STO-MP-AVT-369.
\newblock \urlprefix\url{https://www.sto.nato.int/publications/STO%20Meeting%20Proceedings/STO-MP-AVT-369/MP-AVT-369-26P.pdf}.

\bibitem{noauthor_physical_2007}
Physical, {Technical}, {Tactical}, {Mental} – {The} {Next} {Step}, 2007.
\newblock \url{https://www.badmintoncentral.com/forums/index.php?threads/physical-technical-tactical-mental-%E2%80%93-the-next-step.44065/} [Accessed: 2024-07-05].

\bibitem{desmond_why_2022}
Desmond R.
\newblock Why the technical, tactical, physical and psychological sides of football are deeply intertwined, 2022.
\newblock \urlprefix\url{https://themastermindsite.com/2022/07/02/why-the-technical-tactical-physical-and-psychological-sides-of-football-are-deeply-intertwined/}.

\bibitem{okegbile2022human}
Okegbile SD, Cai J, Niyato D et~al.
\newblock Human digital twin for personalized healthcare: Vision, architecture and future directions.
\newblock \emph{IEEE network} 2022; 37(2): 262--269.

\bibitem{lin2024human}
Lin Y, Chen L, Ali A et~al.
\newblock Human digital twin: A survey.
\newblock \emph{Journal of Cloud Computing} 2024; 13(1): 131.

\bibitem{miller2022unified}
Miller ME and Spatz E.
\newblock A unified view of a human digital twin.
\newblock \emph{Human-Intelligent Systems Integration} 2022; 4(1): 23--33.

\bibitem{glascoe_human_2024}
{Glascoe III, William O} and {Wenli Wang}.
\newblock {HUMAN} {DIGITAL} {TWIN} {INITIATIVES}.
\newblock \emph{Issues In Information Systems} 2024; 25(4).
\newblock \doi{10.48009/4_iis_2024_123}.
\newblock \urlprefix\url{https://iacis.org/iis/2024/4_iis_2024_287-298.pdf}.

\bibitem{fawkes_digital_2025}
Fawkes A and Burden D.
\newblock Digital human twins and the military metaverse: opportunities and challenges.
\newblock \emph{AI \& SOCIETY} 2025; \doi{10.1007/s00146-025-02508-2}.
\newblock \urlprefix\url{https://link.springer.com/10.1007/s00146-025-02508-2}.

\bibitem{knds_knds_2024}
{KNDS}.
\newblock {KNDS} - {NumCo} project, 2024.
\newblock \url{https://www.knds.fr/en/our-news/latest-news/experimentation-first-digital-twin-demonstrator-ground-combat-system} [Accessed: 2024-03-28].

\bibitem{eddy_predictive_2024}
Eddy CW, Castanier MP and Wagner JR.
\newblock Predictive {Maintenance} of a {Ground} {Vehicle} {Using} {Digital} {Twin} {Technology}.
\newblock Detroit, Michigan, United States, pp. 2024--01--2867.
\newblock \doi{10.4271/2024-01-2867}.
\newblock \urlprefix\url{https://www.sae.org/content/2024-01-2867}.

\bibitem{song_architecture_2022}
Song M, Shi Q, Hu Q et~al.
\newblock On the {Architecture} and {Key} {Technology} for {Digital} {Twin} {Oriented} to {Equipment} {Battle} {Damage} {Test} {Assessment}.
\newblock \emph{Electronics} 2022; 12(1): 128.
\newblock \doi{10.3390/electronics12010128}.
\newblock \urlprefix\url{https://www.mdpi.com/2079-9292/12/1/128}.

\bibitem{li_preliminary_2020}
Li S, Yang Q, Xing J et~al.
\newblock Preliminary {Study} on the {Application} of {Digital} {Twin} in {Military} {Engineering} and {Equipment}.
\newblock In \emph{2020 {Chinese} {Automation} {Congress} ({CAC})}. Shanghai, China: IEEE.
\newblock ISBN 978-1-72817-687-1, pp. 7249--7255.
\newblock \doi{10.1109/CAC51589.2020.9326911}.
\newblock \urlprefix\url{https://ieeexplore.ieee.org/document/9326911/}.

\bibitem{noauthor_rheinmetall_2018}
Rheinmetall.
\newblock Rheinmetall and {Rohde} \& {Schwarz} move into position to digitize the {Bundeswehr}, 2018.
\newblock \url{https://www.rheinmetall.com/en/media/news-watch/news/2018/2018-10-30_rrs-mitcos-founded} [Accessed: 2024-03-29].

\bibitem{noauthor_e-learning_nodate}
Rheinmetall.
\newblock E-{Learning} - {Army} - {With} digital training to mission readiness, 2024.
\newblock \url{https://www.rheinmetall.com/en/products/simulation-training/simulation-and-training/military-training-solutions/army/e-learning-army} [Accessed: 2024-03-29].

\bibitem{noauthor_knds_2024}
KNDS.
\newblock {KNDS} supports more than 50 armed forces around the world.
\newblock \emph{Armada International} 2024; \urlprefix\url{https://www.armadainternational.com/2024/06/knds-supports-more-than-50-armed-forces-around-the-world/}.

\bibitem{rohdeschwarz_magazine_looking_2024}
{Rohde\&Schwarz Magazine}.
\newblock Looking {Towards} {6G}.
\newblock \emph{Rohde\&Schwarz magazine} 2024; \urlprefix\url{https://www.rohde-schwarz.com/nl/about/magazine/looking-towards-6g_256451.html}.

\bibitem{tan_construction_2024}
Wang F, Ye L, Zheng S et~al.
\newblock Construction of {Digital} {Twin} {Battlefield} with {Command} and {Control} as the {Core}.
\newblock In Tan Y and Shi Y (eds.) \emph{Data {Mining} and {Big} {Data}}, volume 2018. Singapore: Springer Nature Singapore.
\newblock ISBN 978-981-9708-43-7 978-981-9708-44-4, 2024.
\newblock pp. 103--114.
\newblock \doi{10.1007/978-981-97-0844-4_8}.
\newblock \urlprefix\url{https://link.springer.com/10.1007/978-981-97-0844-4_8}.
\newblock Series Title: Communications in Computer and Information Science.

\bibitem{noauthor_mosaic_2022}
{MoSaiC} - {Real}-time {Monitoring} and {Sampling} of {CB} menaces for improved dynamic mapping of threats, vulnerabilities and response capacities, 2022.
\newblock \url{https://defence-industry-space.ec.europa.eu/system/files/2022-09/Factsheet_EDF21_MOSAIC_updated.pdf} [Accessed: 2024-06-17].

\bibitem{zhang_digital_2024-1}
Zhang Y.
\newblock Digital {Twin} for {Aerial}-{Ground} {Networks}.
\newblock In \emph{Digital {Twin}}, volume~16. Cham: Springer Nature Switzerland.
\newblock ISBN 978-3-031-51818-8 978-3-031-51819-5, 2024.
\newblock pp. 87--103.
\newblock \doi{10.1007/978-3-031-51819-5_7}.
\newblock \urlprefix\url{https://link.springer.com/10.1007/978-3-031-51819-5_7}.
\newblock Series Title: Simula SpringerBriefs on Computing.

\bibitem{choi_digital_2022}
Choi J, Moon S and Min S.
\newblock Digital {Twin} {Simulation} {Modeling} {Process} with {System} {Dynamics}: {An} application to {Naval} ship operation.
\newblock preprint, Preprints, 2022.
\newblock \doi{10.22541/au.166939381.12328549/v1}.
\newblock \urlprefix\url{https://www.authorea.com/users/355212/articles/605073-digital-twin-simulation-modeling-process-with-system-dynamics-an-application-to-naval-ship-operation?commit=14874d15f1b3d954e51c773a907d86a623130b63}.

\bibitem{noauthor_prime_2022}
Navantia.
\newblock Prime {Minister} {Pedro} {Sánchez} presides the cutting of the first steel plate of {F}-110 class frigate for {Spanish} {Navy}, 2022.
\newblock \url{https://www.navantia.es/en/news/press-releases/prime-minister-pedro-sanchez-presides-the-cutting-of-the-first-steel-plate-of-f-110-class-frigate-for-spanish-navy/} [Accessed: 2024-04-22].

\bibitem{major_real-time_2021}
Major P, Li G, Zhang H et~al.
\newblock Real-{Time} {Digital} {Twin} {Of} {Research} {Vessel} {For} {Remote} {Monitoring}.
\newblock In \emph{{ECMS} 2021 {Proceedings} edited by {Khalid} {Al}-{Begain}, {Mauro} {Iacono}, {Lelio} {Campanile}, {Andrzej} {Bargiela}}. ECMS.
\newblock ISBN 978-3-937436-72-2, pp. 159--164.
\newblock \doi{10.7148/2021-0159}.
\newblock \urlprefix\url{http://www.scs-europe.net/dlib/2021/2021-0159.htm}.

\bibitem{blachnik_development_2023}
Blachnik M, Przyłucki R, Golak S et~al.
\newblock On the {Development} of a {Digital} {Twin} for {Underwater} {UXO} {Detection} {Using} {Magnetometer}-{Based} {Data} in {Application} for the {Training} {Set} {Generation} for {Machine} {Learning} {Models}.
\newblock \emph{Sensors} 2023; 23(15): 6806.
\newblock \doi{10.3390/s23156806}.
\newblock \urlprefix\url{https://www.mdpi.com/1424-8220/23/15/6806}.

\bibitem{noauthor_superior_nodate}
Dynamics G.
\newblock Superior situational awareness for maritime security and safety with {BlueSHIELD}., 2024.
\newblock \url{https://gd-ms.it/systems/blueshield/} [Accessed: 2024-03-29].

\bibitem{lunsford_evaluation_2021}
Lunsford I and Bradley TH.
\newblock Evaluation of unmanned aerial vehicle tactics through the metrics of survivability.
\newblock \emph{The Journal of Defense Modeling and Simulation} 2022; 19(4): 855--864.
\newblock \doi{10.1177/15485129211031672}.
\newblock \urlprefix\url{https://doi.org/10.1177/15485129211031672}.
\newblock \eprint{https://doi.org/10.1177/15485129211031672}.

\bibitem{panagiotopoulou_samas_2024}
Panagiotopoulou V, Sbarufatti C and Giglio M.
\newblock {SAMAS} 2: {Structural} health and ballistic impact monitoring and prognosis on a military helicopter.
\newblock \emph{Procedia Structural Integrity} 2024; 54: 482--489.
\newblock \doi{10.1016/j.prostr.2024.01.110}.
\newblock \urlprefix\url{https://linkinghub.elsevier.com/retrieve/pii/S2452321624001100}.

\bibitem{kraft_air_2016}
Kraft EM.
\newblock The {Air} {Force} {Digital} {Thread}/{Digital} {Twin} - {Life} {Cycle} {Integration} and {Use} of {Computational} and {Experimental} {Knowledge}.
\newblock In \emph{54th {AIAA} {Aerospace} {Sciences} {Meeting}}. San Diego, California, USA: American Institute of Aeronautics and Astronautics.
\newblock ISBN 978-1-62410-393-3.
\newblock \doi{10.2514/6.2016-0897}.
\newblock \urlprefix\url{https://arc.aiaa.org/doi/10.2514/6.2016-0897}.

\bibitem{agrawal_deep_2024}
Agrawal A, Farid F and Vyas N.
\newblock Deep {Learning}-{Based} {Digital} {Twining} {Models} for {Inter} {System} {Behavior} and {Health} {Assessment} of {Combat} {Aircraft} {Systems}.
\newblock In \emph{{SAE} {Techinical} {Paper} {Series}}.
\newblock \urlprefix\url{https://www.sae.org/publications/technical-papers/content/2024-26-0478/}.

\bibitem{noauthor_leonardo_2022}
Leonardo.
\newblock Leonardo - {Digitalisation} technological development, 2022.
\newblock \url{https://www.leonardo.com/en/focus-detail/-/detail/digitisation-technological-development} [Accessed: 2024-03-28].

\bibitem{noauthor_technology_2022}
Eurofighter.
\newblock Technology is future for eurofighter support.
\newblock \emph{Eurofighter News} 2022; \url{https://www.eurofighter.com/news/futuresupport} [Accessed: 2024-04-22].

\bibitem{wang_research_2021}
Wang Yc, Zhang N, Li H et~al.
\newblock Research on {Digital} {Twin} {Framework} of {Military} {Large}-scale {UAV} {Based} on {Cloud} {Computing}.
\newblock \emph{Journal of Physics: Conference Series} 2021; 1738(1): 012052.
\newblock \doi{10.1088/1742-6596/1738/1/012052}.
\newblock \urlprefix\url{https://iopscience.iop.org/article/10.1088/1742-6596/1738/1/012052}.

\bibitem{kapteyn_predictive_2022}
Kapteyn M and Willcox KE.
\newblock Predictive {Digital} {Twins} as a {Foundation} for {Improved} {Mission} {Readiness}.
\newblock \emph{NATO} 2022; .

\bibitem{pinello_preliminary_2024}
Pinello L, Hassan O, Giglio M et~al.
\newblock Preliminary {Nose} {Landing} {Gear} {Digital} {Twin} for {Damage} {Detection}.
\newblock \emph{Aerospace} 2024; 11(3): 222.
\newblock \doi{10.3390/aerospace11030222}.
\newblock \urlprefix\url{https://www.mdpi.com/2226-4310/11/3/222}.

\bibitem{darema_hardware_2020}
Salinger SJ, Kapteyn MG, Kays C et~al.
\newblock A {Hardware} {Testbed} for {Dynamic} {Data}-{Driven} {Aerospace} {Digital} {Twins}.
\newblock In Darema F, Blasch E, Ravela S et~al. (eds.) \emph{Dynamic {Data} {Driven} {Applications} {Systems}}, volume 12312. Cham: Springer International Publishing.
\newblock ISBN 978-3-030-61724-0 978-3-030-61725-7, 2020.
\newblock pp. 37--45.
\newblock \doi{10.1007/978-3-030-61725-7_7}.
\newblock \urlprefix\url{http://link.springer.com/10.1007/978-3-030-61725-7_7}.
\newblock Series Title: Lecture Notes in Computer Science.

\bibitem{shen_multi-uav_2023}
Shen G, Lei L, Zhang X et~al.
\newblock Multi-{UAV} {Cooperative} {Search} {Based} on {Reinforcement} {Learning} {With} a {Digital} {Twin} {Driven} {Training} {Framework}.
\newblock \emph{IEEE Transactions on Vehicular Technology} 2023; 72(7): 8354--8368.
\newblock \doi{10.1109/TVT.2023.3245120}.
\newblock \urlprefix\url{https://ieeexplore.ieee.org/document/10045049/}.

\bibitem{middeldorp_quanitfying_2023}
Middeldorp L, Malone K and Noordkamp W.
\newblock Quantifying the robustness of a bayesian belief network in the context of unmanned aerial system threat prediction.
\newblock \emph{The Journal of Defense Modeling and Simulation} 2025; 22(4): 359--371.
\newblock \doi{10.1177/15485129231206825}.
\newblock \urlprefix\url{https://doi.org/10.1177/15485129231206825}.
\newblock \eprint{https://doi.org/10.1177/15485129231206825}.

\bibitem{ji_digital_2021}
Ji G, Hao Jg, Gao Jl et~al.
\newblock Digital {Twin} {Modeling} {Method} for {Individual} {Combat} {Quadrotor} {UAV}.
\newblock In \emph{2021 {IEEE} 1st {International} {Conference} on {Digital} {Twins} and {Parallel} {Intelligence} ({DTPI})}. Beijing, China: IEEE.
\newblock ISBN 978-1-66543-337-2, pp. 1--4.
\newblock \doi{10.1109/DTPI52967.2021.9540131}.
\newblock \urlprefix\url{https://ieeexplore.ieee.org/document/9540131/}.

\bibitem{keskerian_slingshot_2022}
Keskerian K.
\newblock Slingshot {Aerospace} {Announces} {Industry}'s {First} {Digital} {Space} {Twin}, 2022.
\newblock \url{https://www.slingshot.space/news/slingshot-aerospace-stratfi} [Accessed: 2024-03-28].

\bibitem{shangguan_digital_2020}
Shangguan D, Chen L and Ding J.
\newblock A {Digital} {Twin}-{Based} {Approach} for the {Fault} {Diagnosis} and {Health} {Monitoring} of a {Complex} {Satellite} {System}.
\newblock \emph{Symmetry} 2020; 12(8): 1307.
\newblock \doi{10.3390/sym12081307}.
\newblock \urlprefix\url{https://www.mdpi.com/2073-8994/12/8/1307}.

\bibitem{shtofenmakher_adaptation_2024}
Shtofenmakher A and Shao G.
\newblock Adaptation of {ISO} 23247 to {Aerospace} {Digital} {Twin} {Applications}-{On}-{Orbit} {Collision} {Avoidance} and {Space}-{Based} {Debris} {Detection}.
\newblock In \emph{{AIAA} {SCITECH} 2024 {Forum}}. Orlando, FL: American Institute of Aeronautics and Astronautics.
\newblock ISBN 978-1-62410-711-5.
\newblock \doi{10.2514/6.2024-0275}.
\newblock \urlprefix\url{https://arc.aiaa.org/doi/10.2514/6.2024-0275}.

\bibitem{lei_digital_2024}
Lei H, Fanli Z, Wei W et~al.
\newblock Digital twin method and application practice of spacecraft system driven by mechanism data.
\newblock \emph{Digital Twin} 2024; 4: 2.
\newblock \doi{10.12688/digitaltwin.17913.1}.
\newblock \urlprefix\url{https://digitaltwin1.org/articles/4-2/v1}.

\bibitem{noauthor_digital_2020}
of~Aeronautics AI and Astronautics.
\newblock Digital {Twin}: {Definition} \& {Value} – {An} {AIAA} and {AIA} {Position} {Paper}.
\newblock Technical report, American Institute of Aeronautics and Astronautics, 2020.
\newblock \url{https://www.aia-aerospace.org/wp-content/uploads/Digital-Twin-Institute-Position-Paper-December-2020-1.pdf} [Accessed: 2024-05-24].

\bibitem{noauthor_automated_2023}
Automated {Creation} of {Network} {Digital} {Twins} - {EXata}, 2023.
\newblock \url{https://www.keysight.com/us/en/assets/3122-1411/white-papers/Automated-Creation-of-Network-Digital-Twins.pdf} [Accessed: 2024-03-28].

\bibitem{avrilionis_towards_2021}
Avrilionis D and Hardjono T.
\newblock Towards {Blockchain}-enabled {Open} {Architectures} for {Scalable} {Digital} {Asset} {Platforms}, 2021.
\newblock \doi{10.48550/ARXIV.2110.12553}.
\newblock \urlprefix\url{https://arxiv.org/abs/2110.12553}.
\newblock Version Number: 1.

\bibitem{maathuis_decision_2021}
Maathuis C, Pieters W and Van Den~Berg J.
\newblock Decision support model for effects estimation and proportionality assessment for targeting in cyber operations.
\newblock \emph{Defence Technology} 2021; 17(2): 352--374.
\newblock \doi{10.1016/j.dt.2020.04.007}.
\newblock \urlprefix\url{https://linkinghub.elsevier.com/retrieve/pii/S2214914719309250}.

\bibitem{barker_digital_2021}
Vielberth M, Glas M, Dietz M et~al.
\newblock A {Digital} {Twin}-{Based} {Cyber} {Range} for {SOC} {Analysts}.
\newblock In Barker K and Ghazinour K (eds.) \emph{Data and {Applications} {Security} and {Privacy} {XXXV}}, volume 12840. Cham: Springer International Publishing.
\newblock ISBN 978-3-030-81241-6 978-3-030-81242-3, 2021.
\newblock pp. 293--311.
\newblock \doi{10.1007/978-3-030-81242-3_17}.
\newblock \urlprefix\url{https://link.springer.com/10.1007/978-3-030-81242-3_17}.
\newblock Series Title: Lecture Notes in Computer Science.

\bibitem{shafik_utilising_2022}
Sani S, Schaefer D and Milisavljevic-Syed J.
\newblock Utilising {Digital} {Twins} for {Increasing} {Military} {Supply} {Chain} {Visibility}.
\newblock In Shafik M and Case K (eds.) \emph{Advances in {Transdisciplinary} {Engineering}}. IOS Press.
\newblock ISBN 978-1-64368-330-0 978-1-64368-331-7, 2022.
\newblock \doi{10.3233/ATDE220595}.
\newblock \urlprefix\url{https://ebooks.iospress.nl/doi/10.3233/ATDE220595}.

\bibitem{adami_strategic_2023}
Adami M, Nowak M, Toelen M et~al.
\newblock The {Strategic} {Promise} of {Digital} {Twins} to {Enhance} {Supply} {Chain} {Resilience}.
\newblock \urlprefix\url{https://www.researchgate.net/profile/Panagiotis-Kikiras/publication/371167018_European_Defence_Agency_Scientific_2023_Scientific_Paper_Awards/links/647712706a3c4c6efbf4d459/European-Defence-Agency-Scientific-2023-Scientific-Paper-Awards.pdf#page=120}.

\bibitem{winbush_gcss-army_2021}
Winbush J.
\newblock {GCSS}-{ARMY} {Global} {Combat} {Support} {System}, 2021.
\newblock \url{https://www.eis.army.mil/programs/gcss-army} [Accessed: 2024-03-28].

\bibitem{liu_research_2024}
Liu QS.
\newblock Research on the application of digital twin technology in air defense system.
\newblock In \emph{2024 5th International Conference on Computer Engineering and Application (ICCEA)}. IEEE, pp. 286--292.

\bibitem{glaessgen_digital_2012}
Glaessgen E and Stargel D.
\newblock The {Digital} {Twin} {Paradigm} for {Future} {NASA} and {U}.{S}. {Air} {Force} {Vehicles}.
\newblock \emph{53rd AIAA/ASME/ASCE/AHS/ASC Structures, Structural Dynamics and Materials Conference; 20th AIAA/ASME/AHS Adaptive Structures Conference; 14th AIAA} 2012; Honolulu, Hawaii.
\newblock \doi{10.2514/6.2012-1818}.
\newblock \urlprefix\url{http://arc.aiaa.org/doi/abs/10.2514/6.2012-1818/}.

\bibitem{zhang_actor_2022}
Zhang J and Xue Q.
\newblock Actor–critic-based decision-making method for the artificial intelligence commander in tactical wargames.
\newblock \emph{The Journal of Defense Modeling and Simulation} 2022; 19(3): 467--480.
\newblock \doi{10.1177/1548512920954542}.
\newblock \urlprefix\url{https://doi.org/10.1177/1548512920954542}.
\newblock \eprint{https://doi.org/10.1177/1548512920954542}.

\bibitem{parathyras_prospects_2025}
Parathyras K, Papapostolou C and Kanetaki Z.
\newblock Prospects and challenges of integrating digital technologies in the military supply chain.
\newblock In \emph{Novel \& Intelligent Digital Systems Conferences}. Springer, pp. 284--298.

\bibitem{alcaraz_digital_2022}
Alcaraz C and Lopez J.
\newblock Digital {Twin}: {A} {Comprehensive} {Survey} of {Security} {Threats}.
\newblock \emph{IEEE Communications Surveys \& Tutorials} 2022; 24(3): 1475--1503.
\newblock \doi{10.1109/COMST.2022.3171465}.
\newblock \urlprefix\url{https://ieeexplore.ieee.org/document/9765576/}.

\end{thebibliography}

\begin{biogs}

\textbf{Marco Giberna} is a PhD candidate at the Interdisciplinary Centre for Security, Reliability and Trust (SnT), University of Luxembourg. His research focuses on robotic situational awareness in dynamic and challenging environments. He received his MSc in Electronic and Computer Engineering from the University of Trieste.

\textbf{Holger Voos} is a Full Professor at the University of Luxembourg and Head of the Automation and Robotics Research Group at the Interdisciplinary Centre for Security, Reliability and Trust (SnT). His research focuses on situational awareness, motion planning, and networked control for autonomous robots and vehicles.

\textbf{Paulo Tavares} is a Research Coordinator in Advanced Monitoring and Structural Integrity at INEGI, holding a PhD in Mechanical Engineering and an MSc in Lasers and Optoelectronics from the University of Porto. His expertise spans optical metrology, image processing, and failure analysis, complemented by a decade of industrial experience at Texas Instruments and his current role as a non-governmental expert (CnGE) for the European Defence Agency.

\textbf{João Nunes} is a Research and Development Engineer at INEGI, where he focuses on the implementation of Industrial Internet of Things (IIoT) solutions and machine interoperability. He holds an MSc in Biomedical Engineering from the University of Minho.

\textbf{Tobias Sorg} is a Business Development Manager for Research, Technology and Innovation at HENSOLDT Sensors GmbH, and an accredited expert in extended reality (XR) technologies. During his time at IBM Labs and Airbus, he conducted research on digital twin technologies and has developed extensive expertise in this field.

\textbf{Andrea Masini} is the CEO of FlySight, where he leverages over 18 years of experience in the Flyby Group to lead the development of advanced C4ISR and mission support systems. He holds a PhD in Remote Sensing from the University of Pisa, with a research focus on real-time data processing, 3D reconstruction, and situational awareness technologies for the defense and security sectors.

\textbf{Jose Luis Sanchez-Lopez} is a Permanent Research Scientist and Co-Head of the Automation and Robotics Research Group at the University of Luxembourg, holding a PhD in Robotics from the Technical University of Madrid. His research focuses on advancing situational awareness, task-level autonomy, and human-robot interaction for mobile ground and aerial platforms in complex environments.

\end{biogs}

\end{document}